\documentclass[useAMS,usenatbib]{mn2e}

\usepackage{subfigure}
\usepackage{graphicx}

\title[The mass ratio and formation mechanisms of Herbig Ae/Be star binary systems]{The mass ratio and formation mechanisms of Herbig Ae/Be star binary systems\thanks{Based on observations made with the William Herschel Telescope and the Isaac Newton Telescope operated on the island of La Palma by the Isaac Newton Group in the Spanish Observatorio del Roque de los Muchachos of the Instituto de Astrof\'{i}sica de Canarias.}}

\author[H. E. Wheelwright, R. D. Oudmaijer and S. P. Goodwin]{H. E. Wheelwright$^{1}$\thanks{E-mail:
pyhew@leeds.ac.uk}, R. D.
Oudmaijer$^{1}$ and S. P. Goodwin$^{2}$\\
$^{1}$School of Physics and Astronomy, EC Stoner Building, University of Leeds, Leeds, LS2 9JT, UK\\
$^{2}$Department of Physics and Astronomy, Hicks Building, University of Sheffield, Hounsfield Road, Sheffield, S3 7RH, UK\\}

\begin{document}

\date{Accepted. Received ; in original form: }

\pagerange{\pageref{firstpage}--\pageref{lastpage}} \pubyear{2009}

\maketitle

\label{firstpage}

\begin{abstract}

We present \textit{B} and \textit{R} band spectroastrometry of a
sample of 45 Herbig Ae/Be stars in order to study their binary
properties. All but one of the targets known to be binary systems with
a separation of $\rm \sim0.1-2.0$~arcsec are detected by a distinctive
spectroastrometric signature. Some objects in the sample exhibit
spectroastrometric features that do not appear attributable to a
binary system. We find that these may be due to light reflected from
dusty halos or material entrained in winds. We present 8 new binary
detections and 4 detections of an unknown component in previously
discovered binary systems. The data confirm previous reports that
Herbig Ae/Be stars have a high binary fraction, $\rm{74\pm6}$~per cent
in the sample presented here. We use a spectroastrometric
deconvolution technique to separate the spatially unresolved binary
spectra into the individual constituent spectra.  The separated
spectra allow us to ascertain the spectral type of the individual
binary components, which in turn allows the mass ratio of these
systems to be determined. In addition, we appraise the method used and
the effects of contaminant sources of flux. We find that the
distribution of system mass ratios is inconsistent with random pairing
from the Initial Mass Function, and that this appears robust despite a
detection bias. Instead, the mass ratio distribution is broadly
consistent with the scenario of binary formation via disk
fragmentation.

\end{abstract}


\section{Introduction}

Our understanding of the formation and early evolution of massive
stars ($\rm M_* \ga 8M_{\odot}$) is much less complete than in the
case of low mass stars. The scenario of low mass star formation has
been relatively well studied, and a broadly consistent
observational and theoretical picture has now emerged. The
various phases of low mass star formation include: cloud collapse,
proto-stellar creation and a subsequent contraction of Pre Main
Sequence (PMS) objects towards the Zero Age Main Sequence (ZAMS). This
later stage, the T Tauri phase, is easy to observe and therefore relatively
well understood \citep{Bouvier2007}. In the case of more massive stars
the situation is much less clear. Such stars do not experience an
optically visible PMS phase, evolve on a much more rapid timescale,
and are considerably more luminous than low mass stars. Early
studies on the effects of radiation pressure and the considerable
ionising output of massive young stars prompted speculation that
massive star formation might proceed in a different manner to that of
low mass stars \citep{Larson1971, Kahn1974}. For example, it has been
suggested that the most massive stars form via stellar mergers or
competitive accretion \citep{JBally2005}.

\smallskip

However, recent work, on both the observational and theoretical front,
suggests that massive star formation may not be dissimilar to low mass
star formation. As an example of observational results,
\citet{Pateletal2005} report the detection of a massive disk around a
15$\rm M_{\odot}$ protostar, indicating that massive stars may form
via monolithic accretion. On the theoretical front, recent work
indicates accretion onto a massive protostar is not impeded by
radiation pressure
\citep{YorkeandSonnhalter2002,Turner2007,Krumholz2009}. However, while
significant progress has been made, there remain many unaddressed
questions related to the formation and evolution of massive stars
\citep{ZinneckerandYorke2007}. As observations of massive young stars
are challenging, the full extent of the differences and similarities
between low and high mass star formation are still unknown.

\smallskip

Between the two extremes of mass lie the Herbig Ae/Be (HAe/Be) stars
\citep{Herbig1960}. These stars represent the most massive of objects
to experience an optically visible PMS evolutionary phase. Therefore,
HAe/Be stars offer an opportunity to study the early evolution of
stars more massive than the sun. Spectropolarimetry indicates that
Herbig Ae stars may undergo a PMS phase similar to that of the T Tauri
stars, while Herbig Be stars may evolve via disk accretion, rather
than magnetospheric accretion
\citep{JSVink2002,JSVink2005a,JCMottram2007}. Therefore, it appears
that a transition in formation mechanisms occurs across the HAe/Be
mass boundary \citep{JCMottram2007}. However, the critical mass has
not yet been established.

\smallskip

To examine the similarities and differences between low mass T Tauri
stars, HAe/Be stars and the optically invisible Massive Young Stellar
Objects (MYSOs), study of the circumstellar environment at small
angular scales is required. This is not trivial, requiring
observations with high angular resolution
\citep{Mannings1997,Fuente2006,Grady2007,Kraus2008}. Despite the
progress in the field, a full understanding of HAe/Be stars is
hampered by the small sample sizes involved. By way of contrast,
\citet{DB2006} utilised spectroastrometry to study a large sample of
HAe/Be stars with milli-arcsecond (mas) precision. Despite this
resolution \citet{DB2006} did not detect any accretion disks around
HAe/Be stars. However, they did find that the majority,
$\rm{68\pm11}$~per cent, of HAe/Be stars reside in relatively wide
(probably a few-hundred~au, see Section \ref{sep}) binary systems.

\smallskip

The binary fraction reported by \citet{DB2006} is greater than that of
T Tauri stars at similarly wide separations, which in turn is greater
than that of Main Sequence G-dwarfs at the same separations
\citep{DuquennoyandMayor1991,Reipurth1993,Ghez1993}. Indeed, this high
binary fraction is approaching that of more massive stars
\citep{Preibisch1999}. However, little is known about the properties
of such binary systems. The properties of the binary components and
configurations of such systems are of interest as they can constrain
the binary formation mechanism. The seminal study to date is that by
\citet{BC2001}, who used Adaptive Optics assisted observations to
construct Spectral Energy Distributions (SEDs) for each component in a
number of HAe/Be binary systems. The drawback of SED fitting is that
PMS stars, as young stars, are inevitably associated with dusty,
obscured environments. Therefore, the brightness ratio of a binary
determined by SED fitting can occasionally be ambiguous. However, very
few HAe/Be binary systems have been studied with spatially resolved
spectroscopy, and thus far such studies have been conducted with
seeing limited resolution \citep{ACarmona2007,Hubrig2007}.

\smallskip

The position angles of HAe/Be binary systems seem to be preferentially
aligned with the spectropolarimetrically detected circumprimary disks
\citep{DB2006}. This already places constraints on the formation modes
of these stars, in that it seems the systems formed via fragmentation
of a molecular core or disk. This had already been suggested for lower
mass binaries \citep{SWolf2001,Kroupa2001}, but little is known about
the formation mechanisms of more massive stars. This paper describes a
spectroastrometric follow-up of the work of \citet{DB2006} with
dedicated observations to study both components of binary systems. The
objective is to determine the properties of these binary systems and
thus place stronger, more quantitative, constraints on the formation
of stars of intermediate mass. We do this by determining the mass
ratio of these binary systems. This is done using a spectroastrometric
technique to disentangle the constituent spectra of unresolved binary
systems, allowing the spectral type, and hence mass, of each component
to be determined. Spectroastrometry itself is a relatively simple
technique that extracts the spatial information present in
conventional longslit spectra. Crucially, spectroastrometry can probe
changes in flux distributions with a typical precision of a mas or
less \citep{JBailey1998a}, which is required to study unresolved
binary systems. Typically the minimum separation probed is of the
order 100~mas, as the signature of a binary system is dependant upon
the system brightness ratio and separation.

\smallskip
  
This paper is structured as follows: in Section \ref{obsanddatred} we
present our sample selection, observation method and data reduction
procedures. In Section \ref{results_spec_ast} we discuss the
spectroastrometric signatures observed. In Section
\ref{spec_split_app} we present the method of splitting unresolved
binary spectra and in Section \ref{results_spec} we review the results
of separating binary spectra into their constituent spectra. In
Section \ref{discussion} we discuss our results. Finally, we conclude
this paper in Section \ref{conclusions} by summarising the salient
points raised.

\section[]{Observations and data reduction}
\label{obsanddatred}
\subsection{Observations}

The data presented consist of long-slit spectra in the \textit{B} band
($\rm 4200-5000 \rm{\AA}$) and/or the \textit{R} band ($\rm 6200-7000
\rm{\AA}$) of 45 HAe/Be stars, and 2 emission line objects which are
possible HAe/Be stars. The objects were chosen from the catalogs of
\citet{Theetal1994},\,\citet{Vieira2003}\,\&\,\citet{JHernandez2004},
and were selected to be reasonably bright (\textit{V} $\rm \leq $
12-13). Some objects previously observed by \citet{DB2006} were
observed to provide a consistency check on the spectroastrometric
signatures. Given the small population of HAe/Be stars, the objects
observed constitute a representative sample of HAe/Be stars, albeit
brightness limited.

\smallskip

The data were obtained using the 4.2m William Herschel Telescope (WHT)
and the 2.5m Isaac Newton Telescope (INT). At the WHT, data were
obtained on the 6th \& 7th of October 2006, using the Intermediate
Dispersion Spectrograph and Imaging System (ISIS)
spectrograph. Spectra of 20 objects were taken simultaneously in the
\textit{B} and \textit{R} bands using the dichroic slide of ISIS. In
most cases a slit $\rm{5}$~arcsec wide was used to ensure all the
light from a given binary system entered the slit, even in poor
seeing. This allows us to study the individual binary components,
unlike \citet{DB2006}, who used a slit of 1~arcsec. The R1200B and
R1200R gratings were used and the resulting spectral resolving power
was found to be $\rm \sim 3500$, corresponding to $\rm
85\, km\,s^{-1}$. The angular pixel size was $\rm{0.20}$ and
$\rm{0.22}$~arcsec in the \textit{B} band and \textit{R} band
respectively, which means that the spatial profile of the
longslit spectra was well sampled (average FWHM 1.9~arcsec). At the
INT data were obtained using the 235mm camera and the Intermediate
Dispersion Spectrograph (IDS). Observing was conducted from the 27th
of December 2008 to the 3rd of January 2009. The spectra of 32 objects
were obtained, despite adverse weather conditions preventing observing
for the better part of three nights. As at the WHT the slit width was
generally $\rm5$~arcsec. The R1200R and R1200B gratings were used and
the resulting spectral resolution was found to be $\rm \sim 3800 $,
or $\rm80 \,km\,s^{-1}$. The angular size of the pixels was
$\rm{0.4}$~arcsec, which fully sampled the average spatial profile of
the spectra (1.8 arcsec).

\smallskip

Multiple spectra were taken at four position angles (PA) on the
sky. The PAs selected always comprised of two perpendicular sets of
two anti-parallel angles, e.g. $\rm 0\degr$, $90^{\circ}$,
$180^{\circ}$ and $270^{\circ}$. Dispersion calibration arcs were made
using CuNe and CuAr lamps. Table \ref{logofobs} presents a summary of
the observations.

\begin{center}
\begin{table*}  
\begin{minipage}{\textwidth}
\caption{\label{logofobs}\small{Log of the observations, column 1
lists the objects observed, column 2 denotes the spectral type of the
objects, column 3 lists the \textit{V} band magnitudes of the sample,
and column 4 designates which telescope the object in question was
observed with. Columns 5 and 6 list the average seeing conditions,
columns 7 and 8 list the total exposure times and column 9 denotes the
slit width used. Column 10 lists the total Signal to
Noise Ratios, and finally, column 11 presents the date(s) each object
was observed. Information on the objects is taken from SIMBAD
(simbad.u-strasbg.fr) unless otherwise stated.}}
\begin{center}

\begin{tabular}{p{1.725cm} p{1.25cm} p{0.7cm} p{0.95cm} p{0.90cm}  p{0.90cm} p{0.550cm} p{0.550cm} p{1.1cm} p {1.35cm} p{3.15cm}}
\hline
Object & Spec type & \hspace*{2mm}\textit{V}  &Telescope & $\rm \overline{FWHM}^a$ & $\rm \overline{FWHM}^b$& $\rm{t_{blue}}$ &$\rm{t_{red}}$ & Slit & SNR & Date\\
    &  &    &      &(\arcsec)& (\arcsec)   & (s)  & (s) & (\arcsec) \\
\hline
\hline
VX Cas & $\rm{A0e}$  & 11.3 &  WHT & 1.3 &1.2  & 4800 & 4800  & 5.0&$\rm{600_B}$,$\rm{570_R}$ &07/10/2006 \\
VX Cas  & $\rm{A0e}$  & 11.3 &  INT & 1.1 & 1.7 & 2800 &  3600 & $\rm 2.5_{\rm B},5.0_{\rm R}$ & $\rm{370_B}$,$\rm{370_R}$&28/12/2008,31/12/2008\\
V594 Cas    & Be & 10.6 & INT &1.3 & -- & 3200 &-- &5.0 &610 &01/01/2009\\
V1185 Tau  & A1 & 10.7 & INT&  1.7 & --& 3200 &-- & 5.0 & 430&03/01/2009 \\
IP Per  &  $\rm{A3}$ & 10.3 & INT & 1.2 & 1.4 &2000 & 2400 &$\rm 2.5_{\rm B},5.0_{\rm R}$ & $\rm{110_B}$,$\rm{320_R}$&28/12/2008,31/12/2008\\
AB Aur & A0Vpe & 7.1  & WHT & 1.9 & 1.9 & 330  & 320  & 5.0  &$\rm{110_B}$,$\rm{650_R}$ &06/10/2006\\
MWC 480 &   $\rm{A3pshe}$& 7.7 &  WHT & 2.0 & 2.1 & 960  & 640 & 5.0 & $\rm{1100_B}$,$\rm{800_R}$&06/10/2006\\
UX Ori &  $\rm{A3e}$  & 9.6 &  WHT & 2.4 &2.4 & 3600 & 3600 & 5.0 & $\rm{1200_B}$,$\rm{940_R}$&06/10/2006\\
V1012 Ori   & $\rm{Be^c}$ &12.1  &INT &2.1& -- &  4800 & --& 5.0 &150 &02/01/2009\\
V1366 Ori    & A0e& 9.8& INT& 1.3 &--  & 2400 & --& 3.0 &570 &31/12/2008 \\
V346 Ori  &   $\rm{A5III}$&10.1 & INT & 1.5 & -- & 3600 &-- & 5.0 & 200&01/01/2009\\
HD 35929  & A5 & 8.1 &   WHT & 1.9  & 1.7 & 2060 & 1470 & 5.0 & $\rm{40_B}$,$\rm{900_R}$&07/10/2006\\ 
V380 Ori &  A0& 10.7 & INT  & 1.5 &1.6 & 3600 & 2940 & $\rm 3.0_{\rm B},5.0_{\rm R}$& $\rm{100_B}$,$\rm{200_R}$&28/12/2008,31/12/2008\\
MWC 758 &   A3e & 8.3 &  WHT & 1.4 & 1.3&1080 & 960& 5.0  &$\rm{50_B}$,$\rm{660_R}$ &07/10/2006\\
HK Ori  & A4pev &11.9 & INT & 2.4 & -- & 4800 & --& 5.0 & 200&02/01/2009\\
HD 244604  & A3 & 9.4 &  WHT & 1.7  & 1.6 & 3180 & 3660 & 5.0 &$\rm{100_B}$,$\rm{720_R}$ & 07/10/2006\\
V1271 Ori  &  A5 & 10.0  & INT &1.6&-- & 2460 & --& 5.0 & 410& 01/01/2009\\
T Ori      & A3 & 9.5 & INT &1.9 & -- & 3200& -- & 5.0 & 300 &03/01/2009 \\
V586 Ori & $\rm{A2V}$ &9.8 & INT & 3.3 & -- & 2940  & --& 5.0 & 650&02/01/2009\\
HD 37357 & A0e &8.8 & INT &  1.4 & 1.4 &2060 & 1470& $\rm 3.0_{\rm B},5.0_{\rm R}$ & $\rm{200_B}$,$\rm{350_R}$ & 28/12/2008,31/12/2008\\
V1788 Ori  &  B9Ve & 9.9 & INT& 1.7 &-- & 1350  &-- & 5.0 & 450 &01/01/2009\\
HD 245906   & B9IV & 10.7 & INT & 1.8 & --  &2800  & --& 5.0  &100 &03/01/2009\\
RR Tau       & A2II-IIIe & 10.9& INT & 1.7 & -- &2800 &-- & 5.0 & 250&03/01/2009 \\
V350 Ori     & A0e & 10.4(\textit{B})& INT &  1.9 & -- &4800 &-- & 5.0 & 130& 03/01/2009 \\
MWC 120 &  A0 & 7.9 &  WHT & 2.1  & 1.9 & 480 & 480 &5.0 &$\rm{1500_B}$,$\rm{690_R}$ & 06/10/2006\\
MWC 120  &  A0 & 7.9 &  INT & 1.4& 1.6  & 2460 & 1250 & $\rm 3.0_{\rm B},5.0_{\rm R}$ & $\rm{1200_B}$,$\rm{560_R}$&28/12/2008,31/12/2008\\
MWC 790     & Be & 12.0 & INT & 3.1 & --  & 4050  &-- & 5.0 &200 & 02/01/2009 \\
MWC 137 & Be & 11.2& INT& -- &  2.1 &-- & 4560& 5.0 & 200 & 28/12/2008 \\
HD 45677 &  $\rm{Bpshe}$ & 8.0  & WHT &  2.0  & 1.9 &360  &240  &5.0 & $\rm{900_B}$,$\rm{550_R}$& 06/10/2006\\
LkH$\rm \alpha$ 215 &   B7.5e  & 10.6 & INT & 1.5 & 2.3 &3600 & 3600& $\rm 5.0_{\rm B},4.0_{\rm R}$ & $\rm{300_B}$,$\rm{360_R}$ & 27/12/2008,31/12/2008\\
MWC 147 &   $\rm{B6pe}$ & 8.8 & WHT & 1.8 & 1.5 & 3000 & 1700 & 5.0 & $\rm{1200_B}$,$\rm{500_R}$& 07/10/2006\\
MWC 147   &  $\rm{B6pe}$ & 8.8 & INT & 1.3 & --  & 2400 & --& 5.0 & 620& 01/01/2009\\
R Mon &  B0 & 10.4& INT & 4.2 & 2.5  & 3600 & 3600 & 5.0& $\rm{100_B}$,$\rm{240_R}$& 28/12/2008,02/01/2009\\
V590 Mon  &  B8pe & 12.9  & INT & 1.5 & -- & 2670 & --& 5.0 & 200 & 01/01/2009\\
V742 Mon  & B2Ve & 6.9 & INT & 1.4  & 2.8& 1740 & 2535 & 5.0 & $\rm{400_B}$,$\rm{800_R}$ & 30/12/2008,31/12/2008\\
OY Gem       & Bp[e] & 11.1 & INT &1.8 & -- & 2880 & --& 5. 0 & 100 & 03/01/2009\\
GU CMa&  $\rm{B2Vne}$ & 6.6  & WHT & 2.5  & 2.4 & 360 & 360 & 5.0 & $\rm{1500_B}$,$\rm{900_R}$&06/10/2006 \\
GU CMa  &  $\rm{B2Vne}$ & 6.6  & INT &1.8 & -- & 720 &--& 5.0 & 1400 &03/01/2009 \\
MWC 166 &  B0IVe & 7.0  & WHT &  2.5 & 2.3& 210& 120  &5.0 & $\rm{1200_B}$,$\rm{800_R}$& 06/10/2006\\
HD 76868 & B5  & 8.0 & INT &1.5 & 2.4 & 4830 & 2100& 5.0 & $\rm{100_B}$,$\rm{100_R}$ & 30/12/2008,01/01/2009\\                    
 HD 81357   & B8 & 8.4 & INT & 1.7 & --  & 4800 & -- & 5.0& 100 &03/01/2009\\ 
MWC 297 & $\rm{Be}$  & 12.3 & WHT &  1.4 &  1.2 &4100 & 3120 &5.0 & $\rm{100_B}$,$\rm{500_R}$&07/10/2006\\
HD 179218 &  B9e  & 7.2  & WHT & 2.6 & 2.4 &  2100& 1200 &1.0/1.5 & $\rm{2300_B}$,$\rm{940_R}$&06/10/2006\\
HD 190073 & $\rm{A2IVpe}$ & 7.8  & WHT & 1.5 & 1.2 &540  & 360&5.0 &$\rm{600_B}$,$\rm{800_R}$ &07/10/2006\\
BD +40 4124&B2 & 10.7  & WHT & 2.1 & 1.9 &  600 &  660 & 4.0 & $\rm{500_B}$,$\rm{370_R}$ &07/10/2006\\
MWC 361  &  $\rm{B2Ve}$ & 7.4  & WHT & 1.7  & 1.7 & 1350 & 960 &2.5/4.0 & $\rm{1400_B}$,$\rm{1400_R}$ &06/10/2006\\
SV Cep  &  $\rm{Ae}$& 10.1(\textit{B}) & INT & 1.6& --& 3000& -- & 5.0 & 600 &02/01/2009\\
MWC 655     &  B1IVnep & 9.2 & INT & 1.7 & --  & 2400 & -- & 5.0 & 400&03/01/2009\\
Il Cep  &  $\rm{B2IV/Ve}$  & 9.3  & WHT & 1.4 & 1.2 &3500 & 3000&5.0& $\rm{800_B}$,$\rm{500_R}$&07/10/2006\\
BHJ 71&B4e & 10.9 &  WHT &  1.8  & 1.8 & 1200 & 1080 &4.0 & $\rm{500_B}$,$\rm{340_R}$ &06/10/2006\\
BHJ 71  & B4e & 10.9 &  INT &  1.7 & -- &4200 & --& 5.0 & 500 &01/01/2009\\
MWC 1080 &  $\rm{B0}$ & 11.6 &  WHT & 2.0 & 2.0 &  3300 & 4170 & 5.0 &$\rm{200_B}$,$\rm{500_R}$& 06/10/2006\\
\hline

 \end{tabular}

\end{center}
\renewcommand\footnoterule{}
\footnotetext{{$\rm^a$ Average seeing in the blue spectral region, approximated by the average of the individual median FWHM, where necessary averaged\\ \hspace*{3.4mm} over multiple slit widths.}}
\footnotetext{{$\rm^b$ Average seeing in the red spectral region, approximated by the average of the individual median FWHM, where necessary averaged \\ \hspace*{3.4mm}over multiple slit widths.}}
\footnotetext{{$\rm^c$ \citet{Theetal1994}}}
\end{minipage}
 \end{table*}
\end{center}

\subsection{Data reduction}
\label{data_red}

Data reduction was conducted using the Image Reduction and Analysis
Facility (IRAF)\footnote{IRAF is distributed by the National Optical
Astronomy Observatories, which are operated by the Association of
Universities for Research in Astronomy, Inc., under cooperative
agreement with the National Science Foundation \citep{IRAF}.} and
routines written in Interactive Data Language (IDL). Initial data
reduction consisted of bias subtraction and flat field division. The
total intensity spectra were then extracted from the corrected data in
a standard fashion. Wavelength calibration was conducted using the arc
spectra, and the wavelength calibration solution had a precision of
the order $\rm{<0.1\AA}$.

\smallskip

Spectroastrometry was performed by fitting Gaussian functions to the
spatial profile of the long-slit spectra at each dispersion
pixel. This resulted in a positional spectrum, the centroid of the
Gaussian as a function of wavelength, and a Full Width at Half Maximum
(FWHM) spectrum, the FWHM as a function of wavelength. Spot checks
were used to ensure that a Gaussian was an accurate representation of
the data. The continuum position exhibited a general trend across the
CCD chip: of the order of 10 pixels in the case of the ISIS data and 2
pixels in the IDS data. This was removed by fitting a low order
polynomial ($\rm{4^{th}}$ or $\rm{5^{th}}$ order) to the continuum
regions of the spectrum.

\smallskip

All intensity, positional and FWHM spectra at a given PA were combined
to make an average spectrum for each PA. A correction for slight
changes in the dispersion across PAs was determined by
cross-correlating average intensity spectra obtained at different
PAs. The correction was then applied to the average intensity,
positional and FWHM spectra.  The average positional spectra for
anti-parallel PAs were then combined to form the average,
perpendicular, position spectra, for example: ($\rm 0^{\circ} -
180^{\circ}$)/2 and ($\rm 90^{\circ} - 270^{\circ}$)/2. This procedure
eliminates instrumental artifacts as real signatures rotate by $\rm
180^{\circ}$ when viewed at the anti-parallel PA, while artifacts
remain at a constant orientation. In addition, all positional spectra
were visually inspected for artifacts not fully removed by this
procedure. As with the positional spectra, the FWHM spectra at
anti-parallel PAs were also combined to make two averaged,
perpendicular spectra. While FWHM features do not rotate across
different PAs, the features observed at anti-parallel PAs were used to
exclude artifacts via a visual comparison. All conditions being
constant, a real FWHM signature should not change from one PA to the
opposite angle at $\rm +180^{\circ}$.

\section{Spectroastrometric signatures}
\label{results_spec_ast}

\subsection{Binary spectroastrometric signatures over H\,{\sc i} lines}

An unresolved binary system, in which each component has a unique
spectrum, displays a clear signature in the behaviour of the spectral
photocentre. As the spatial profile is the sum of the two stars
convolved with the seeing, the peak is not located at the position of
either star, but somewhere between the components. The exact location
of the photo-centre depends on the intensity ratio and the separation
of the two components. Over spectral lines the binary flux ratio
changes from its continuum value, which results in the peak position
shifting towards the dominant component. Therefore, unresolved binary
systems are revealed by a displacement in the positional spectrum over
spectral line. In addition, an unresolved binary system is also
revealed by a change in the FWHM over lines in the spectrum. Again,
this is because the spatial profile of an unresolved binary system is
dependent upon the binary flux ratio, which changes from its continuum
value across certain lines. As the error in the centre of the Gaussian
profile is governed by photon statistics, changes of mas scales can be
traced. This allows binary systems with separations as small as
$\rm{\sim0.1}$~arcsec and differences in brightness up to 5 magnitudes
to be studied \citep{DB2006}.

\smallskip

To illustrate the detection of a binary system the
observations of GU CMa are presented in some detail. GU CMa is known
to be a Herbig Be binary system with a separation of
$\rm\sim0.65$~arcsec, a PA of $\rm\sim195^{\circ}$, a
brightness difference between components of 0.7-1.0 magnitudes in the
optical band and a primary with a spectral type of B1
\citep{FUetal1997,Fabriciusetal2000,BC2001}.

\smallskip

\begin{center}
\begin{figure*}
\begin{center}
{\includegraphics[width=120mm,height=90mm]{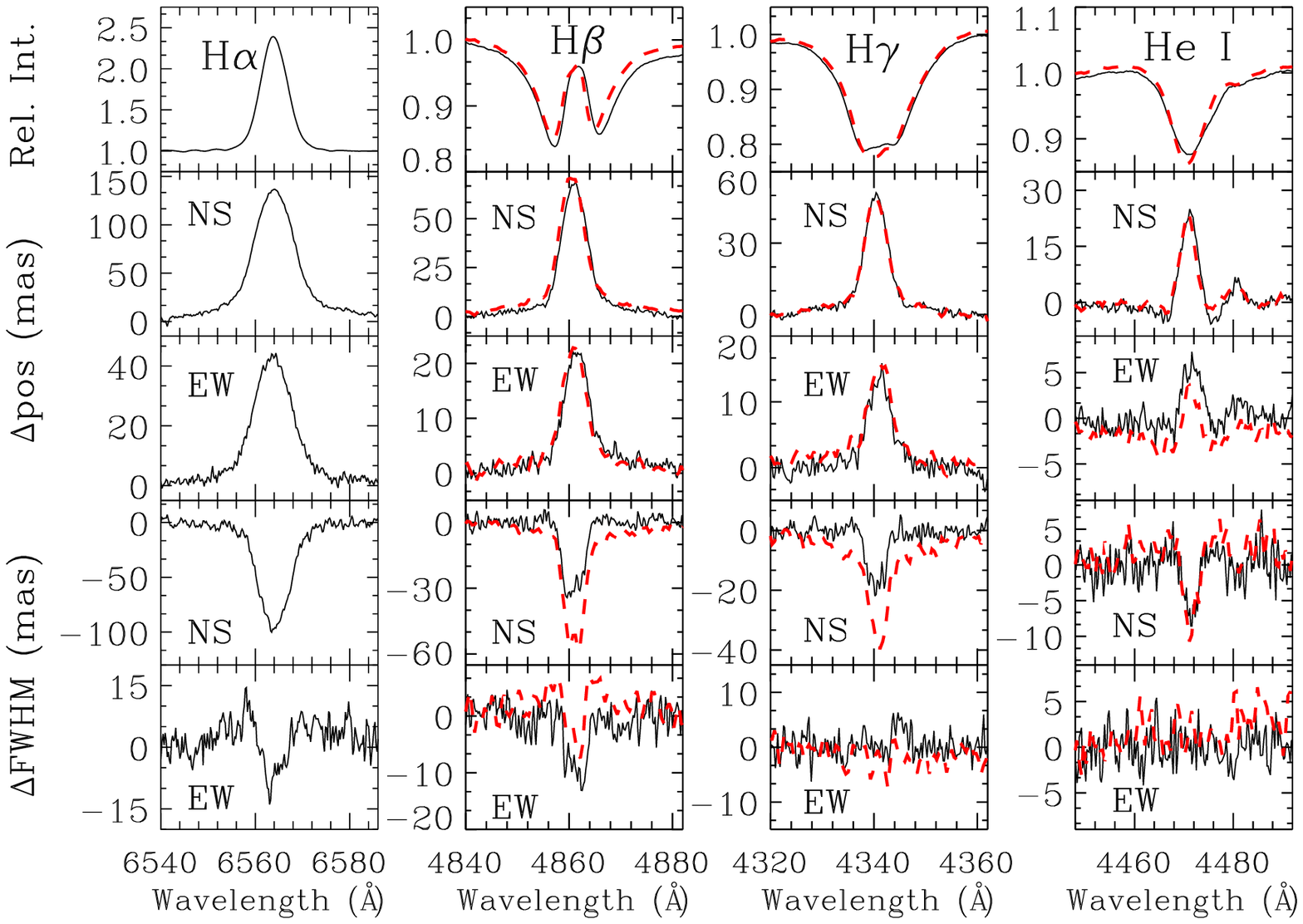}} 
\begin{tabular}{l r}
{\includegraphics[width=60mm,height=60mm]{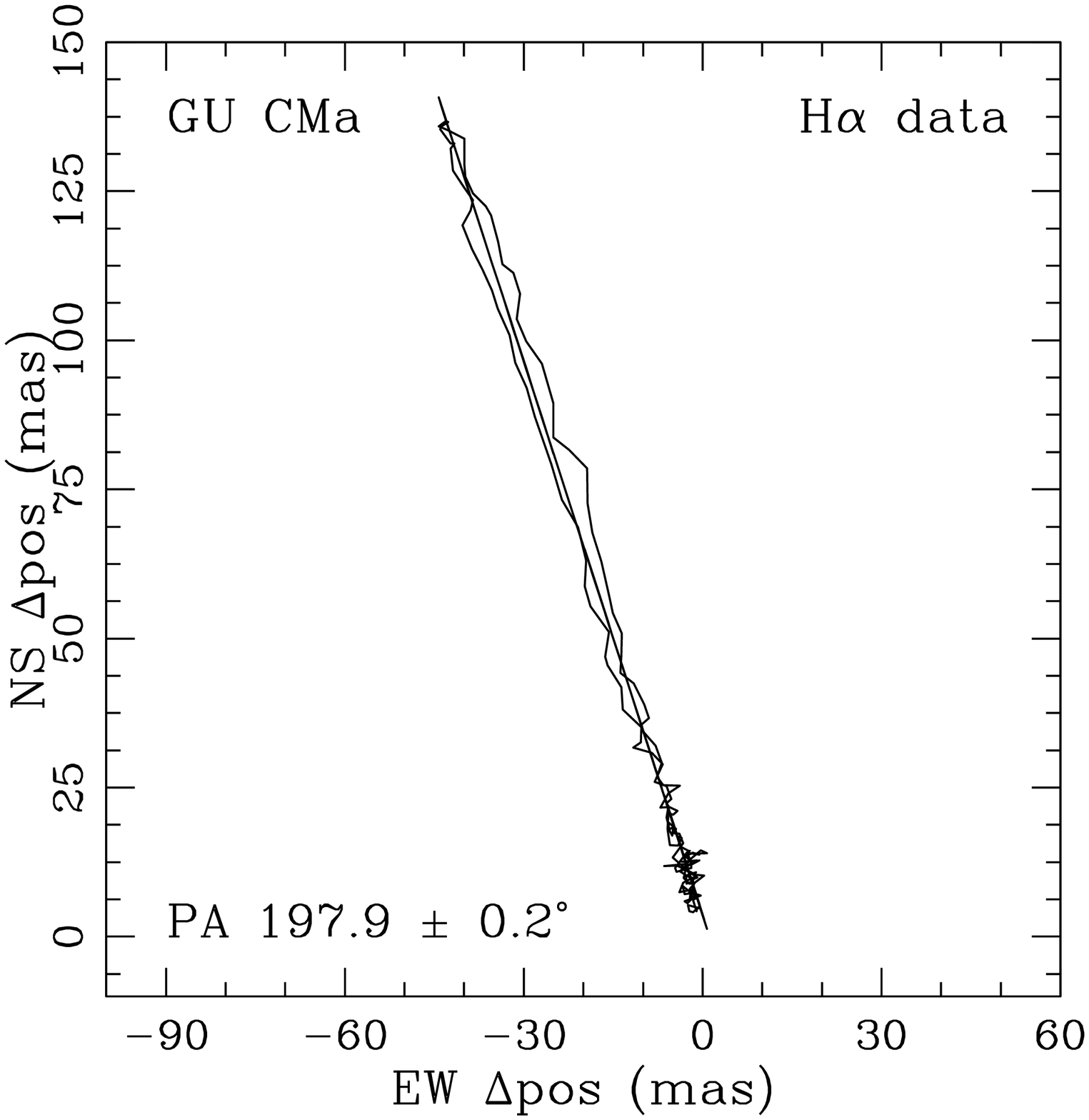}} &
{\includegraphics[width=60mm,height=60mm]{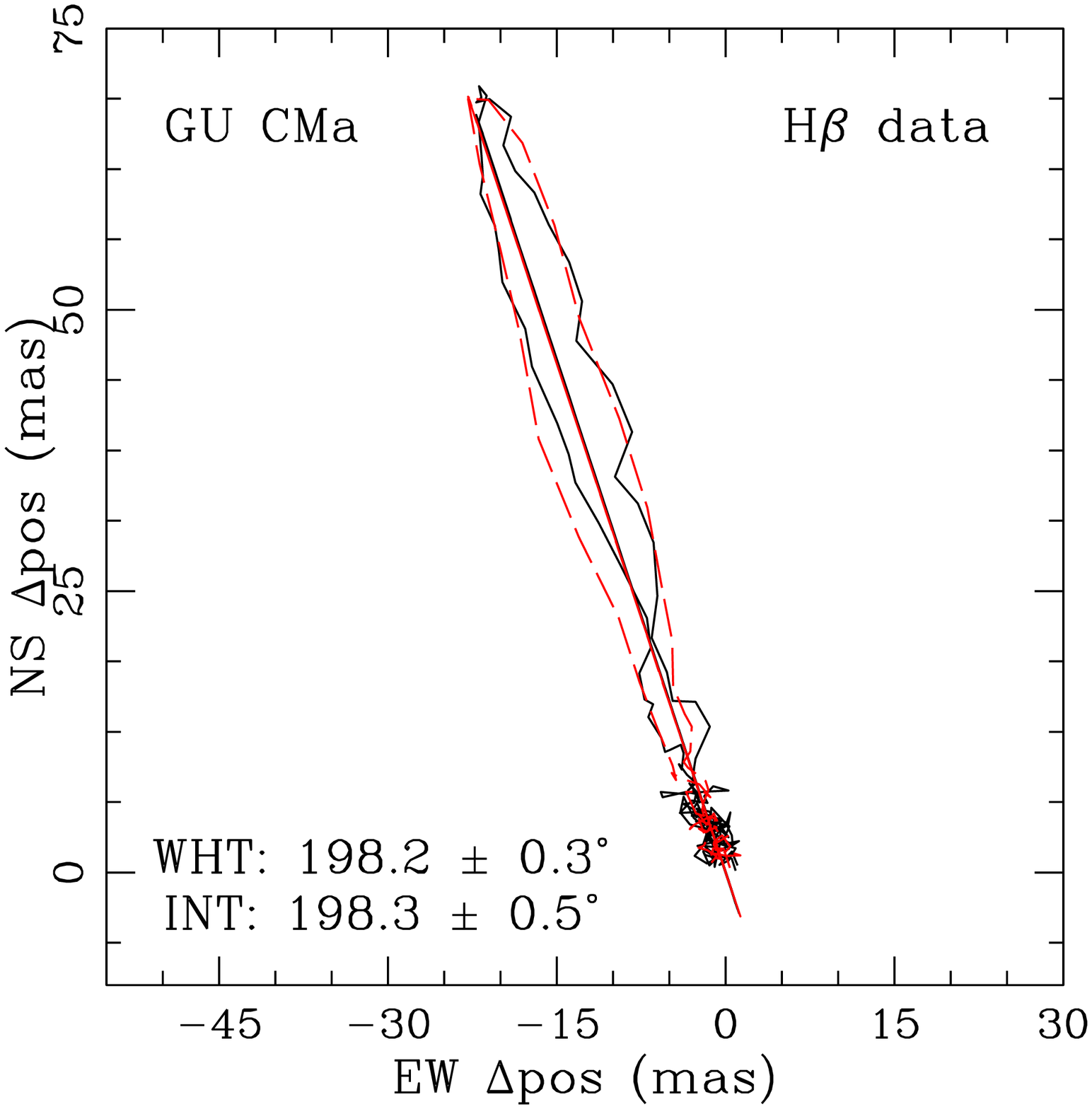}} \\
{\includegraphics[width=60mm,height=60mm]{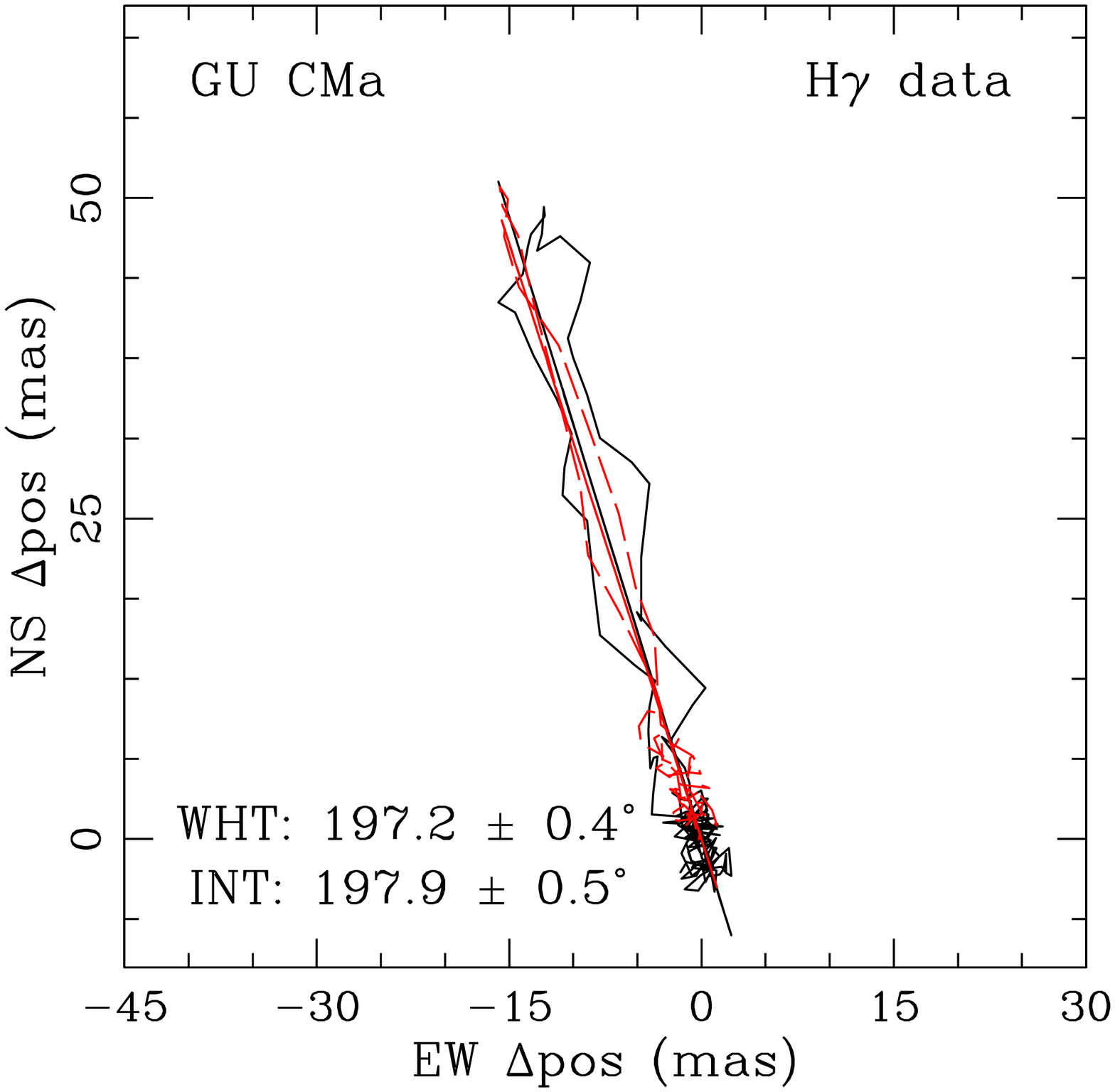}} &
{\includegraphics[width=60mm,height=60mm]{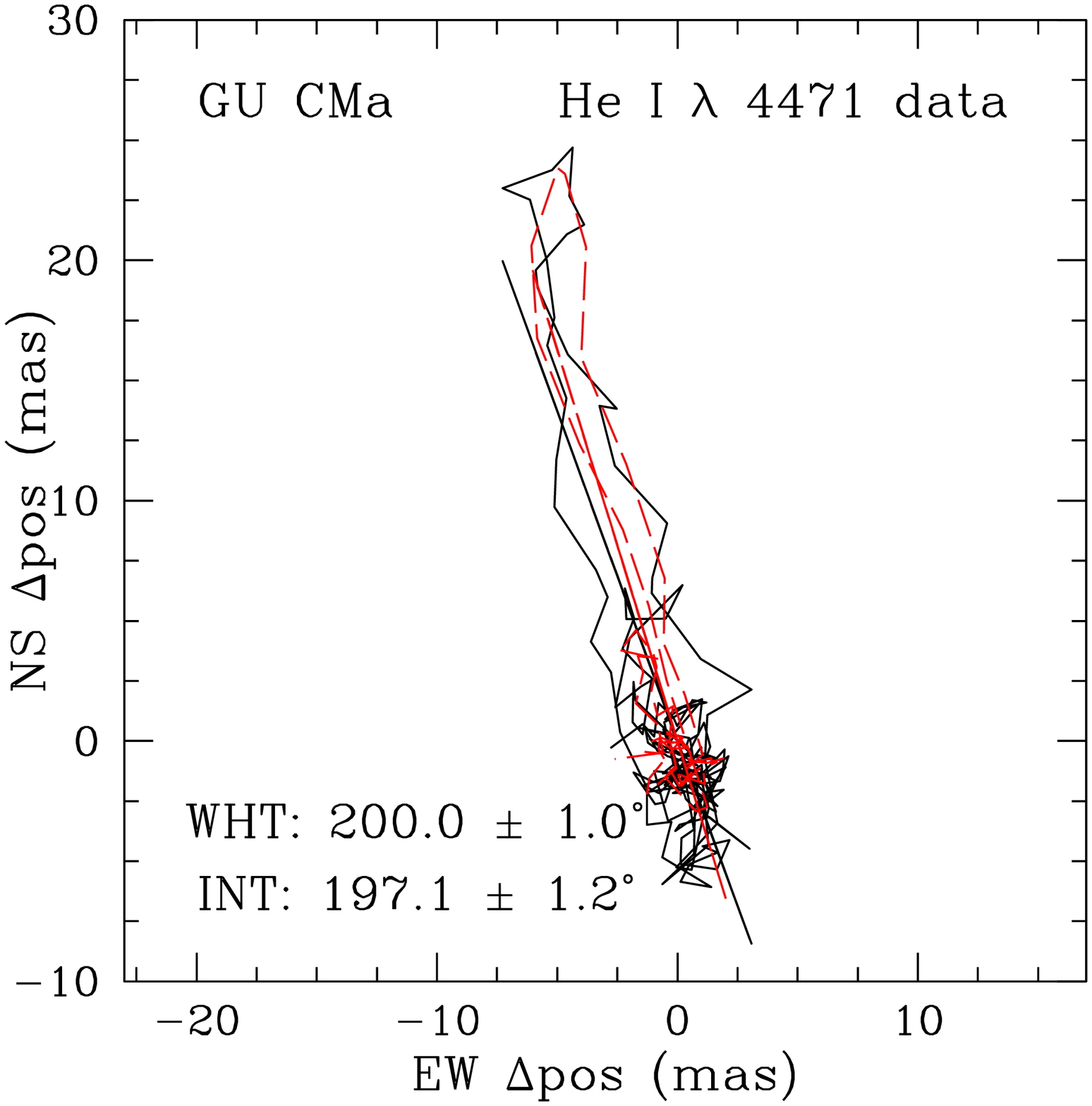}} \\
\end{tabular}

\caption{The spectroastrometric signature of GU CMa. In the
  \textit{top} panel, presented from \textit{left} to \textit{right}:
  the H$\rm \alpha$, H$\rm \beta$, H$\rm \gamma$ and He\,{\sc i}
  $\rm \lambda 4471$ spectral profiles and associated
  spectroastrometric signatures. In the spectroastrometric signatures
  North and East are positive. In the \textit{lower} section of the
  figure we present the associated XY-plots of the H$\rm \alpha$,
  H$\rm \beta$, H$\gamma$, and by way of contrast, the He\,{\sc i}
  $\rm \lambda$ 4471 spectroastrometric displacements. In the
  XY-plots North is \textit{up} and East is to the \textit{left}. The
  data from the WHT is represented by the \textit{solid} lines, while
  the data obtained at the INT is represented by the \textit{dashed}
  lines. Note the consistency between the two datasets.}
\label{gucma_spec_ast}
\end{center}
\end{figure*}
\end{center}

GU CMa presents a very clear binary signature in the
spectroastrometric observations (Fig. \ref{gucma_spec_ast}). Across
the H\,{\sc i} lines the photo-centre of the spectrum clearly shifts
towards the North-East. This demonstrates that the primary, the
component brightest in the continuum, dominates the emission
spectrum. It also indicates that the secondary, the component least
bright in the continuum, has the larger absorption profile of H$\rm
\gamma$. As the photo-centre shifts to the North-East the FWHM of the
spectrum is seen to decrease. This also indicates that the primary
dominates the spectrum at these particular wavelengths. The
photo-centre is also observed to shift to the North-East across the He
\,{\sc i} lines. This again indicates that it is the primary that
dominates the binary flux over this line.

All the spectroastrometric excursions across the H\,{\sc i} lines
produce a PA consistent to within $\rm\sim1^{\circ}$. The agreement
between the spectroastrometric displacements across different lines is
an important consistency check. In addition, the close agreement
between the data gathered at different telescopes provides compelling
evidence that the signatures observed are real and not contaminated by
instrumental affects. The difference in the FWHM changes can be
explained by the difference in the seeing between
observations. Finally, that our observations concur with the results
of \citet{DB2006}, further proves that these signatures are real and
no instrumental effect. As demonstrated by the spectroastrometric
signature associated with the He\,{\sc{i}} line such features do not
solely occur across lines with an emission component. This is
important to note as it means the spectra splitting method does not
require emission lines to separate spectra.

\smallskip

In the case of some stars significant FWHM changes are observed which
are not accompanied by a change in the spectral
photo-centre. \citet{DB2006} regard such a signature as a possible
binary detection. This is substantiated as \citet{DB2006} demonstrate
that the spectroastrometric signature of a binary system with a
separation of greater than half the slit width exhibits larger FWHM
than positional features. However, in the data presented here, this
scenario is unlikely. As a wide slit was used, a binary with a
separation of half the slit width would be resolved, even in seeing
conditions of 2 arcsec. As a resolved system does not exhibit a
spectroastrometric signature we suspect there may be an alternative
explanation to the FWHM features not accompanied by positional
features. We note that such features are not instrumental as some
stars exhibit no change in FWHM over spectral lines. In many cases the
large FWHM features occur over absorption features in the emission
profiles. This suggests that these features may trace an extended
structure which scatters the line profile, rather than being an
intrinsic source. If the scattering media were close to being
symmetrically distributed around the central star it could generate a
large FWHM increase while not resulting in a positional
signature. Such sources of flux could be a disk/stellar wind
\citep{Azevedo2007}, the halos reported by \citet{Leinert2001} and
\citet{Monnieretal2006}, or nebulosity.  This topic will be returned to
in Section \ref{halos}.

\smallskip

The observational results naturally fall into three categories: clear
binary signatures, possible binary/other signatures and null
detections. We present a summary of the detections in our results in
Table \ref{binaries}, in which we separate known binary systems and
new spectroastrometric detections. It is important to note that it is
not only binaries that are detected by spectroastrometry, optical
outflows and disks can also result in a spectroastrometric feature
\citep{JBailey1998b,MTakami2001}. However, there are no disk
signatures, and only a few detections of outflows, in the data
presented here.

\smallskip

There are 29 stars in our sample that are referred to in the
literature as being part of a binary system. However, we exclude AB
Aur and HD 244604 as the binary nature of these objects is open to
question, section \ref{halos} explains why this is the case. We detect
20 of the 27 previously known binary systems. As six of the seven
undetected systems have separations greater than $\rm{2}$~arcsec
and/or brightness differences as great as 8~magnitudes, we detect all
but one, UX Ori, of the binary systems that we would expect to
detect. Given that the majority of known binary systems are detected,
objects which are not known to be part of a binary system but exhibit
similar spectroastrometric signatures to the known binary systems are
classified as new binary detections. We detect 8 new binary
systems. The raw binary fraction of the sample is 0.60. Including the
non detections of known binary systems the binary fraction of the
sample is 0.74. While these figures are high for a limited separation
range, they are consistent with previous work
\citep{NPirzkal1997,DB2006}.

\smallskip

As expected, we do not detect the binaries with separations greater
than $\rm{\sim2.0}$~arcsec and differences in brightness greater than
$\rm{5}$~magnitudes, e.g. HD 179218 and MWC 297. We note that the wide
companions are not detected in the longslit spectra as distinct
sources. VX Cas, T Ori, LkH$\rm \alpha$ 215 and Il Cep are all known
to be binary systems and all display a binary signature in their
spectroastrometric signatures. Therefore, these stars are classified
above as detections of known binary systems. However, these systems
are wide binaries with separations greater than 5~arcsec. These
companions are clearly resolvable, and thus the spectroastrometric
signatures we observe can not be due to the previously reported
companion. Therefore, we suggest we have detected previously unknown
companions to VX Cas, T Ori, LkH$\rm{\alpha}$ 215 and Il Cep.

\begin{center}
\begin{table*}
\begin{minipage}{\textwidth}
\caption{Previously known binary systems and new detections. Column 1
list the objects in question, columns 2 and 3 list the separation and
PA of the known binaries, taken from the literature. Column 4 contains
difference in brightness between the two binary components. The
amalgamation of data is not complete, if more than one value of binary
parameter is available in the literature, only one is presented for
the sake of clarity.\label{binaries}}
\begin{center}

\begin{tabular}{p{1.75cm} p{5cm} p{5cm} p{3.50cm}}

\hline
Object & Separation & PA& $\rm \Delta flux$ \\
 &  (mas) & ($\rm{^{\circ}}$) & (magnitudes) \\
\hline
\hline

 \underline{\bf{Known binaries detected:}} & & \\
VX Cas & $\rm{5340^{A}}$& $\rm{{165.3}^{A}}$ & \textit{K:}$\rm{4.8}^{A}$\\
V380 Ori &   $\rm{125\pm25^{B}}$&  $\rm{224.0\pm2.0^{B}}$& \textit{K}:$\rm{1.42^{B}}$\\
HK Ori & $\rm{347.7\pm2.5^{B}}$ &  $\rm{41.8\pm0.7^{B}}$& \textit{V}: $\rm{0.87^{B}}$\\
T Ori &   $\rm{7700\pm200^{C}}$ and $\rm{{spectroscopic}^{D}}$& $\rm{72.6^{C}}$&\textit{K}:$\rm{> 4.5^{C}}$\\
V586 Ori & $\rm{990^{A}}$& $\rm{30.3^{A}}$&\textit{K}:$\rm{2.8^{A}}$\\
HD 37357 &  $\rm{186^{E}}$&  $\rm{49.0^{E}}$&\textit{K}:$\rm{1.7^{A}}$\\
V1788 Ori & $\rm{520^{A}}$& $\rm{352.9^{A}}$& \textit{K}:$\rm{3.5^{A}}$\\
HD 245906 & $\rm{130^{A}}$& $\rm{77.1^{A}}$&\textit{K}:$\rm{1.5^{A}}$ \\
V350 Ori & $\rm{290^{A}}$& $\rm{206.8^{A}}$&\textit{K}:$\rm{3.2^{A}}$\\
HD 45677 & & $\rm{150 \pm 17^{F}}$ & \\
LkH$\rm \alpha$ 215 & $\rm{8500^{A}}$ & $\rm{226.6^{A}}$ & \textit{K:}$\rm{4.8^{A}}$ \\
MWC 147 & $\rm{150^{A}}$ & $\rm{55.6^{A}}$& \textit{K}:$\rm{3.8^{A}}$ \\
R Mon &  $\rm{670^{G}}$& $\rm{290.7^{G}}$& \textit{K}:$\rm{{4.9^{A}}}$\\
GU CMa & $\rm{654^{H}}$ &$\rm{194.5^{H}}$ & \textit{V}:$\rm{0.95\pm0.02^{H}}$\\
MWC 166 & $\rm{654^{H}}$ & $\rm{297.8^{H}}$  & \textit{V}:$\rm{1.41^{H}}$ \\
BD +40 4124 & $\rm{720^{A}}$& $\rm{175.1^{A}}$& \textit{K}:$\rm{5.4^{A}}$\\
MWC 361 & $ \rm{2250\pm240^{I}}$ & $\rm{164.0 \pm 1.0^{I}}$ & \textit{K}:$\rm{{4.9}^{I}}$ \\
$\rm{SV\, Cep^{J}}$ & & & \\
Il Cep & $\rm{6960^{I}}$ & $\rm{147.0^{I}}$ & \textit{K}:$\rm{ 0.0^{I}}$ \\
MWC 1080 &  $\rm{760 \pm 2^C}$ &  $\rm{267.0 \pm 1.0^C}$ & \textit{K}:$\rm{3.25 \pm 0.08^4}$ \\
 \underline{\bf{Known binaries not detected:}} & & \\

UX Ori & $\rm{22(min)^{K}}$ & $\rm{257.4 \pm 18.4^{K}}$ & \\
MWC 758 & $\rm{2280^{A}}$& $\rm{311.3^{A}}$&\textit{K}:$\rm{8.3^{A}}$\\
V1271 Ori &  $\rm{8380^{A}}$  & $\rm{294.7^{A}}$  &  \textit{K}$\rm{6.7^{A}}$ \\
V590 Mon & $\rm{5007^{A}}$ &  $\rm{97.1^{A}}$ &  \textit{K}$\rm{6.6^{A}}$ \\ 
MWC 297 & $\rm{3930 \pm 200^{L}}$ & $\rm{313 \pm 2^{L}}$ & \textit{H}:$\rm{{8.5 \pm 0.25}^{L}}$\\
HD 179218 & $\rm{2540^{A}}$& $\rm{140.5^{A}}$& \textit{K}:$\rm{6.6^{A}}$\\
BHJ 71 &$\rm{6170^{A}}$ &$\rm{29.2^{A}}$ &  \textit{K}$\rm{8.3}^{A}$\\
\underline{\bf{New spectroastrometric detections:}} & &\\
\multicolumn{4}{l}{V1366 Ori, HD 35929, RR Tau, MWC 120, V742 Mon, OY Gem, HD 76868 and HD 81357}\\

\hline
\hline
\end{tabular}

\end{center}
\renewcommand{\footnoterule}{}

\begin{center}
\begin{tiny}
\vspace*{-7mm}
\footnotetext{\hspace*{2.0mm}\underline{\textbf{References:}} A) \citet{Thomas2007}; B) \citet{KWSmithetal2005}; C) \citet{CLeinert1997}; D) \citet{Shevchenko1994};\\\hspace*{21.70mm}E) \citet{Hart1996}; F) \citet{DB2006}; G) \citet{Weigelt2002}; H) \citet{Fabriciusetal2000};\\\hspace*{22.0mm}I) \citet{NPirzkal1997}; J) Rodgers et al. priv. com. (2008); K) \citet{CBertout1999}; L) \citet{JSVink2005b}.}
\end{tiny}
\end{center}
\end{minipage}
\end{table*}
\end{center}

Table \ref{halphabeta} summarises the spectroastrometric signatures
over H$\rm \alpha$ and H$\mathrm{\beta}$. We note that similar
behaviour was observed across H$\rm \gamma$ and other lines, but to
keep this paper concise we only present the H$\rm \alpha$ and
H$\mathrm{\beta}$ signatures. Furthermore, the spectroastrometric
signatures of the entire sample over either the H$\rm{\alpha}$ or
H$\rm \beta$ lines are presented in Appendix
\ref{appendix_one}. Spectral variability is a common behaviour of
HAe/Be stars \citep{Rodgers2002,Mora2004}. In the case of the few
objects observed twice, some line profile variations are
seen. However, the spectroastrometric signatures of objects observed
twice are generally consistent, e.g. the example of GU CMa (Fig
\ref{gucma_spec_ast}). In addition, in the case of objects common to
the this sample and that of \citet{DB2006}, the spectroastrometric
signatures presented here are consistent with the previous
results. Therefore, we conclude that spectral variability, on
timescales of years, does not effect the spectroastrometric signatures
observed. No line profile variability on timescales of minutes is
observed.

\smallskip

To summarise, spectroastrometry is a powerful tool with which to study
binary systems, as GU CMa demonstrates. Not only do we clearly detect
a $\rm 0.6$~arcsec binary in seeing as large as $\rm 2.5$~arcsec, we also trace
the PA of the system with a precision of $\rm 1^{\circ}$ or
less. Indeed, spectroastrometry detects all but one of the known
binary systems with separations less than $\rm{\sim2}$~arcsec and
differences in brightness of less than 5~magnitudes. In addition, the
PAs of these systems are all traced with a precision of the order of
$\rm{1^{\circ}}$, and are generally consistent with literature
values to within $\rm{\sim5^{\circ}}$. Most importantly, the
spectroastrometric displacements contain information as to which
component of the binary system dominates the flux over certain
spectral features. This information can be used to separate the
constituent spectra.

\subsection{Artifacts}

\label{artifacts}

Several spectroastrometric signatures presented in Table
\ref{halphabeta} are referred to as artifacts. An artifact is defined
as a spectroastrometric signature which does not rotate by
$\rm{180^{\circ}}$ when viewed at two anti-parallel position
angles. Artifacts in spectroastrometric data can arise from a number
of sources. Instrumental effects include: the misalignment of the
dispersion axis with the CCD columns, a change in focus along the
slit, curvature of the spectrum and any departure of the CCD array
from a regular grid \citep{JBailey1998b}. As discussed in Section
\ref{data_red}, such artifacts are readily identified and negated. The
observation of unresolved lines can also result in false signatures
\citep{JBailey1998a}. Here, the data were obtained with a resolution
sufficient to resolve most spectral lines. However, the narrow
absorption troughs in many of the double-peaked H\,{\sc i} emission
profiles may cause some of the artifacts as these features are often
barely resolved.

\smallskip

In addition, \citet{EBrannigan2006} report an artifact that
is a consequence of image distortion, regardless of whether
the spectral lines are well resolved. An offset between the image
centre and the centre of the slit results in a change in the angle of
incidence of light onto the grating, and thus a slightly blue or red
shifted image. This causes a wavelength dependant change in position,
and therefore an artificial spectroastrometric signature. As we use a
wide slit of 5~arcsec this effect is likely to be origin of many of
the artifacts present in our data.

\smallskip

The empirical finding is that it is crucial to obtain multiple
spectra, comprising of anti-parallel sets of data. Such data will
identify artifacts regardless of their cause, and can also be used to
remove systematic effects.

\section{Splitting binary spectra}

\label{spec_split_app}

As spectroastrometric signatures trace changes in flux distributions
such signatures can be used to disentangle a convolved binary
spectrum into its constituent spectra. Two approaches can be used. One
method relies upon \textit{a priori} information while the other does
not.

\smallskip

The first approach was pioneered by \citet{JBailey1998a}, and later
used by \citet{MTakami2003}. The positional signature of a binary
system is directly proportional to the system separation and continuum
flux ratio. Therefore, if these properties of a binary system are
known, the intensity and positional spectra observed can be used to
disentangle the individual fluxes of the two components.

\smallskip

In contrast, the spectra splitting method of \citet{JMPorter2004} does
not require any prior knowledge to separate binary
spectra. \citet{JMPorter2004} present a series of simulations, in
which the dependence of spectroastrometric observables on the flux
ratio and separation of a binary system are investigated. Using
relationships established by the models of \citet{JMPorter2004}, and
the three spectroastrometric observables (the centroid, total flux and
width), the individual fluxes of the binary components can be
recovered. For the details of the method the reader is referred to
\citet{JMPorter2004}. Essentially a model binary system is considered
with a range of separations. For each separation the continuum flux
ratio is estimated, from the observed width of the spectral profile,
$\rm \sigma$, in the continuum. Then, using the positional excursions
observed, the binary $\rm \sigma$ spectrum is predicted.  This is then
compared to the observed $\rm \sigma$ distribution. The best fit
allows the binary separation to be estimated. Once the binary
separation, and the associated continuum flux ratio, have been
determined the approach used is essentially the same as that of
\citet{JBailey1998a} and \citet{MTakami2003}. We discuss the use of
this method in more detail in \citet{Wheelwright2009}. Here we attempt
to apply the method of \citet{JMPorter2004} in the red region, as the
H$\rm \alpha$ line is often associated with the largest features. We
then use the determined properties of the binary system with the
method of \citet{JBailey1998a} to separate the binary spectra in the
blue region

\subsection{Separated binary spectra}

\label{results_spec}

To separate unresolved binary spectra into the constituent
spectra it is required that prominent spectroastrometric signatures
are observed across photospheric lines in the \textit{B} region. Also,
we only attempt to separate component spectra when the
spectroastrometric signatures trace a linear excursion in the XY
plane, as opposed to a loop. If a spectroastrometric signature is
solely due to a binary system the signature will trace a linear
excursion in the XY plane, as demonstrated by the example of GU CMa
(Fig. \ref{gucma_spec_ast}). Therefore, this criterion should exclude
contaminated binary signatures and signatures not due to binary
systems, issues discussed in Sections \ref{halos} and
\ref{spec_split_problems}.  As a result of these criteria it was not
possible to separate the constituent spectra of all the binary systems
detected.

\smallskip

We separate the unresolved binary spectra of 9 systems into the
constituent spectra (we present the binary properties used/established
in Table \ref{bin_prop}). Spectral types for each spectra were
determined by comparing the spectra to that of Morgan-Keenan standard
stars and comparing ratios of key diagnostic lines. We present the
results of assessing the spectral type of each component in Table
\ref{spec_types}. In addition, we determine the system mass ratios by
assessing the mass of each component from its spectral type, using the
data of \citet{Harmanec1988}. In some cases the spectral types of each
component of a binary system had already been estimated, e.g. GU CMa
and MWC 166. The spectral types determined using the
spectroastrometrically split spectra are in good agreement with
previous results \citep{BC2001}. This provides an important check on
the validity of the spectroastrometric procedure. In addition, the
spectral types determined for the primary components generally agree
with previous classifications of the composite spectra, which also
provides a consistency check. 

\begin{center}
  \begin{table}
    \begin{center}
      \caption{A summary of the binary properties used/established
      when separating the unresolved spectra. Column 1 presents the
      binary systems for which the constituent spectra were separated,
      column 2 lists the binary separations used or established
      and column 3 presents the binary separations in the
      literature. Column 4 presents the binary PAs determined from
      the data discussed here, and column 5 lists the binary PAs from
      the literature. Finally, column 6 contains the difference in
      brightness between the two components that was used. References
      for the literature values are presented in Table
      \ref{binaries}.}
      \label{bin_prop}
      \begin{tabular}{l c c r r c}
	\hline
	Binary & $d$ & $d_{\rm{lit}}$ & \it{PA}&$PA_{\rm{lit}}$& ${\Delta}B$\\
               &  $\rm{\arcsec}$ &$\rm{\arcsec}$  & $\mathrm{(^{\circ})}$ & $\mathrm{(^{\circ})}$ & (magnitudes)\\
	\hline
	\hline
	HK Ori & 0.36 & 0.35 & $\mathrm{46.9 \pm 3.1}$ & $\mathrm{41.8}$ &     1.0\\
	T Ori & 0.84 & --& $\mathrm{107.2 \pm 2.5}$ & -- &     2.5\\
	V586 Ori &1.00 & 0.99 & $\mathrm{216.8 \pm 3.3}$ & $\mathrm{30.3 }$ &        3.5\\
	HD 37357 & 0.14& 0.19 & $\mathrm{61.5 \pm 4.1}$ & $\mathrm{49.0}$ &       1.75\\
	V1788 Ori & 0.69& 0.52 & $\mathrm{131.3 \pm 6.6}$ & $\mathrm{352.9}$ &       3.5\\
	HD 245906 & 0.13& 0.13 & $\mathrm{81.9 \pm 3.1}$ & $\mathrm{77.1}$ &   2.5\\
	GU CMa & 0.65& 0.65 & $\mathrm{197.9 \pm 0.2}$ & $\mathrm{ 194.5 }$ &         1.1\\
	MWC 166 & 0.52& 0.65 & $\mathrm{298.3 \pm 0.7}$ & $\mathrm{297.8 }$ &       1.2\\
	Il Cep & 0.44& --& $\mathrm{54.3 \pm 2.0}$ & -- &      3.5\\
	\hline
      \end{tabular}
    \end{center}
  \end{table}
\end{center}

\smallskip

\begin{figure*}

\begin{tabular}{l c r}
{\includegraphics[width=55mm,height=55mm]{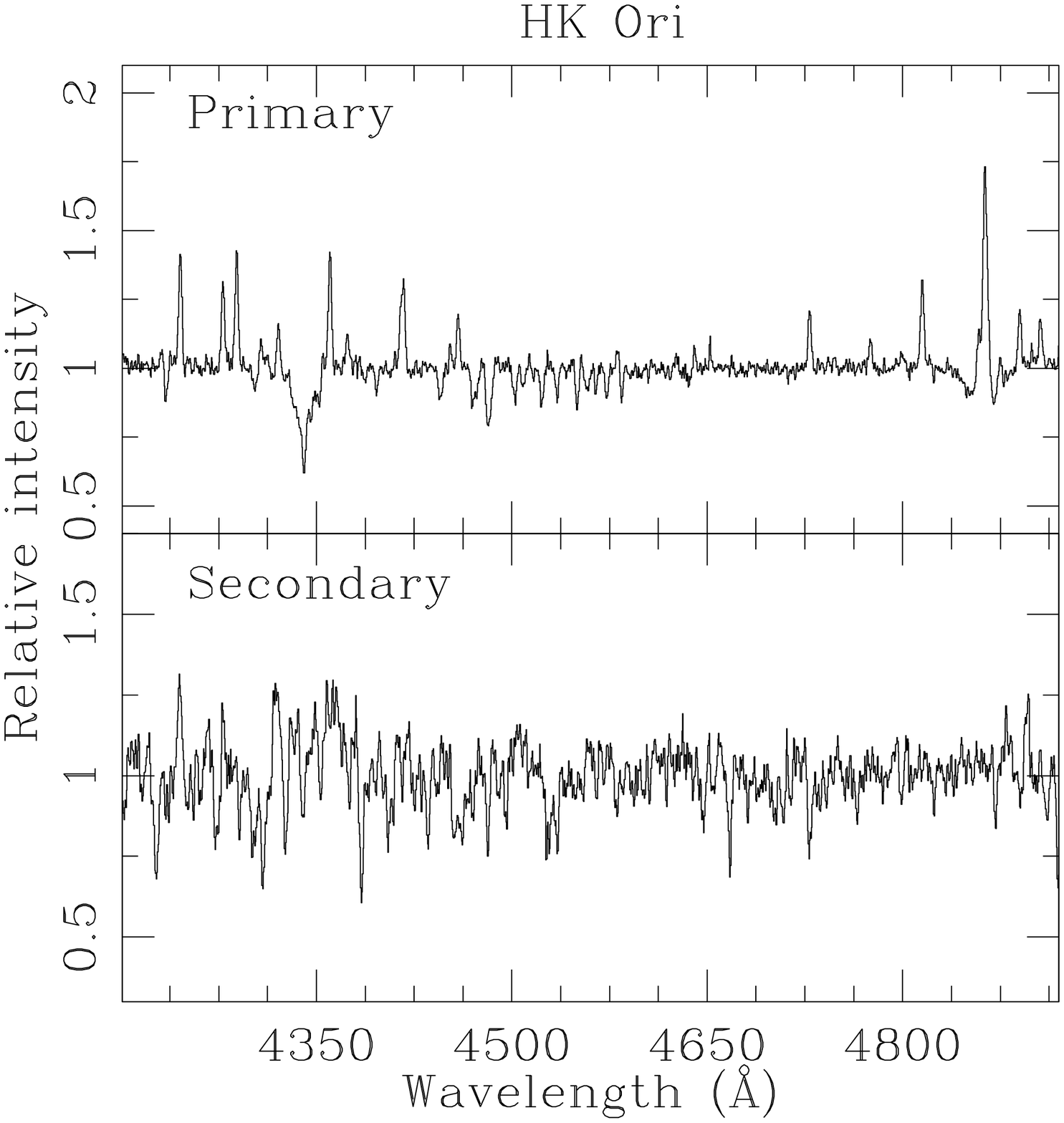}}  & 
{\includegraphics[width=55mm,height=55mm]{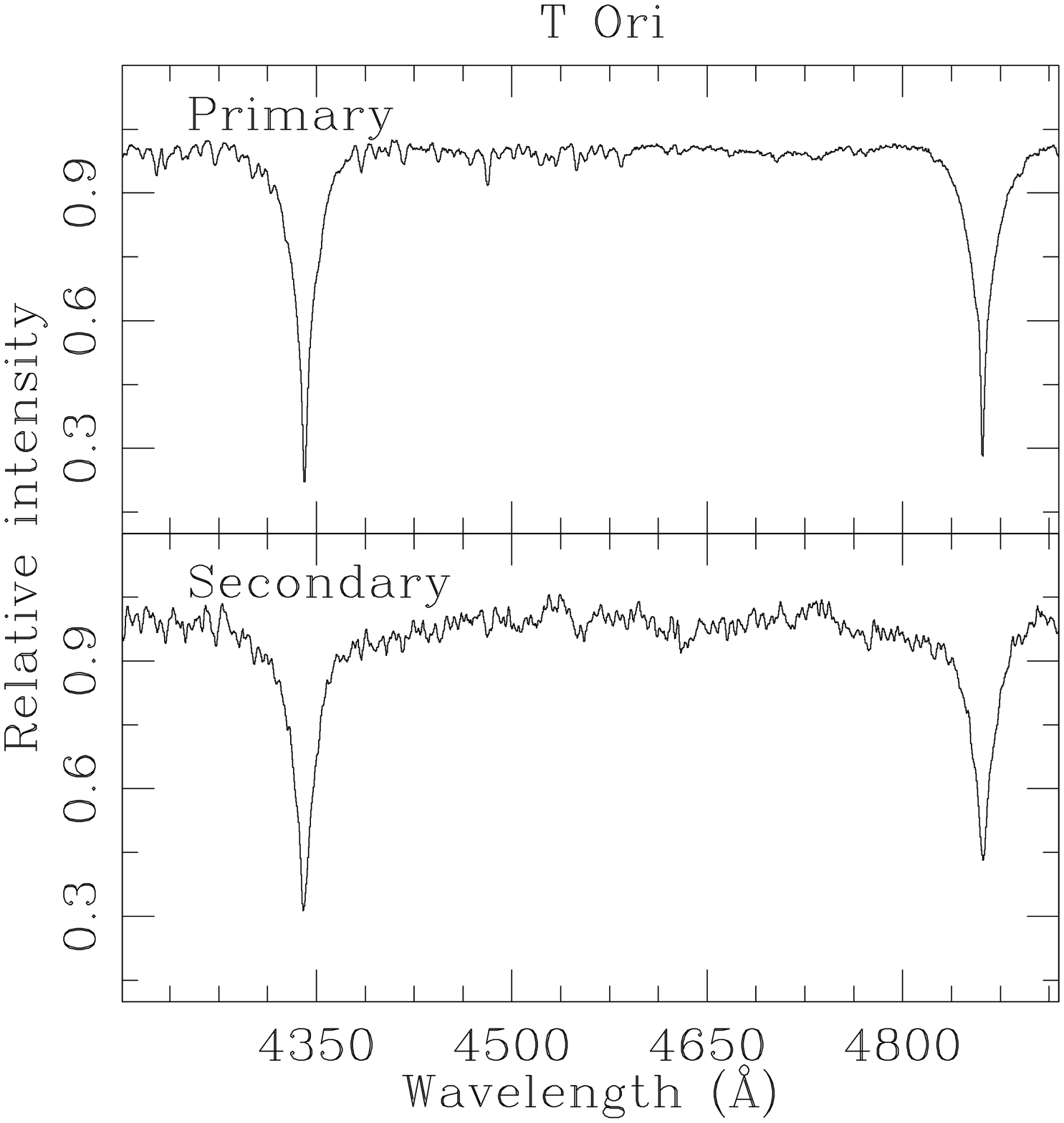}}  & 
{\includegraphics[width=55mm,height=55mm]{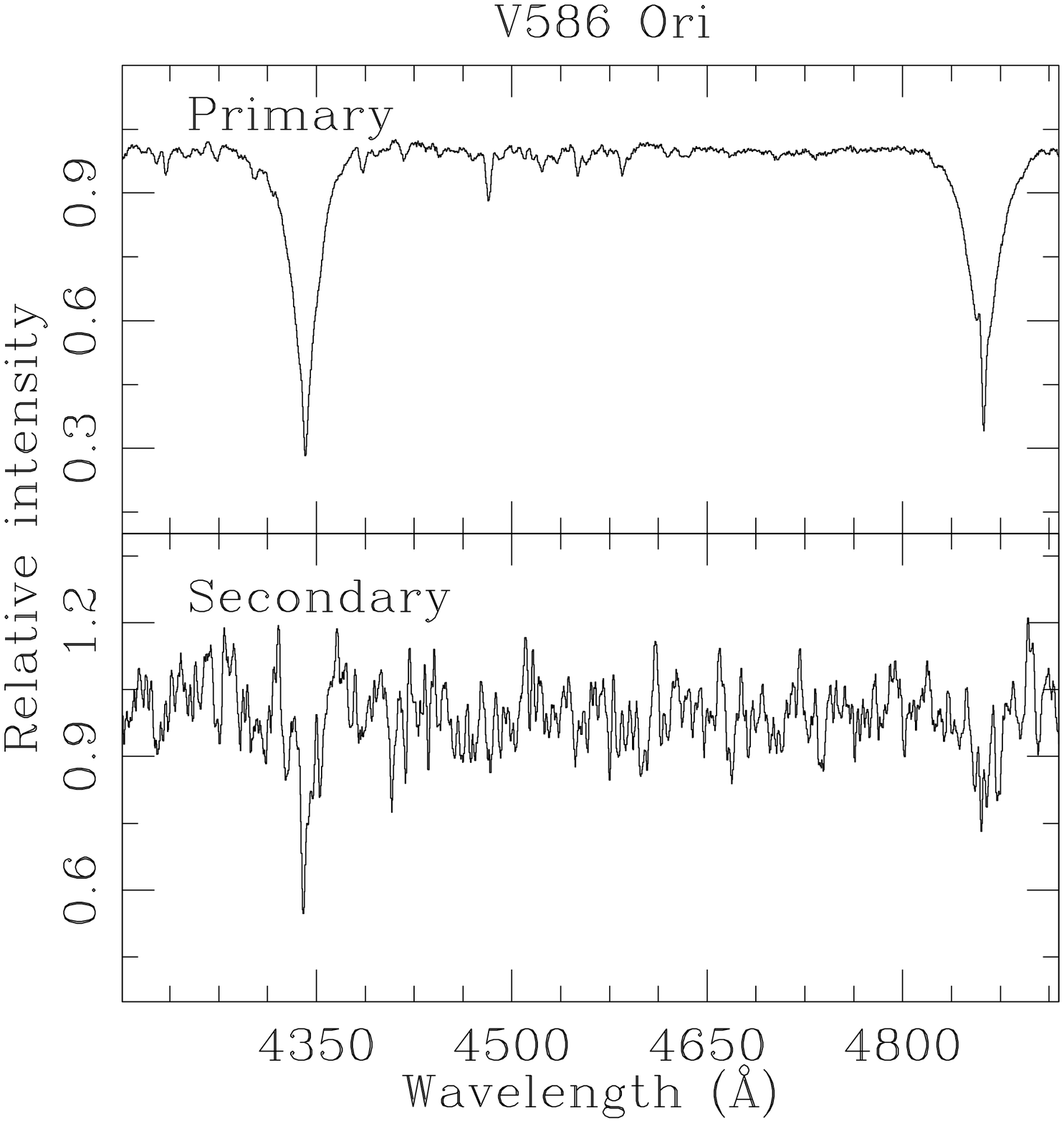}}  \\
{\includegraphics[width=55mm,height=55mm]{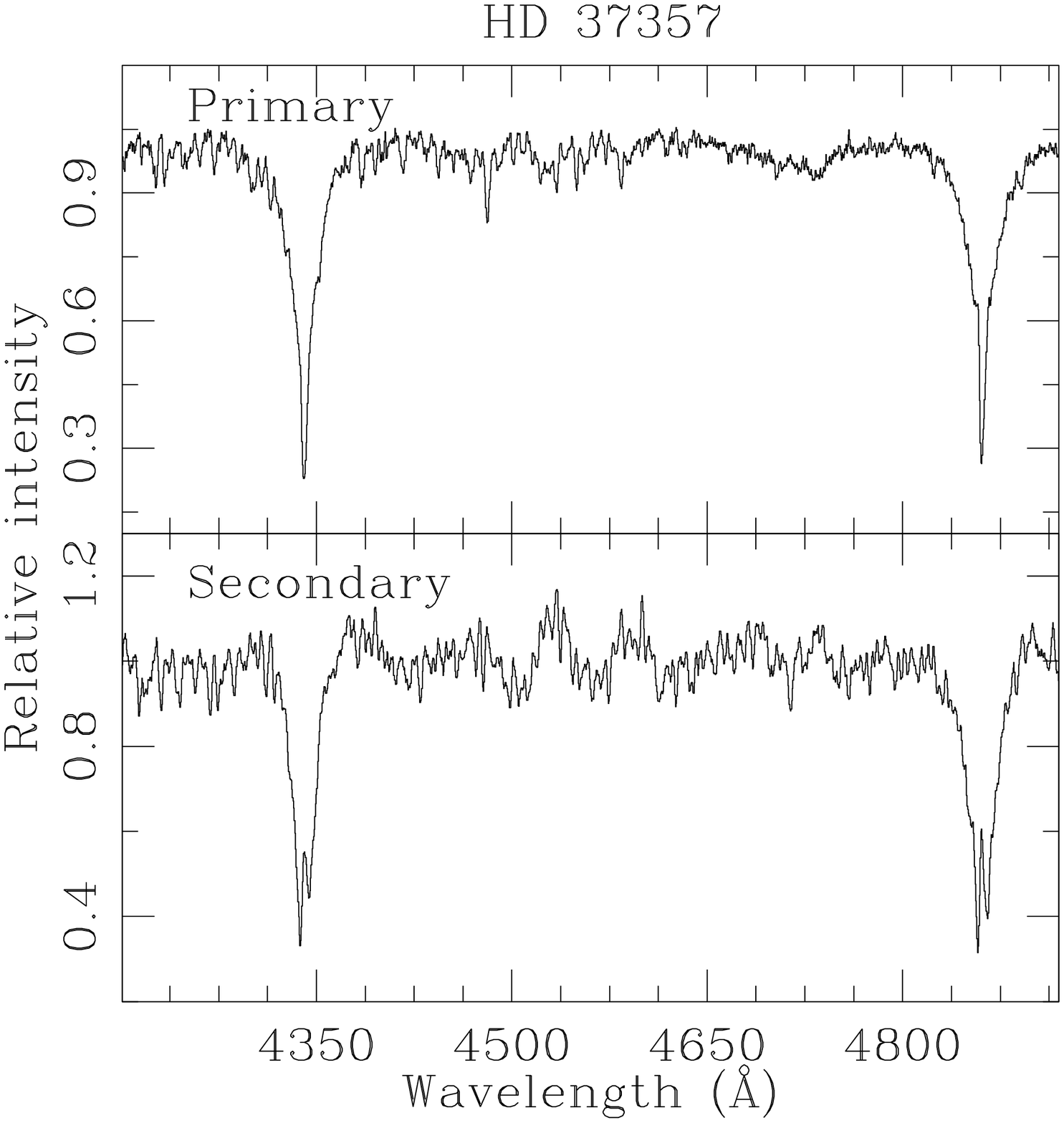}}  & 
{\includegraphics[width=55mm,height=55mm]{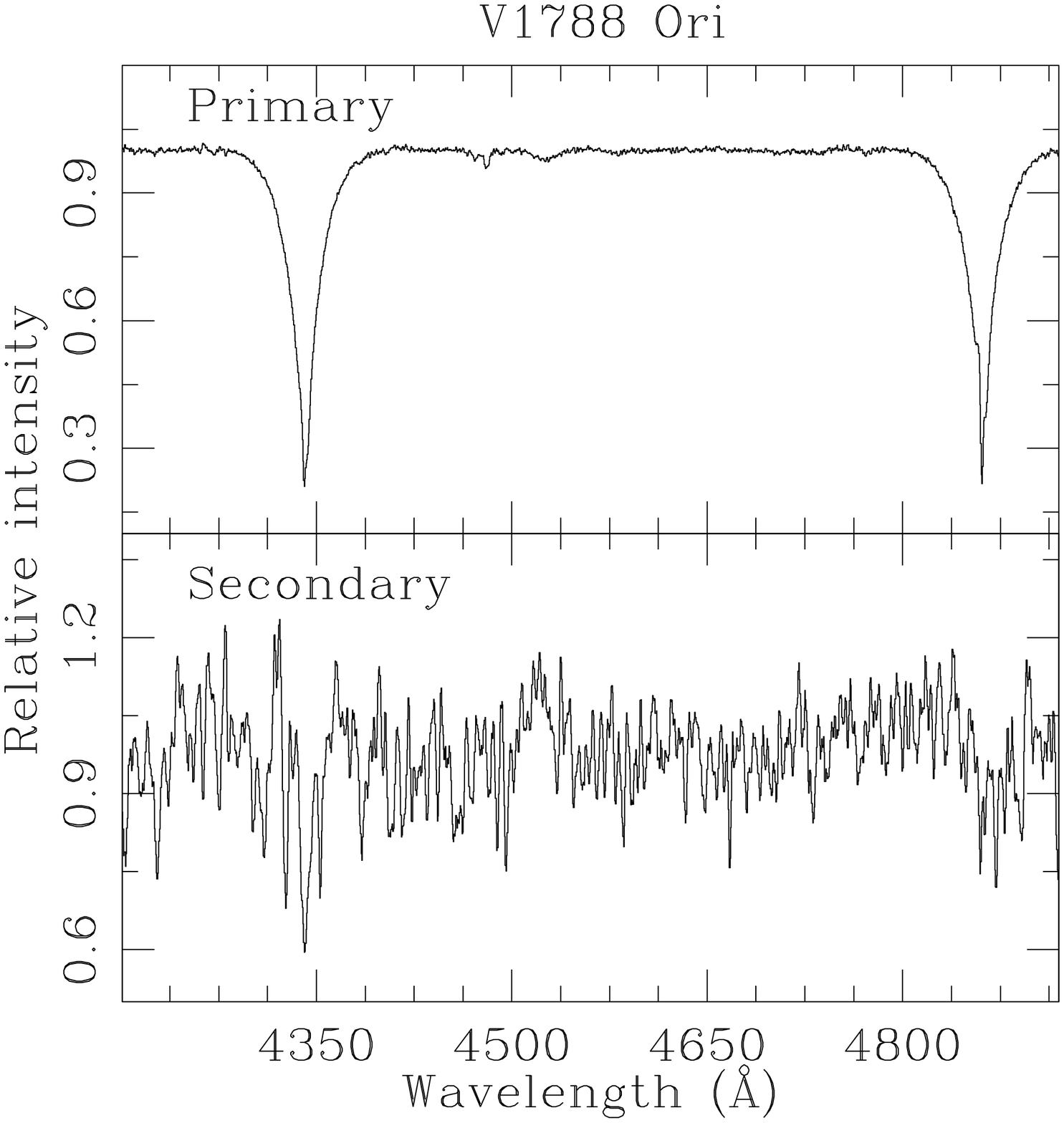}}  & 
{\includegraphics[width=55mm,height=55mm]{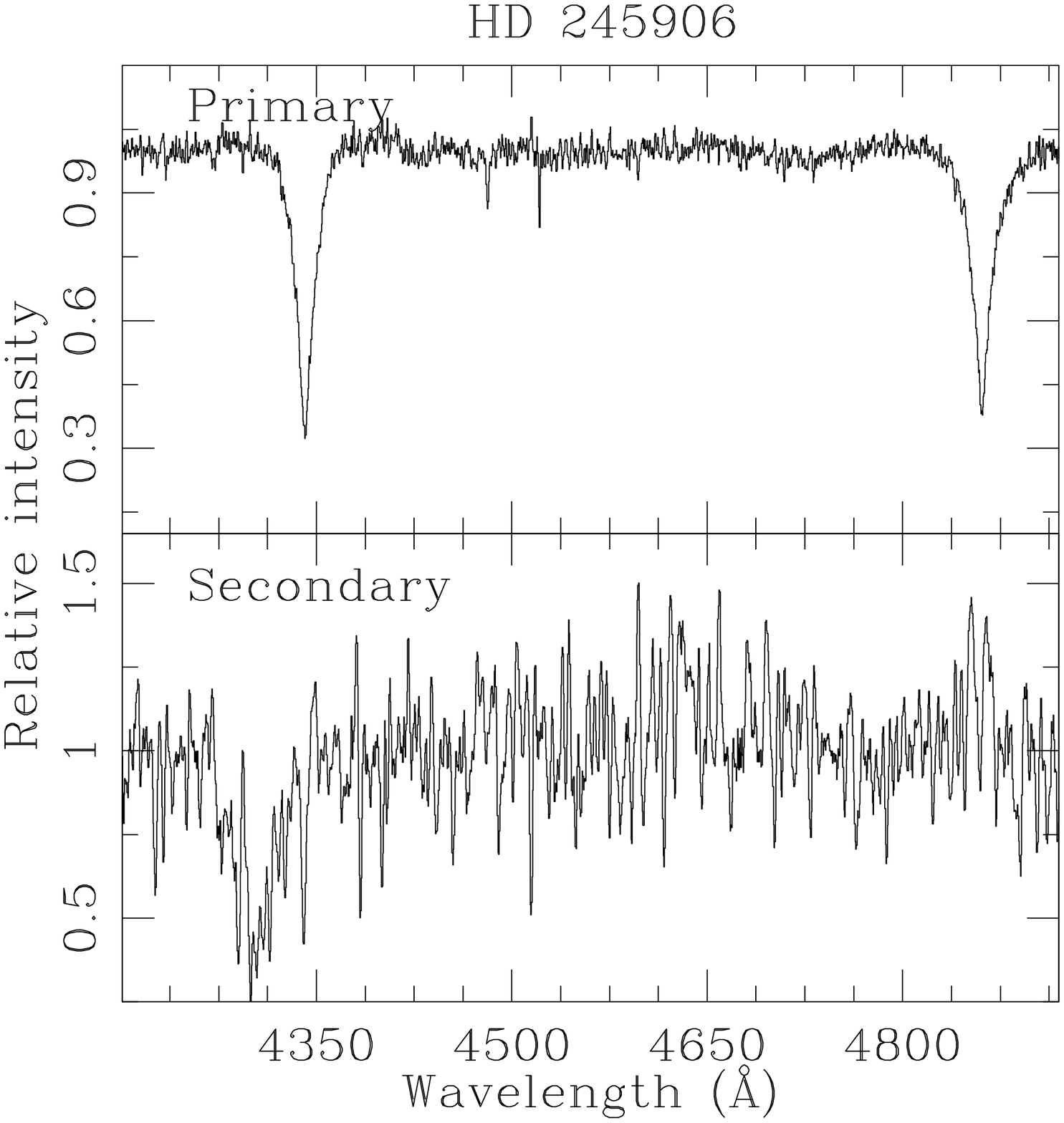}}  \\
{\includegraphics[width=55mm,height=55mm]{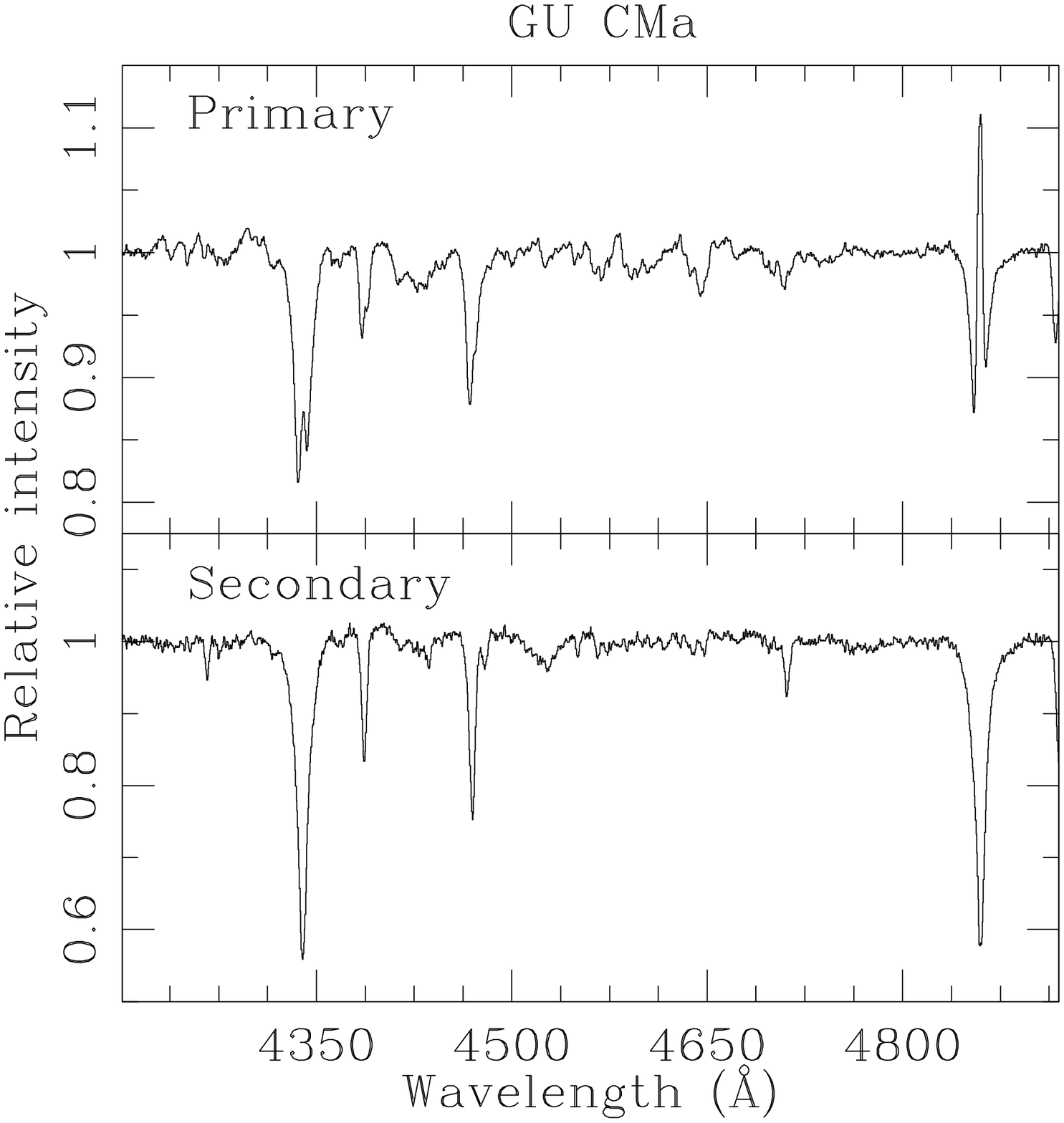}}  & 
{\includegraphics[width=55mm,height=55mm]{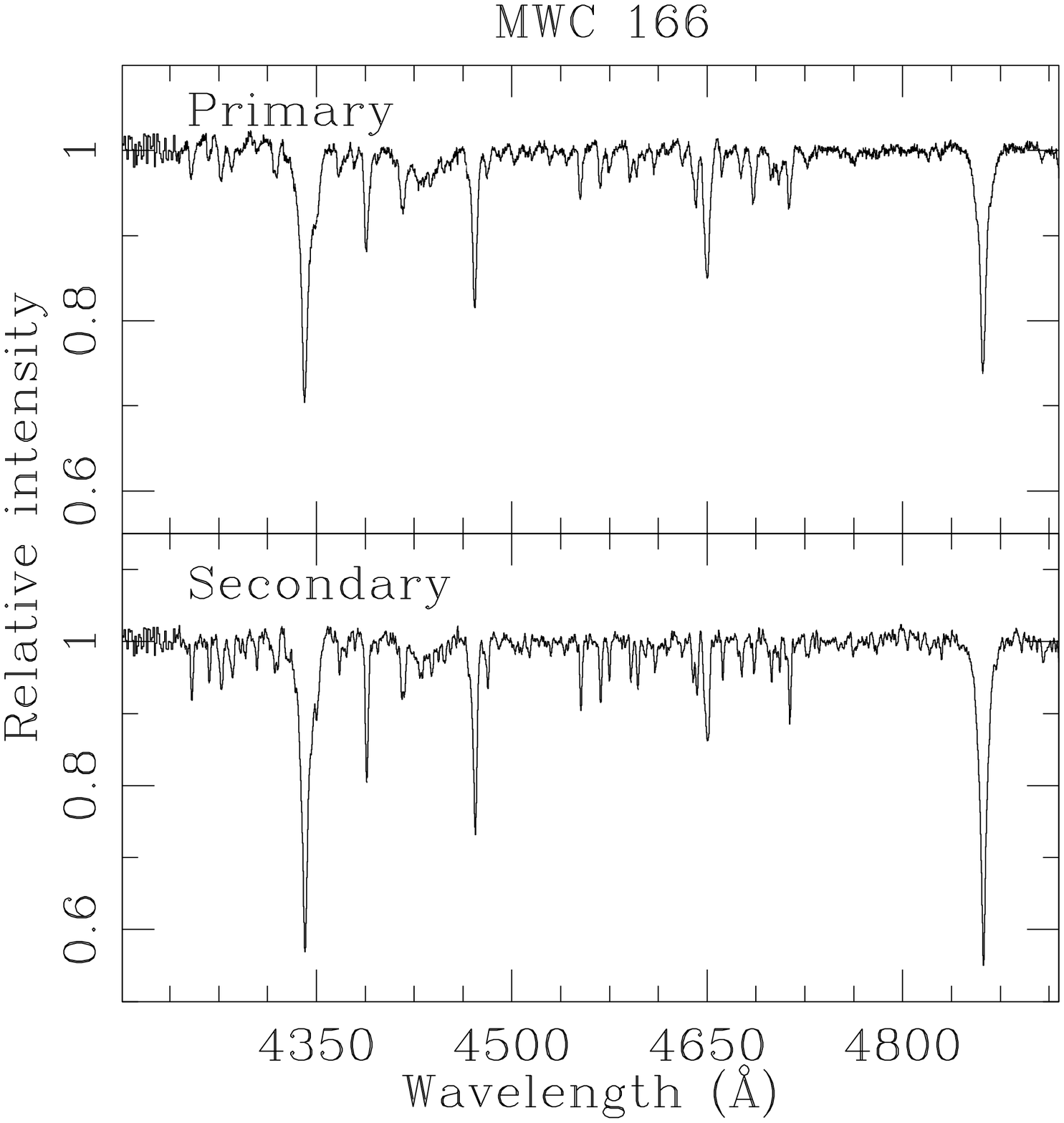}}  & 
{\includegraphics[width=55mm,height=55mm]{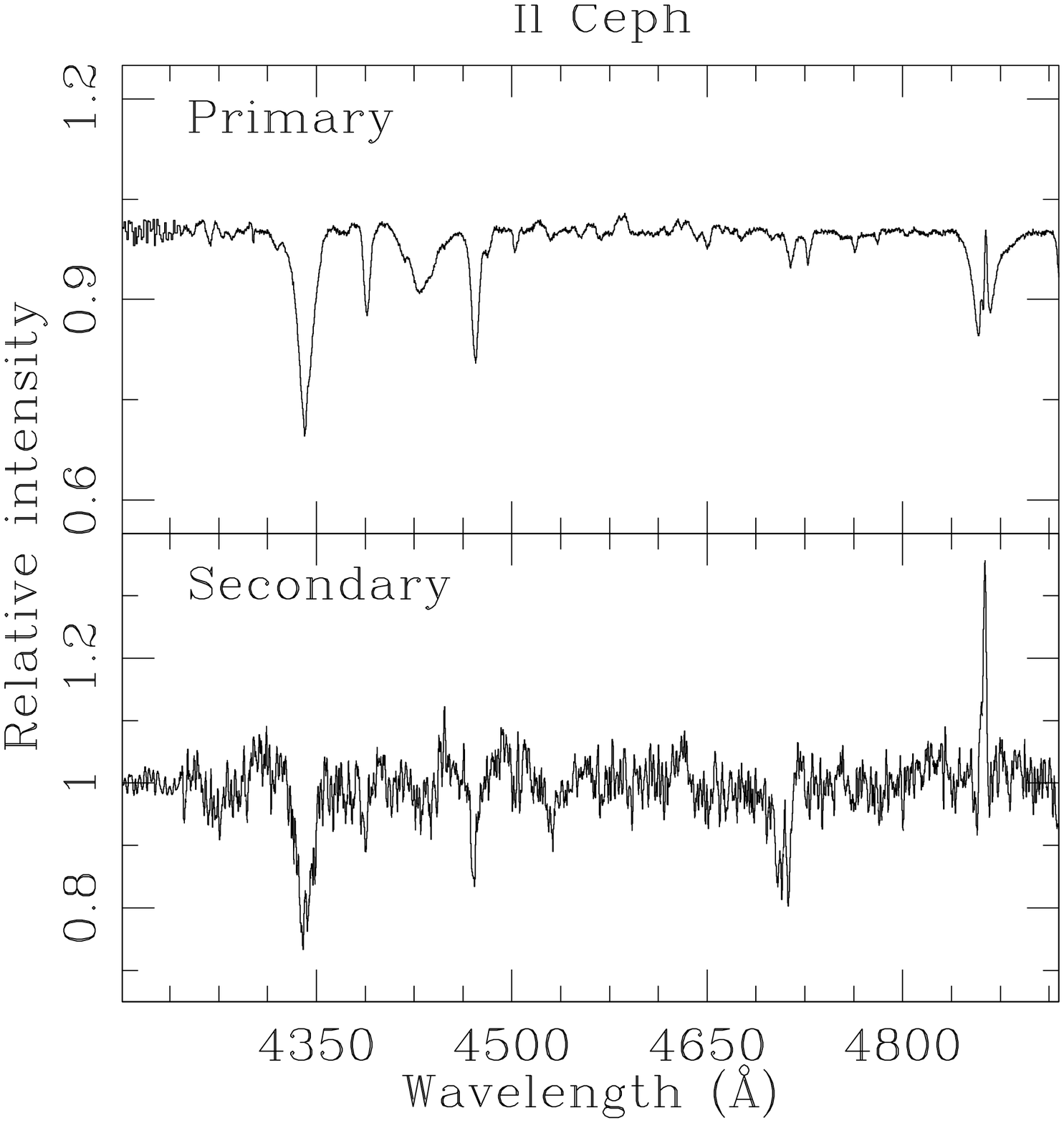}}  \\
\end{tabular}

\caption{The separated binary spectra in the \textit{B} band. From
  \textit{top} to \textit{bottom} and \textit{left} to \textit{right}:
  HK Ori, T Ori, V586 Ori, HD 37357, V1788 Ori, HD 245906, GU CMa, MWC
  166 and Il Cep. For each system the spectrum of the primary
  component is shown above the spectrum of the secondary component.}
\label{spec_split}
\end{figure*}

The separated spectra are presented in Fig. \ref{spec_split}. From
examination of the spectra (Fig. \ref{spec_split} and spectra split in
the \textit{R} band) it is clear that in some cases only the primary
component is responsible for the emission lines seen in the composite
spectrum. This is in agreement with the finding of \citet{BC2001}, who
report that in many HAe/Be binary systems only the primary exhibits a
significant NIR excess, i.e possess circumstellar material.

\begin{center}
\begin{table*}
\caption{The results of separating binary spectra into the two
constituent spectra. Column 1 lists the objects in
question and column 2 denotes the spectral type of the primary. Column 3
contains the uncertainty in the spectral type of the primary while
column 4 lists the spectral type of the secondary and column 5
presents the uncertainty in the spectral type of the secondary. The
spectral types of these objects taken from the literature are listed
in column 6. The resulting mass ratio is presented in column
7. Finally, the predicted mass ratio of the system, if the secondary
were drawn at random from the IMF (see text for explanation), is
listed in column 8.}
\label{spec_types}
\begin{center}
\begin{minipage}{\textwidth}
\begin{center}
\begin{tabular}{l l c l c l c c}
\hline
Binary & $\rm{Type_{\rm{1}}}$ & $\rm{\Delta Type_{\rm{1}}}$ & $\rm{Type_{\rm{2}}}$ & $\rm{\Delta Type_{\rm{2}}}$ & $\mathrm{Spec\, Type_{\rm{lit}}}$&$\rm{q_{ob}}$ & $\rm{q_{pred}}$\\
 & & (sub-types) & & (sub-types) \\
\hline
\hline
HK Ori & A0 & 2 & K3 & 3& $\mathrm{G1Ve^{A}}$,$\mathrm{A4pev^{B}}$ & 0.33 & 0.07\\
 T Ori & A2 & 1 & A2 & 2& $\mathrm{A3IVev^{A}}$,$\mathrm{A0^{C}}$  & 1.00&  0.07\\
V586 Ori & A2 & 1 & F5 & 5 & $\mathrm{A2V^{D}}$&0.65 & 0.07 \\
HD 37357 & A2 & 1 & A4 & 2& $\mathrm{A0Ve^{E}}$&0.94 & 0.07 \\
V1788 Ori & A2 & 1 & F5 & 3 &  $\mathrm{B9Ve^{E}}$&0.65& 0.07\\
HD 245906 & A1 & 1 & G5 & 5&$\mathrm{B8e^{F}}$ & 0.52 & 0.07\\
GU CMa & B1 & 1 & B2 & 1&$\mathrm{B2vne^{G}}$  & 0.78 & 0.02\\
MWC 166 & B0 & 1 & B0 & 1 & $\mathrm{B0IVe^{H}}$&1.00 & 0.01\\
Il Cep & B3 & 2 & B4 & 2& $\mathrm{B2pe^{H}}$ &0.84 &  0.03\\
\hline
\end{tabular}
\end{center}

  \renewcommand{\footnoterule}{}
\footnotetext{\textbf{References:} A) \citet{AMora2001}, B) \citet{Bidelman1954}, C) \citet{JHernandez2004}, D) \citet{Smith1972}, E) \citet{Theetal1994}, F) \citet{HerbigandBell1988}, G) \citet{Guetter1968}, H) \citet{Hiltner1956}.}

\end{minipage}
\end{center}
\end{table*}
\end{center}

\smallskip

\begin{center}
\begin{figure}
\begin{center}
{\includegraphics[width=70mm,height=70mm]{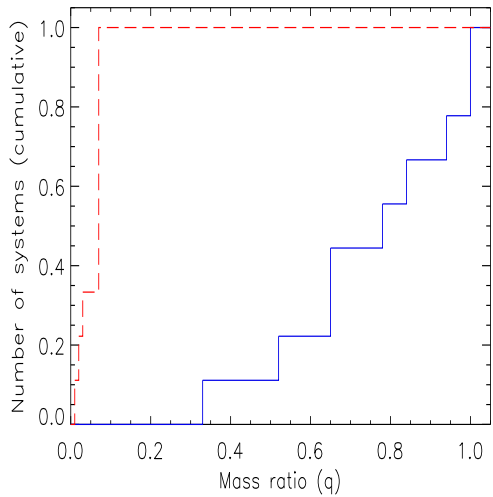}}
\end{center}
\caption{The cumulative distribution of: the observed binary mass
ratio (\textit{solid} line) and the mass ratio distribution
predicted by random sampling of the IMF, from \citet{KroupaIMF}
(\textit{long dashed} line).}
\label{KS2}

\end{figure}
\end{center}

We compare the binary mass ratio distribution observed with that
predicted assuming the secondary mass is drawn at random from the
Initial Mass Function (IMF). For the determined mass of each primary
we randomly draw a companion mass from the IMF given by
\citet{KroupaIMF}. To estimate the most probable companion mass we do
so 10,000 times and use the resultant average mass. Table
\ref{spec_types} compares the observed and the predicted mass
ratio. The predicted mass ratio distribution peaks at a relatively low
values, and no systems are predicted to have a mass ratio greater than 0.1. In contrast, the observed mass ratio distribution is noticeably
skewed towards higher values, see Fig. \ref{KS2}.

We assess how different the two distributions are using the one sample
Kolmogorov--Smirnov (KS) Test. According to the KS test the scenario
that the secondary mass is randomly selected from the IMF may be
rejected with almost 100~per cent confidence. Thus it appears the mass
ratio distribution of the binary systems is almost certainly not
determined by random sampling from the IMF. Clearly, this finding only
retains its statistical significance if we can detect all mass ratios
equally well. However, the lowest detectable mass is, on average, $\rm
{\sim 0.9M_{\rm{\odot}}}$. This is as the sensitivity of
spectroastrometry is limited by the relative brightness of binary
components. If the primary component of a binary system is more than 5
magnitudes brighter than the secondary the system will probably not be
detected by spectroastrometry. As a result the lowest detectable mass
ratio is $\rm{\sim0.3}$, greater than the location of the peak of the
mass ratio distribution predicted. Therefore, it is possible that a
large number of low mass ratio systems are undetected, which would
introduce a bias to the mass ratio observed, skewing the distribution
to high values.

\smallskip

To quantify the effect this may have, we consider the case in which
this bias has the largest effect possible. We assume every star that
is a non detection \emph{is} a binary system which we do not detect,
due to a large difference in brightness between the two components. We
assign each fictional system a mass ratio determined by the IMF, which
is generally 0.1 and below our detection limits, and add these systems
to our sample. Using the KS test we find that the hypothesis that the
new `observed' mass ratio originates from randomly sampling the
secondary mass from the IMF may still be rejected with 99.55~per cent confidence. In this case the observed mass ratio distribution appears
incompatible with the secondary mass being selected at random from the
IMF at almost a $\rm 3\sigma$ level. Therefore, the result appears
robust, despite our detection limit.

\section{Discussion}
\label{discussion}
\subsection{On the large FWHM features unaccompanied by positional features}
\label{halos}

Many stars in the sample, such as AB Aur, present spectroastrometric
signatures in which the FWHM features are much more prominent than any
positional excursions. \citet{DB2006} suggest that these features are
due to wide binary systems, where wide refers to a separation greater
than half the slit width. In the case of AB Aur we detect a similar
spectroastrometric signature over H$\rm \alpha$ to
\citet{DB2006}. However, a `wide' binary would be resolved in the
data, as we use a slit of 5~arcsec. The longslit spectra were visually
checked for evidence of a resolved companion, none was
found. Therefore, as a resolved system does not create a
spectroastrometric signature, there must be another, as yet
unconsidered, source of the spectroastrometric signature. It is
plausible that light from nebulosity could have distorted the
spectroastrometric signatures. Extended emission is noticeable in many
longslit spectra. However, it was found that masking the nebulosity
had no effect on the spectroastrometric signature observed.

\smallskip

As an alternate explanation we consider the suggestion of
\citet{Monnieretal2006} that some spectroastrometric features could be
caused by the presence of dusty halos around HAe/Be
stars. \citet{Monnieretal2006} found that many HAe/Be stars, including
AB Aur, are surround by extended features of up to
$\rm0.5$~arcsec. These features contribute up to 20 per cent of the
NIR flux detected. Such halos are not well studied, but could
constitute light scattered from the remnant natal envelopes of such
stars, dust entrained in a wind or localised thermal emission a few au
from the central star \citep{Monnieretal2006}. Such extended emission
would be unresolved in a longslit spectrum, and could lead to an
increase in the FWHM while not changing the photo-centre
position. However, this requires that the line profile of the
scattered light is different from the original emission source
profile. As discussed by \citet{Monnieretal2006} this is certainly
plausible. A non-uniform distribution of H$\rm \alpha$ flux and line
of sight dependent absorption would both result in the observer and
the scattering media seeing slightly different line profiles.

\smallskip

\begin{center}
  \begin{figure}
        \begin{center}
      {\includegraphics[width=70mm,height=70mm]{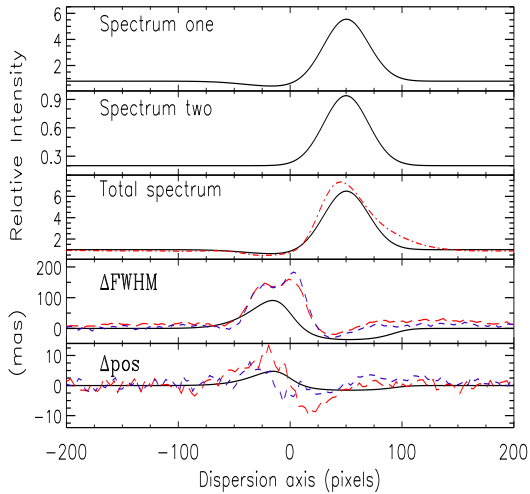}}
\end{center}
      \caption{The input spectra, the resultant spectrum and the
      spectroastrometric signature of the model comprising a star
      surrounded by a halo. The halo is modelled as a uniform ring
      from $\rm{0.45}$~arcsec to $\rm{0.60}$~arcsec, centred on a
      point $\rm{0.075}$~arcsec offset from the star. The seeing used
      was $\rm 1$~arcsec and the halo contributes 20 per cent of the
      total flux. The \textit{long dashed} lines are the
      spectroastrometric signature of AB Aur over the H$\rm \alpha$
      line, in the North-South direction and the \textit{short dashed}
      lines are the spectroastrometric signature of AB Aur in the
      East-West direction. The \textit{dot-dashed} line is the
      averaged spectrum of AB Aur in the H$\rm \alpha$ region (the AB
      Aur data has been rescaled in the dispersion direction).}
    \label{halos_quest}
    \end{figure}
  \end{center}

We explore whether an unresolved, extended halo could result in a
spectroastrometric signature, similar to that observed over the H$\rm
\alpha$ line in the case of AB Aur and other stars, using a simple
model. The model treats a single star as a point source and surrounds
the star with a halo which contributes 20 per cent of the total
flux. Here the halo is offset from the central star position by
approximately $\rm0.50$~arcsec. The flux emanating from the halo has a
uniform distribution in space. A P-Cygni type profile is assigned to
the star and a similar line profile, minus the blue-shifted absorption
component and with a slightly different line to continuum ratio,
assigned to the halo flux. The total flux distribution is mapped onto
an array representing a CCD chip. The array is then convolved with a
Gaussian in the spatial direction to represent the effects of
seeing. Finally, the output spectrum is extracted in a standard
fashion and spectroastrometry is conducted on the artificial
observation. We present the results of this exercise in
Fig. \ref{halos_quest}.

There exists a qualitative similarity between the model and observed
signatures (Fig. \ref{halos_quest}). The model did not completely
recreate the extent of the FWHM feature observed in the case of AB
Aur. However, given the unknowns involved, e.g. the amount of light
scattered and the extent of the halo, this does not exclude this
scenario. Therefore, we conclude that it is likely that FWHM features
accompanied by small or nonexistent positional signatures are due, at
least in part to unresolved, extended, halos. Alternatively, a wind
could also result in a similar positional spectroastrometric
signature, see \citet{Azevedo2007}. This has important implications on
splitting the binary spectra, which are discussed in Section
\ref{spec_split_problems}. We note that some known binary
systems exhibited larger FWHM features than positional features, and
as such this is not a unique diagnostic.
  
\subsection{An evaluation of the method of \citet{JMPorter2004}}
\label{spec_split_problems}

Implicit in both methods of splitting spectra is the assumption that
the system in question comprises of two point sources. In Section
\ref{halos} we demonstrate that dusty halos, which surround some
HAe/Be stars \citep{Leinert2001,Monnieretal2006}, can give rise to
spectroastrometric signatures. It may be expected that if the
spectroastrometric signature of a binary system is contaminated by the
signature of a halo, the spectra splitting method of
\citet{JMPorter2004} will not be able to correctly separate the
constituent spectra. In many situations the method of
\citet{JMPorter2004} failed to fit the observed FWHM spectrum of a
known binary system, even when the separation considered was increased
to many times the binary separation. Here we investigate whether
this could be due to contamination of the binary spectroastrometric
signature by an additional, unresolved source of flux.

\smallskip 

We construct a model of a binary system with a separation of
0.3~arcsec. The binary system has a difference in brightness of 3.5
magnitudes and is surrounded by a halo that extends from $\rm
0.4$~arcsec to $\rm 0.6$~arcsec from the central star. An artificial
longslit spectra is generated and spectroastrometry is applied to the
synthetic data to generate the observables necessary to separate the
constituent spectra. Finally, the method of \citet{JMPorter2004} is
used to attempt to split the unresolved binary spectrum into its
constituent spectra.

\begin{center}
\begin{figure*}
\begin{center}
\begin{tabular}{c c}
 {\includegraphics[width=80mm,height=80mm]{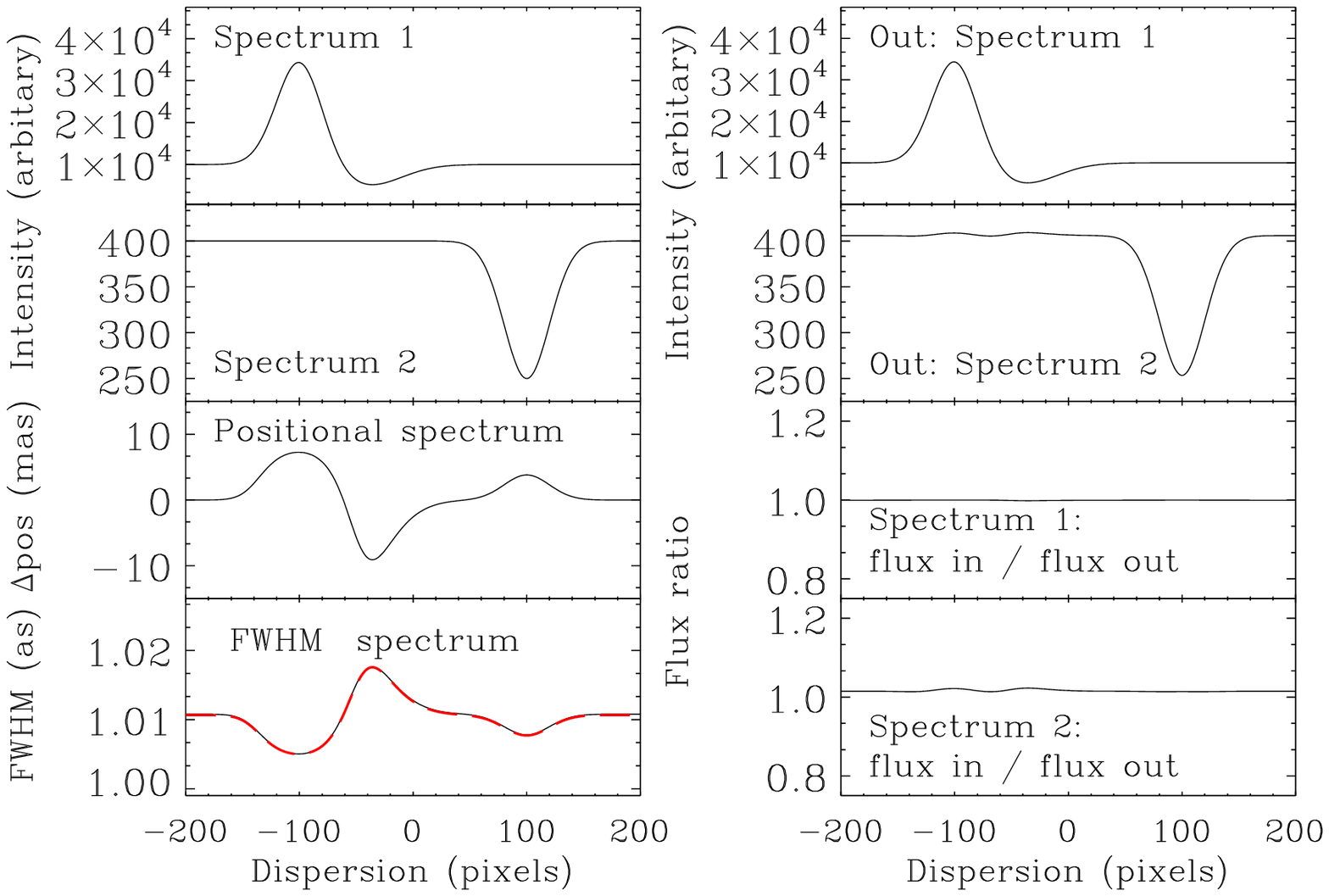}} &
 {\includegraphics[width=80mm,height=80mm]{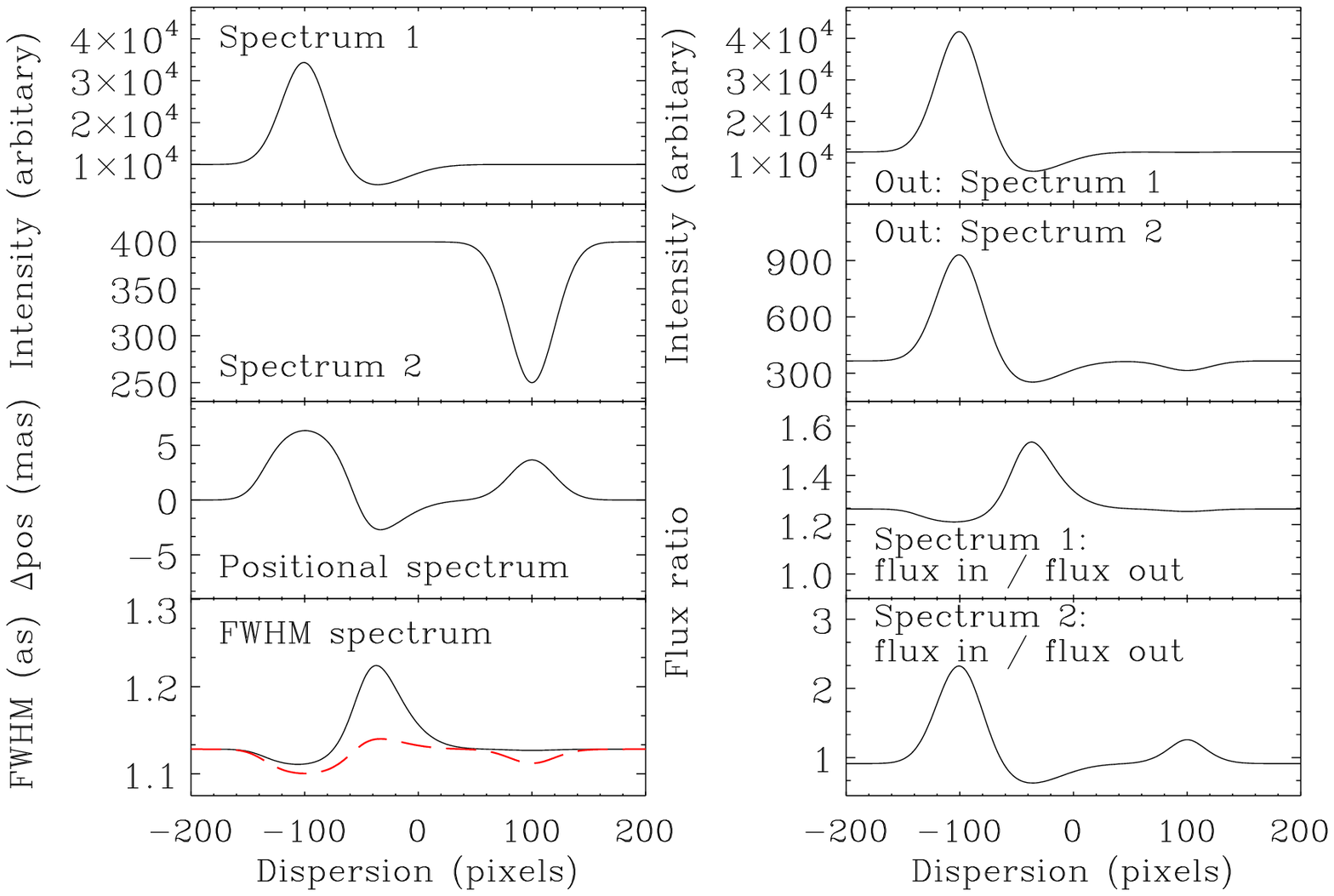}} \\
\end{tabular}
\end{center}
\caption{The results of splitting the spectra of the model of a binary
  system (\textit{leftmost two panels}) and the model of the same
  binary system plus an unresolved halo which contributes 20~per cent of the
  system flux (\textit{rightmost two panels}). In the panels
  presenting the observed $\rm \sigma$ distribution the
  \textit{solid} line is the observed quantity while the
  \textit{dashed} line is the best fit $\rm \sigma$ distribution
  predicted by the method of \citet{JMPorter2004}.}
\label{ext_em_prob_split}

\end{figure*}
\end{center}

The results of first modelling the aforementioned binary system
without an extended halo component, and the results of including an
extended halo component, are displayed in
Fig. \ref{ext_em_prob_split}. When no halo component is added the two
spectra are clearly separated, demonstrating the power of this
approach. The method of \citet{JMPorter2004} fits the observed FWHM
spectrum, and as a consequence splits the binary spectra correctly. In
contrast, when the halo component is added to the binary model, the
method of \citet{JMPorter2004} can no longer correctly separate the
constituent spectra. The method of \citet{JMPorter2004} no longer fits
the observed FWHM signature, as shown, and consequently fails to
separate the two binary spectra correctly. This is only to be expected
as: a) the positional and FWHM features observed are no longer due to
two point sources and b) the method attempts to apportion the observed
flux, which is due to three sources, to only two sources.

\smallskip

This would also be the case if the spectroastrometric
signature observed were due to a triple system. The degree to which a
third component would compromise the spectra splitting procedure would
depend on the relative brightness of the system components. For
example, the least bright component would have to be brighter than 1
per cent of the combined flux emanating from the two brightest
components to contaminate the spectroastrometric signature. A triple
system might be expected to exhibit distinctly different
spectroastrometric signatures over different lines. The norm for this
sample is for the spectroastrometric signatures over different lines
to be consistent, as demonstrated by the example of GU CMa
(Fig. \ref{gucma_spec_ast}). Therefore, if triple systems are present
in the sample it would appear that the tertiary components are not
bright enough to significantly effect the spectroastrometric
signatures observed.

\smallskip

In summary, the method of \citet{JMPorter2004} is compromised if an
additional source of flux, besides the binary system, is present. Such
a source may a tertiary stellar component, a dusty halo, or material
in a wind. We suggest that this is the reason that, more times than
not, the method of \citet{JMPorter2004} clearly does not fit the
observed FWHM features and, as a consequence, fails to separate the
spectra of many unresolved binary systems. If this is the case
for the systems where we apply the spectra splitting procedure, the
returned spectra will not be correctly separated. However, we only
present separated spectra for systems whose spectroastrometric
signature appears solely due to a binary system, with none of the
complications mentioned above (see Section \ref{results_spec} for the
condition used to exclude contaminated signatures).

\subsection{On the separation of HAe/Be binary systems}
\label{sep} 
The previously detected systems in the sample have physical
separations between $\rm{\sim 40}$ and $\rm{1200}$~au.
Unfortunately, this cannot be easily translated into a separation
distribution as the stars are at very different distances (between 143
and 2000~pc), and also due to the various selection effects in
different detection methods. Of the newly discovered binaries the
angular separation to which we are sensitive is in the range
$\rm{0.1}$ -- $\rm{2.0}$ arcsec.  However, the
different distances to each star change the physical separation to
which this corresponds.  For the nearest system at 143~pc, the
separation range is 14 -- 285~au.  For a system at the average
distance of a star at 600~pc it is $\rm{\sim}$60 -- 1000~au.  For
the most distant system at 2000~pc, it is 200 -- 4000~au.

\smallskip

None of the previously detected binaries are closer than 30~au, which
is the peak of the field G-dwarf distribution. Half of all G-dwarf
binary systems have separations less than 30~au
\citep{DuquennoyandMayor1991}. Of the newly discovered binaries, only
one {\em{could}} be closer than 30~au (V1366 Cas at 164~pc), and two
other systems could have separations as small as 30 --
40~au. Therefore, at least 60 per cent of HAe/Be stars have a
companion between about 30 -- 4000~au, and probably in the range 60 --
1000~au. This is significantly greater than the fraction of G-dwarfs at
the same separations (around 40 per cent between 30 -- 4000~au, and 25
per cent between 60 -- 1000 ~au). This overabundance of binaries is not
dissimilar to, but apparently larger than, the overabundance of binary
systems found in young T Tauri stars (see \citet{Duchene2007PPV} and
references therein). 

\smallskip

Thus, unless there is an almost complete lack of
companions $\rm{< 30}$~au, it is difficult to imagine that the
binary fraction of HAe/Be stars is much less than 100 per cent.  If
HAe/Be stars exhibit a similar abundance of companions
$\rm{<30}$~au to G-dwarfs this would suggest that many HAe/Be
stars are triple or higher-order systems. Indeed we present the
detection of four additional components in previously detected binary
systems, meaning that these systems are at least triple systems.

\smallskip

It is worth noting that many of these systems are relatively soft,
with separations greater than a few hundred~au. As a result, these
systems are susceptible to destruction in dense clusters (see
\citet{Parker2009} and references therein).  This suggests that many
of these HAe/Be stars have not spent a significant time in a very
dense environments, e.g. densities of $>10^4$ M$_\odot$ pc$^{-3}$,
which are not unusual in star forming regions of young
clusters. Indeed, \citet{Testietal1999} found that no Herbig Ae stars
are associated with clustered environments, and that while Herbig
Be stars are sometimes situated in a small cluster, the associated
stellar densities are approximately $\rm{10^2-10^3pc^{-3}}$.

\subsection{The mass ratio and formation mechanisms of HAe/Be binary systems}

It appears that the mass ratio of Herbig Ae/Be binary systems is
skewed towards relatively high values, and is inconsistent with random
sampling from the IMF. However, as the spectra splitting technique did
not work in every case, the sample size is too small to attempt to
constrain the underlying distribution of mass ratios. Instead, we
discuss the possible implications this finding has on the formation
mechanisms of intermediate mass stars. We note that random pairing of
binary components has already been excluded in the OB association Sco
OB2 \citep{Kouwenhoven2005}. Here we extend this finding to younger
systems, and higher masses.

\smallskip

\citet{DB2006} found that the circumstellar discs of the components of
Herbig Ae/Be systems are preferentially aligned with the binary
position angle.  This already suggests that the secondary formed by
disc fragmentation, see \citet{Goodwin2007}. As noted by
\citet{Kouwenhoven2009}, disc fragmentation would be expected to
produce stars of roughly similar mass (within a factor of a few). Disc
fragmentation should occur during the earliest phases of star
formation. During such phases there is an abundance of gas to accrete,
and the circumprimary disc is still massive enough to fragment. The
secondary in the disc is able to accrete material from the disk more
easily than the primary, as the angular momentum of the material is
closer to the secondary than the primary \citep{Whitworth1995,
Bate1997}. Therefore, this scenario results in a binary system with a
high mass ratio, higher than if random sampling from the IMF
determined the mass of the secondary. Indeed, recent models of massive
star formation demonstrate that binary systems with high mass ratios
(0.7) and large separations ($\rm \sim1000$~au) can be formed from
disk fragmentation \citep{Krumholz2009}.

\smallskip

The separations of the binary systems in our sample also suggest disc
fragmentation as the mode of binary formation.  \citet{Whitworth2006}
show that a massive disc can fragment beyond a critical radius $R_{\rm
frag}$ which depends on the mass of the primary $M_\star$ as follows:
\[
R_{\rm{frag}} > 150 \left( \frac{M_{\rm{\star}}}{\rm{M_\odot}} \right)^{1/3}
\,\,\,\,{\rm{ au}}
\]
For a Herbig Ae/Be star $M_\star \sim 10 M_\odot$, and so $R_{\rm
frag} > 300$~au, which is a typical separation of the systems in our
sample.

\smallskip

Therefore, the properties of the Herbig Ae/Be binary systems observed
indicate that these systems formed via disk fragmentation. Given that
the sample includes stars as massive as $\mathrm{\sim15M_{\odot}}$,
this favours the core collapse and subsequent monolithic accretion
scenario of massive star formation \citep{Krumholz2009}, as opposed to
the merger and capture scenarios \citep{JBally2005,JBally2007}.

\smallskip

The Herbig Ae/Be stars in this sample are not located in dense
clusters \citep{Testietal1999}. In addition, the wide separations of
the binary systems and their young ages suggest that they formed in
isolation.  Firstly, binaries this wide are relatively soft and could
not have spent a significant amount of time in a dense cluster, see
\citet{Parker2009}.  Secondly, no binary this wide could have survived
ejection from a cluster. Together this suggests that a fairly large
core ($\rm{> 10 M_\odot}$) formed in relative isolation and
produced a massive binary system, rather than a small cluster. That
these HAe/Be stars formed in a massive, isolated core shows that
competitive accretion, e.g. \citet{Bonnell1998}, is not required to
form stars of up to at least $\rm{10M_{\odot}}$, as presumably no
larger reservoir of gas existed beyond the single core.

\smallskip

This cannot be infrequent as we find several Herbig Ae/Be systems
which fit this pattern.  However, the sample we have used is not
complete, the population of HAe/Be is heterogeneous to begin with and the
selection criteria may well impose certain selection effects on
membership of the HAe/Be class. Biases and incompleteness are
impossible to fully quantify, but we can state that a not
insignificant fraction of A/B stars can form in isolation from a
massive core. This is in qualitative agreement with \citet{deWit2005},
who report that even O type stars may form in isolation. In addition,
\citet{Parker2007} also find that a few per cent of massive stars
might form in relative isolation.

\section{Conclusions}

\label{conclusions}

In this paper we present spectroastrometric observations of a
relatively large sample of HAe/Be stars. Here we present the salient
findings of this work:

\begin{itemize}

\item{We find a high binary fraction, $\rm{74\pm6}~per cent$, consistent with
previous studies.}

\item{Using spectroastrometry to separate the unresolved binary spectra we determine spectral types for the components of 9 systems.}

\item{The mass ratios of these systems, determined from the constituent spectral types, are inconsistent with a secondary mass randomly selected from the IMF.}

\item{Although our sample is small this result constrains the mode of binary formation in that the mass ratios and separations of the binary systems observed suggest that the secondary forms via disk fragmentation.} 

\item{The properties of the binary systems observed indicate that these systems have not spent a significant amount of time in dense, clustered environments. Therefore, these systems demonstrate that isolated star formation can produce stars as massive as $\rm{\sim10-15M_{\odot}}$.}

\end{itemize}

\section*{Acknowledgements}

H.E.W gratefully acknowledges a PhD studentship from the Science and
Technology Facilities Council of the United Kingdom (STFC). R.D.O is
grateful for the support from the Leverhulme Trust for awarding a
Research Fellowship. The authors would like to thank the referee,
Dr Paulo Garcia, for insightful comments which helped improve
the paper. The authors would also like to thank Dr Bernadette Rodgers
and collaborators for communicating some of their binary detections
prior to publication.

\bibliographystyle{mn2e}
\bibliography{bib.bib}

\appendix

\section{A summary of the spectroastrometric signatures}

In Table \ref{halphabeta} we present a summary of the
spectroastrometric signatures over the H$\rm\alpha$ or H$\rm \beta$
lines for the 47 stars in the sample. Where possible the signature
over H$\alpha$ is summarised. However, not all the objects in the
sample were observed in the $R$ band. Therefore, in these cases the
properties of the H$\beta$ signature are presented.

\newpage

  \begin{table*}
\begin{center} 
\begin{minipage}{\textwidth}
\begin{center}
    \caption{\label{halphabeta}\small{A summary of the
    spectroastrometric results across $\rm H \alpha$ or
    H$\mathrm{\beta}$. Column 1 lists the objects observed, columns 2
    and 3 contain the continuum uncertainty in the position and FWHM
    spectra respectively, column 4 lists the average equivalent width
    of the line in question (accurate to 10\% on average), column 5
    denotes the emission profile type while columns 6 and 7 list the
    observed change in centroid position and FWHM over the line, and
    column 8 contains the calculated PA of the systems detected.}}

\begin{tabular}{l r r r l r r r}
    \hline

       Object & \hspace*{10mm}$\overline{\rm{pos \; \sigma}}\,$  & $\rm{\overline{FWHM \; \sigma}}$ & $\overline{\rm H \alpha \; W_{\lambda}}$ & $\rm H \alpha$ $\rm{profile^{\dag}}$ & $ \rm \Delta pos\,\,\,\,$ & $\rm \Delta$FWHM & PA$\,\,\,\,\,\,\,\,\,$\\

   &    (mas) &  (mas)$\,\,\,\,\,$ & $ \rm   (\AA)$ $\,\,\,\,$&  & (mas) $\,$ & (mas) $\,\,$ &  $ \rm ({\circ})$$\,\,\,\,\,\,\,\,\,$\\
\hline
\hline

VX Cas (WHT)&  1.2$\,\,\,\,$&  3.0$\,\,\,\,\,\,\,$ & $-$1.7$\,\,\,\,\,$  & IIR &$\rm{Artifact^{\ddag}}$&97 $\rm{\pm}$  3  & \\
VX Cas (INT)& 3.0$\,\,\,\,$& 6.8$\,\,\,\,\,\,\,$& $\rm -1.8$$\,\,\,\,\,$& IIR & Artifact &  61 $\rm{\pm}$ 7& \\
IP Per & 1.8$\,\,\,\,$& 4.4$\,\,\,\,\,\,\,$&  $\rm 2.8$$\,\,\,\,\,$& M & $\mathrm{\leq}$8$\,\,\,\,\,$ &15 $\rm{\pm}$ 5 & \\   
AB Aur & 1.2$\,\,\,\,$& 2.6$\,\,\,\,\,\,\,$  & $-$22$\,\,\,\,\,$ & IVB & 14 $\rm \pm$  1 & 233 $\rm \pm$ 3 & 45.8 $\rm \pm$ 5.1\\
MWC 480 &  1.1$\,\,\,\,$&  2.9$\,\,\,\,\,\,\,$  & $-$18$\,\,\,\,\,$  & IVB &  $\mathrm{\leq}$4$\,\,\,\,\,$& 145 $\rm \pm$ 3 & \\
UX Ori & 1.7$\,\,\,\,$&  4.0$\,\,\,\,\,\,\,$ & 3.0$\,\,\,\,\,$  & IIR & Artifact & 134 $\rm \pm$ 4 & \\
HD 35929 &  0.9$\,\,\,\,$& 3.6$\,\,\,\,\,\,\,$ & 1.8$\,\,\,\,\,$ & I & Artifact & 13 $\rm \pm$ 3 & \\
V380 Ori & 1.9$\,\,\,\,$& 5.3$\,\,\,\,\,\,\,$& $\rm -93$$\,\,\,\,\,$& I & $\rm 37 \pm 2$& 74 $\rm\pm$ 7& 264.8 $\rm\pm$ 2.4\\ 
MWC 758 &  1.0$\,\,\,\,$&  2.3$\,\,\,\,\,\,\,$ & $-$6.3$\,\,\,\,\,$ & I & $\mathrm{\leq}$4$\,\,\,\,\,$ & 21 $\rm \pm$ 2& \\
HD 244604 & 4.1$\,\,\,\,$& 11$\,\,\,\,\,\,\,$ & 1.6$\,\,\,\,\,$ & IVB & Artifact & 80 $\rm \pm$  11& \\
HD 37357 & 1.3$\,\,\,\,$& 2.9$\,\,\,\,\,\,\,$& $\rm 0.7$$\,\,\,\,\,$& IIIB& $\rm 50 \pm 1$ & 45 $\rm{\pm}$ 3 & 234.9$\rm \pm$ 1.0\\
MWC 120 (WHT) & 1.3$\,\,\,\,$ & 3.2$\,\,\,\,\,\,\,$ &  $-$28$\,\,\,\,\,$ &IIIB & Artifact &154 $\rm \pm$  3 & \\
MWC 120 (INT) & 1.1$\,\,\,\,$& 2.7$\,\,\,\,\,\,\,$& $\rm -28$$\,\,\,\,\,$& IIB& $\rm 25 \pm 1$& 123 $\rm\pm$ 3 & 33.7 $\rm\pm$ 1.5\\
MWC 137 & 4.4$\,\,\,\,$& 11$\,\,\,\,\,\,\,$& $\rm -665$$\,\,\,\,\,$& I &Artifact &  94 $\rm \pm$ 12& \\
HD 45677 & 1.5$\,\,\,\,$ & 4.0$\,\,\,\,\,\,\,$ & $-$235$\,\,\,\,\,$ & IIB & Artifact & 59 $\rm \pm$ 4 & \\
LkH$\rm \alpha$ 215 & 2.3$\,\,\,\,$& 5.1$\,\,\,\,\,\,\,$& $\rm -30$$\,\,\,\,\,$& IIR & $\rm 25 \pm 3$& 220 $\rm \pm$ 5& 230.7 $\rm \pm$ 6.0\\
MWC 147 &  1.3$\,\,\,\,$ & 3.2$\,\,\,\,\,\,\,$ &$-$71$\,\,\,\,\,$  & II &  Artifact & 50 $\rm \pm$  3 & \\
R Mon & 7.8$\,\,\,\,$& 22$\,\,\,\,\,\,\,$& $\rm -91$$\,\,\,\,\,$& IIIB& $\rm160 \pm 8$& Artifact & 279.3 $\rm\pm$ 3.7\\
V742 Mon & 1.0$\,\,\,\,$& 2.0$\,\,\,\,\,\,\,$& $\rm -43$$\,\,\,\,\,$& I& $\rm 76 \pm 1$& 204 $\rm \pm $ 2&  47.0 $\rm\pm$ 0.4\\
GU CMa & 1.4$\,\,\,\,$ &  2.9$\,\,\,\,\,\,\,$ & $-$10$\,\,\,\,\,$ & I & 144 $\rm \pm$ 1 &  101 $\rm \pm$ 3 & 197.9 $\rm \pm$ 0.2\\
MWC 166 & 1.9$\,\,\,\,$  &  4.8$\,\,\,\,\,\,\,$ & $\; \; \;$1.3$\,\,\,\,\,$ &Ab&  49 $\rm \pm$ 2 & 29 $\rm \pm$  5  &  298.3 $\rm \pm$ 0.7\\
HD 76868 & 0.9$\,\,\,\,$& 3.3$\,\,\,\,\,\,\,$& $\rm -11$$\,\,\,\,\,$& I &  $\rm 33 \pm 1$ &  100 $\rm \pm$ 4 &  51.4 $\rm\pm$ 1.1\\ 
MWC 297 & 2.0$\,\,\,\,$ &  4.9$\,\,\,\,\,\,\,$ & $-$537$\,\,\,\,\,$  & I & Artifact  & 34 $\rm \pm$  6 &  \\
HD 179218 & 1.5$\,\,\,\,$ & 3.6$\,\,\,\,\,\,\,$ & $-$3.5$\,\,\,\,\,$ & M & $\mathrm{\leq}$5$\,\,\,\,\,$ & $\mathrm{\leq}$12$\,\,\,\,\,$& \\
HD 190073 & 1.2$\,\,\,\,$ & 2.9$\,\,\,\,\,\,\,$ & $-$27$\,\,\,\,\,$& IVB & Artifact & 108 $\rm \pm$  3 & \\
BD +40 4124 & 3.0$\,\,\,\,$  &  8.4$\,\,\,\,\,\,\,$ & $-$147$\,\,\,\,\,$ & IIB &  9 $\rm \pm$ 3 & 89 $\rm \pm$ 8 & $\sim$0\\
$\rm{MWC\, 361_{({2.5''} \space {slit})}}$ & 0.9$\,\,\,\,$ &  2.4$\,\,\,\,\,\,\,$ & $-$62$\,\,\,\,\,$& II &  Artifact & 35 $\rm \pm$  2& \\

$\rm{MWC\, 361_{({4''} \space {slit})}}$ & 0.9$\,\,\,\,$ & 2.6$\,\,\,\,\,\,\,$ &$-$62$\,\,\,\,\,$ & II & Artifact  & 43 $\rm \pm$  3 & \\
Il Cep & 0.7$\,\,\,\,$ & 2.4$\,\,\,\,\,\,\,$ &  $-$18$\,\,\,\,\,$& I & 11 $\rm \pm$ 1 & 49 $\rm \pm$  2 & 234.3 $\rm \pm$ 2.0\\
BHJ 71 & 2.7$\,\,\,\,$ & 7.0$\,\,\,\,\,\,\,$ & $-$58$\,\,\,\,\,$& IIR &  Artifact & 61 $\rm \pm $ 7 &  \\
MWC 1080 &  1.9$\,\,\,\,$ &  4.8$\,\,\,\,\,\,\,$ & $-$112$\,\,\,\,\,$ & IVB & 109 $\rm \pm$ 2 & 586 $\rm \pm$  5 &  269.2 $\rm \pm$ 1.5\\

\hline

       Object & \hspace*{10mm}$\overline{\rm{pos \; \sigma}}\,$  & $\rm{\overline{FWHM \; \sigma}}$ & $\overline{\rm H \beta \; W_{\lambda}}$ & $\rm H \beta$ $\rm{profile^{\dag}}$ & $ \rm \Delta pos\,\,\,\,$ & $\rm \Delta$FWHM & PA$\,\,\,\,\,\,\,\,\,$\\

   &    (mas) &  (mas)$\,\,\,\,\,$ & $ \rm   (\AA)$ $\,\,\,\,$&  & (mas) $\,$ & (mas) $\,\,$ &  $ \rm ({\circ})$$\,\,\,\,\,\,\,\,\,$\\
\hline
\hline
V594 Cas     & 1.6 & 3.3 & $\rm{-4.2}$& IVB & Artifact& $\rm{233 \pm 4}$ & \\
V1185 Tau    & 1.8 & 4.2 & 16 & Ab  &$\mathrm{\leq}$8$\,\,\,\,\,$&  $\mathrm{\leq}$19$\,\,\,\,\,$ & \\
V1012 Ori    & 3.3 & 8.3 & 9.5 & Ab & Artifact  & $\rm{256 \pm 8}$& \\
V1366 Ori    & 1.3 & 2.8 & 18 & Ab & Artifact&  $\rm{25 \pm 2}$ & \\
V346 Ori     & 1.3 & 3.8 & 14 & Ab & Artifact&  $\rm{39 \pm 4}$ & \\
HK Ori       & 5.0 & 10 & $\rm{-0.7}$& IIIB & $\mathrm{40 \pm 5}$& $\rm{135 \pm 10}$& $\mathrm{46.9 \pm 3.1}$\\
V1271 Ori    & 1.6 & 3.5 & 14 & IIB & $\mathrm{\leq}$5$\,\,\,\,\,$& $\rm{40 \pm 3}$& \\
T Ori        & 1.8 & 4.1 & 14 & Ab & $\mathrm{47 \pm 2}$& $\rm{117 \pm 4}$& $\mathrm{107.2 \pm 2.5}$\\
V586 Ori     & 2.5 & 5.8 & 16 & Ab & $\mathrm{55 \pm 3}$& $\rm{246 \pm 6}$& $\mathrm{216.8 \pm 3.3}$\\
V1788 Ori    & 1.9 & 4.7 & 18 & Ab & $\mathrm{65 \pm 2}$& $\rm{114 \pm 4}$& $\mathrm{131.3 \pm 6.6}$\\
HD 245906    & 2.3 & 6.1 & 12 & Ab & $\mathrm{24 \pm 3}$& $\rm{39 \pm 6}$& $\mathrm{81.9 \pm 3.1}$\\
RR Tau       & 3.8 & 8.6 & 6.7 & IIR & $\mathrm{21 \pm 4}$&$\rm{187 \pm 8}$ & \\
V350 Ori     & 2.0 & 4.8 & 17 & Ab & Artifact& $\rm{124 \pm 5}$& \\
MWC 790      & 10 & 28 & $\rm{-23}$& I & $\mathrm{\leq}$44$\,\,\,\,\,$ &$\rm{191 \pm 26}$ & \\
V590 Mon     & 4.6 & 12 & 4.5 & IIB & $\mathrm{\leq}$20$\,\,\,\,\,$ & $\mathrm{\leq}$45$\,\,\,\,\,$ & \\
OY Gem       & 3.2 & 8.3 & $\rm{-100}$& I & $\mathrm{27 \pm 3}$& $\rm{215 \pm 9}$& $\mathrm{157.5 \pm 8.0}$\\
HD 81357     & 0.6 & 2.1 & 6.5 & IIR & Artifact& $\rm{47 \pm 2}$& \\
SV Cep       & 1.9 & 4.3 & 15 & Ab & Artifact &$\rm{111 \pm 5}$ & \\
MWC 655      & 1.4 & 3.3 & $\rm{-0.3}$& II &  $\mathrm{\leq}$6$\,\,\,\,\,$& $\mathrm{\leq}$14$\,\,\,\,\,$  & \\
  \end{tabular}

  \renewcommand{\footnoterule}{} \footnotetext{{$\rm{\dag}$ Profile
  classification from \citet{Reipurth1996} (I: symmetric emission, II:
  double peaked emission where the secondary peak is
  \hspace*{5mm}greater than half the intensity of the primary peak,
  III: double peaked emission where the weaker peak is less than half
  the intensity \hspace*{5mm}of the stronger peak, IV: P-Cygni
  profile. The position of the weaker peak (or absorption component)
  with respect to the central \hspace*{5mm}wavelength is indicated by
  R or B. Absorption profiles are designated by Ab. Profiles with
  multiple absorption components are\\ \hspace*{4mm} designated by M.)}}
  \footnotetext{{\hspace*{0.0mm}$\rm{\ddag}$ Artifacts are artificial
  signatures, see Section \ref{artifacts}.}}
\end{center}
\end{minipage}
\end{center}
\end{table*}

\section{H{\textsc{i}} spectroastrometric signatures}

\label{appendix_one}

In Fig. \ref{spec_ast_fig} the H$\rm\alpha$ or H$\rm \beta$
profiles and the associated spectroastrometric signatures of the 47
stars in the sample are presented. For each object the average,
normalised, intensity spectrum is presented alongside the position
spectra in the two perpendicular directions observed. In addition, we
also present the FWHM spectra, also in these two directions. To keep
the appendix concise only one profile per observation is presented,
where possible the H$\alpha$ signature, and where not the H$\beta$
signature. Artifacts are indicated by dashed lines.

\newpage

  \begin{figure*}
     \vspace*{9mm}
      \begin{tabular}{c c c}
	\includegraphics[width=60mm,height=70mm]{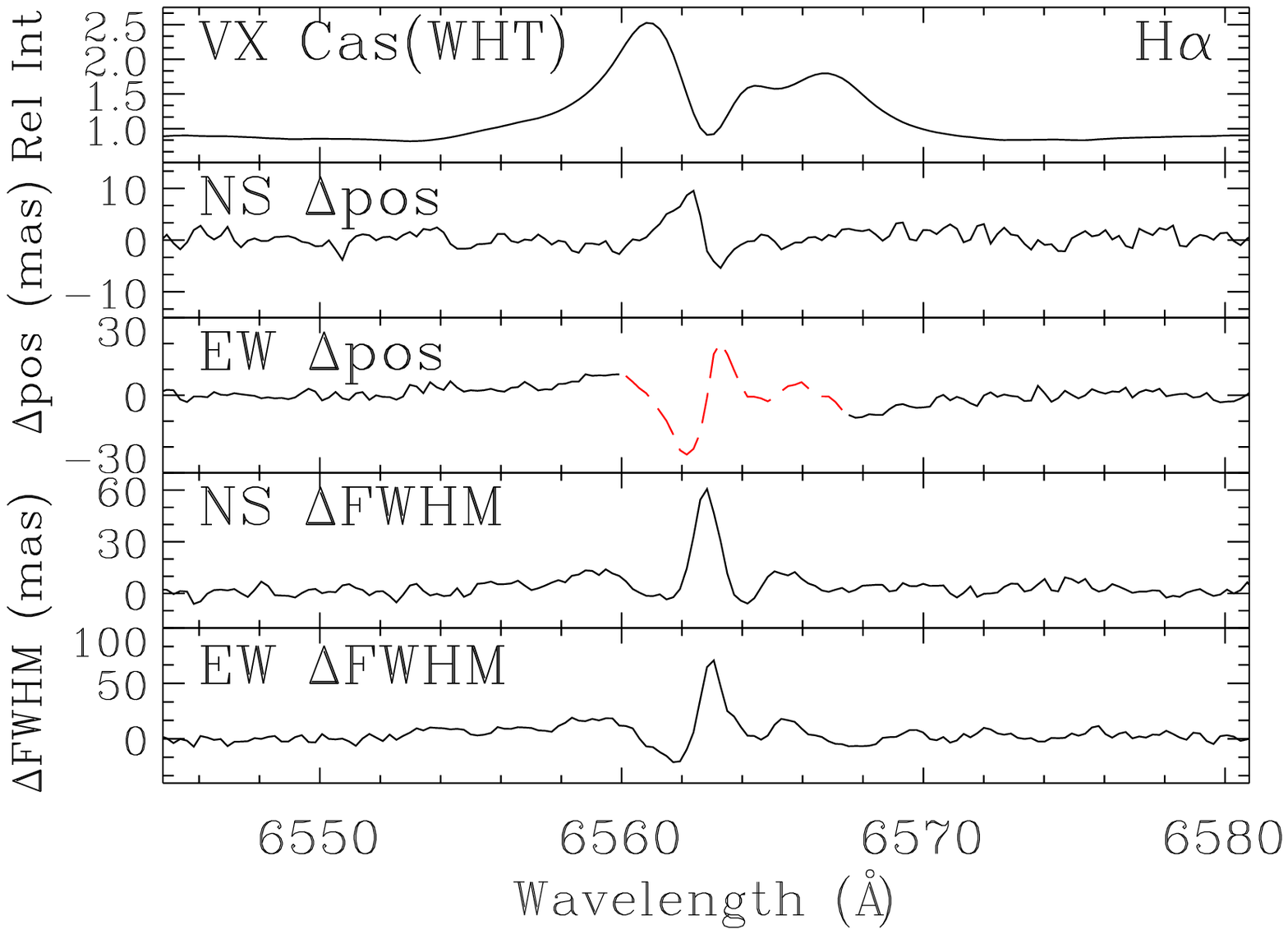} & 
	\includegraphics[width=60mm,height=70mm]{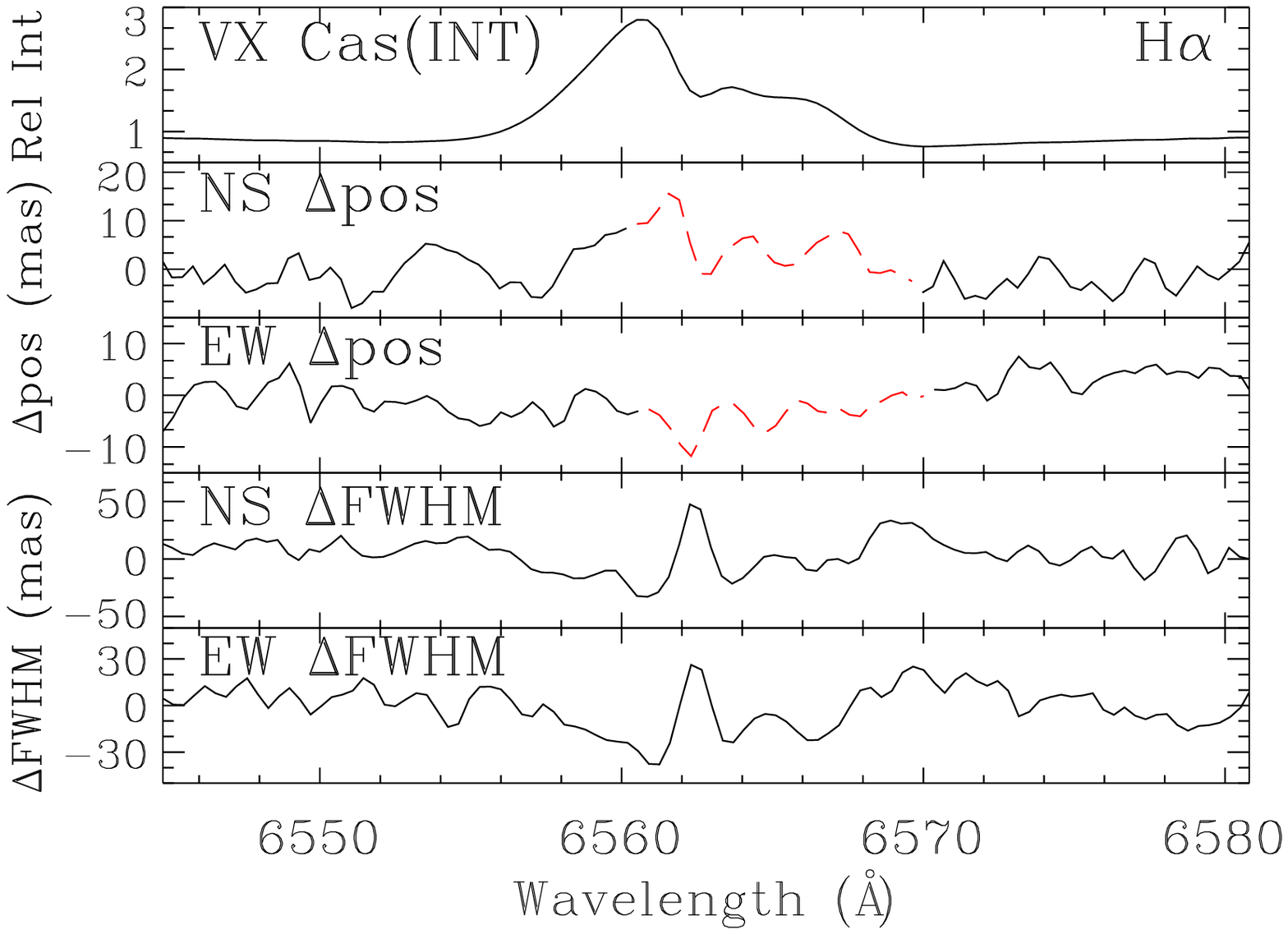} & 	\includegraphics[width=60mm,height=70mm]{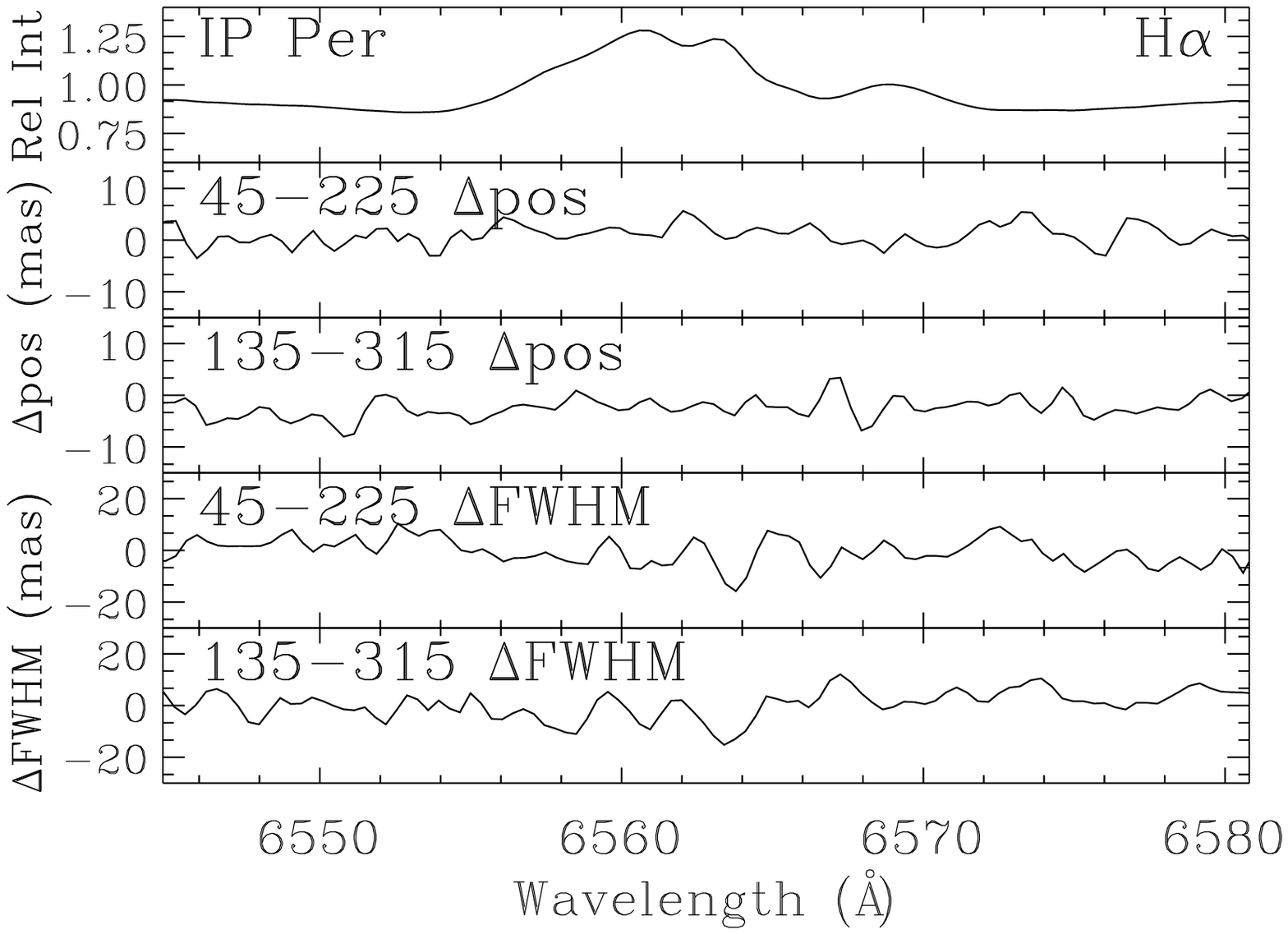}\\ 

      	\includegraphics[width=60mm,height=70mm]{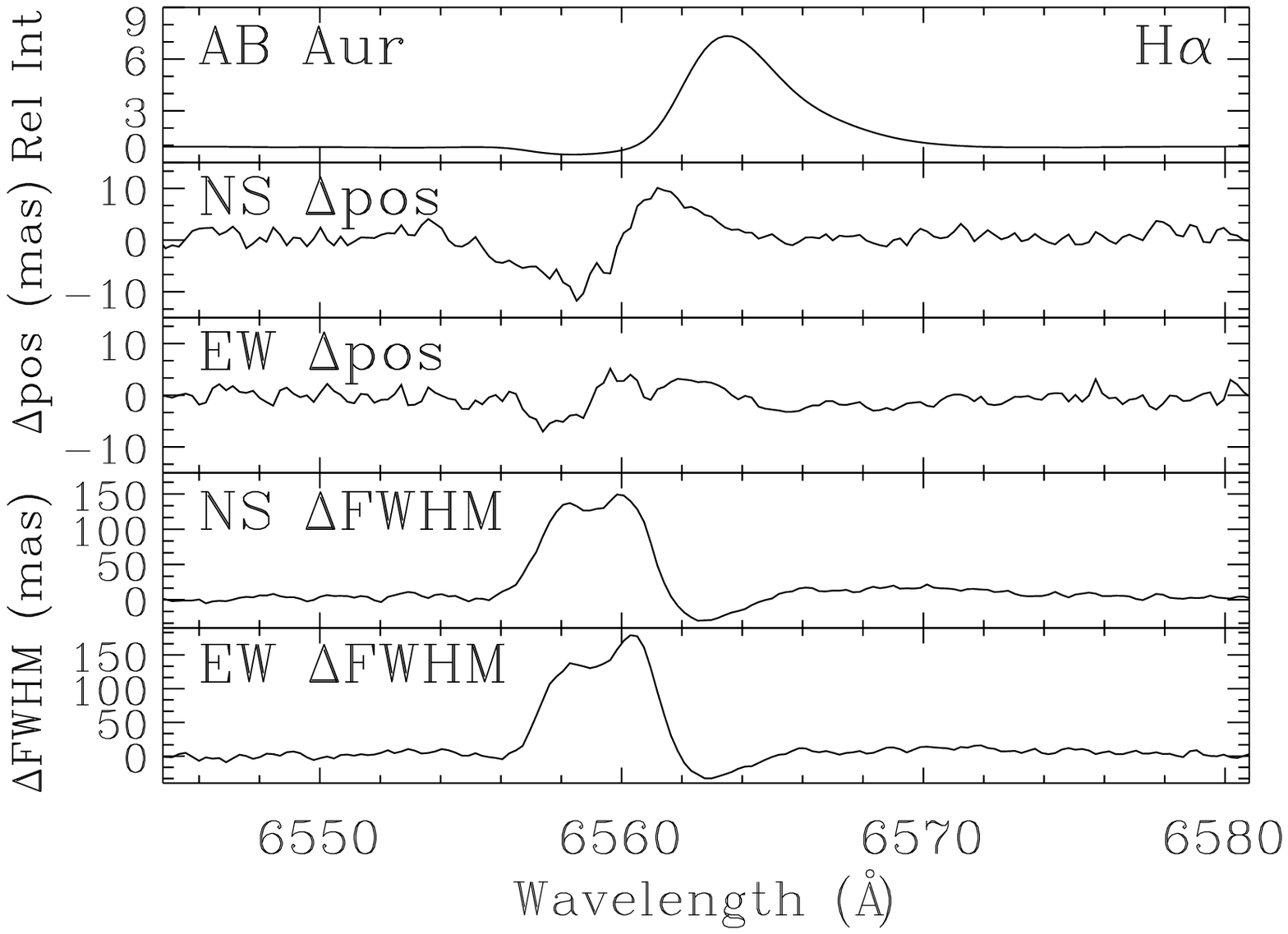} & 
	\includegraphics[width=60mm,height=70mm]{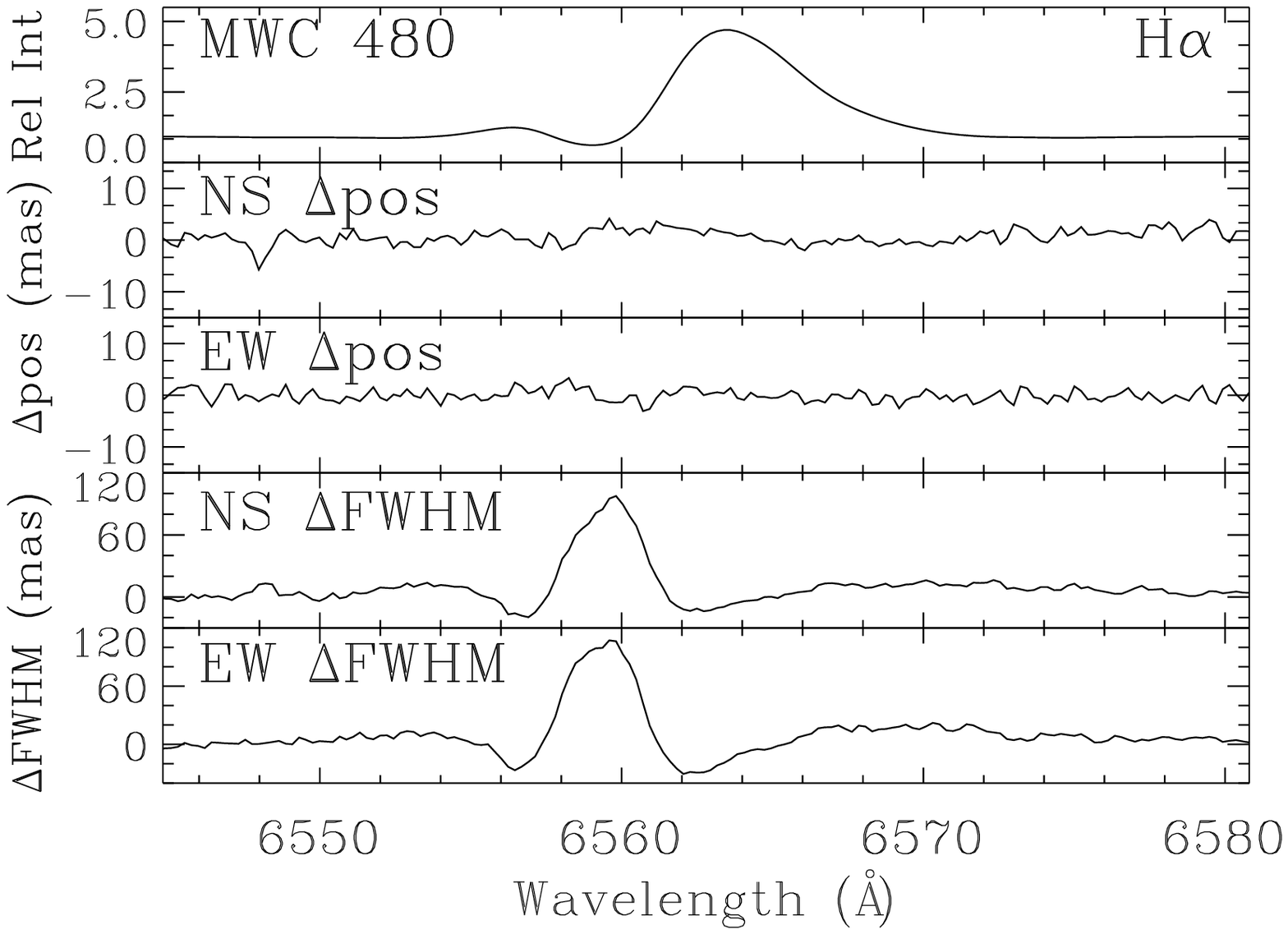} &
        \includegraphics[width=60mm,height=70mm]{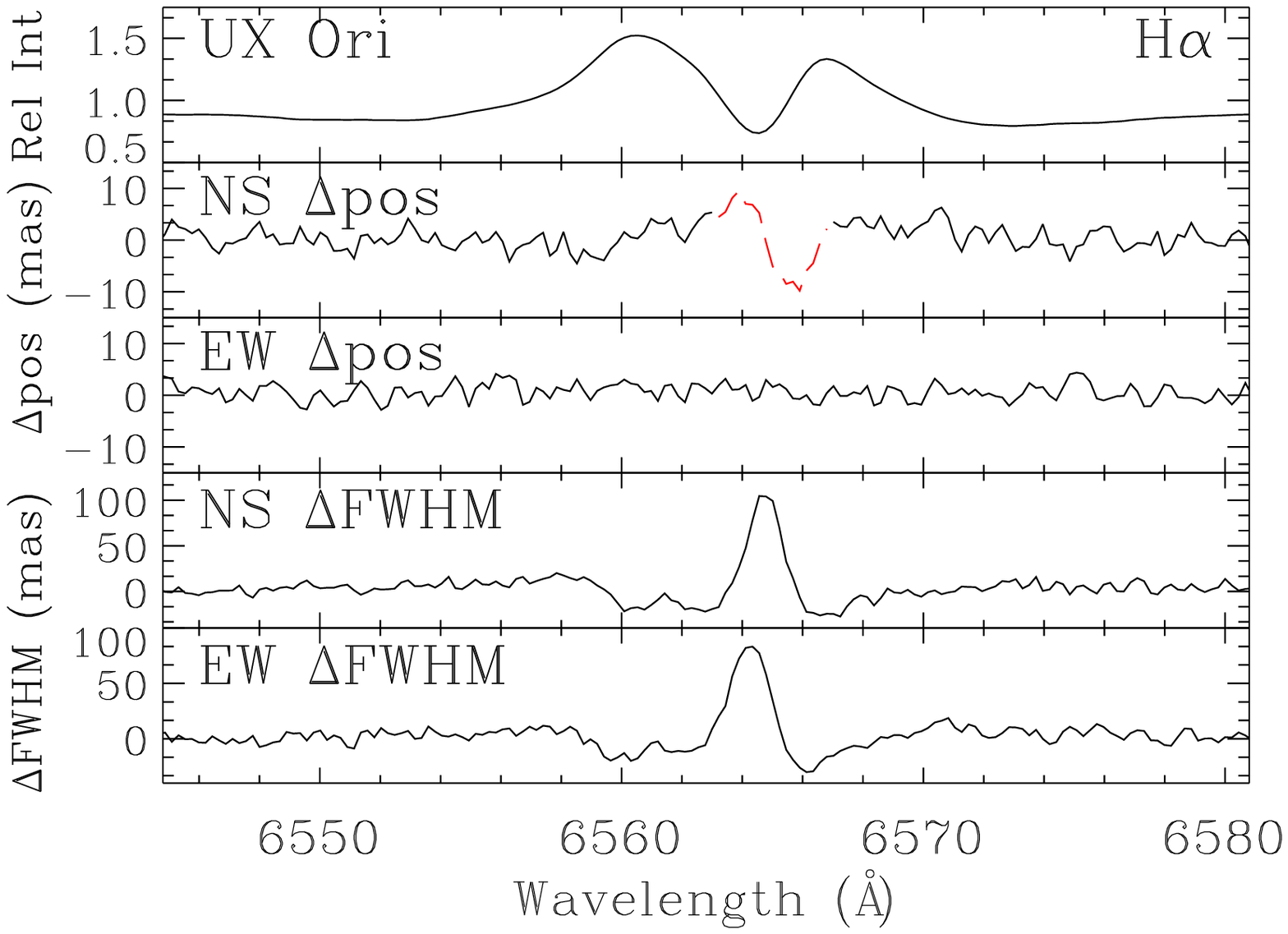} \\

	\includegraphics[width=60mm,height=70mm]{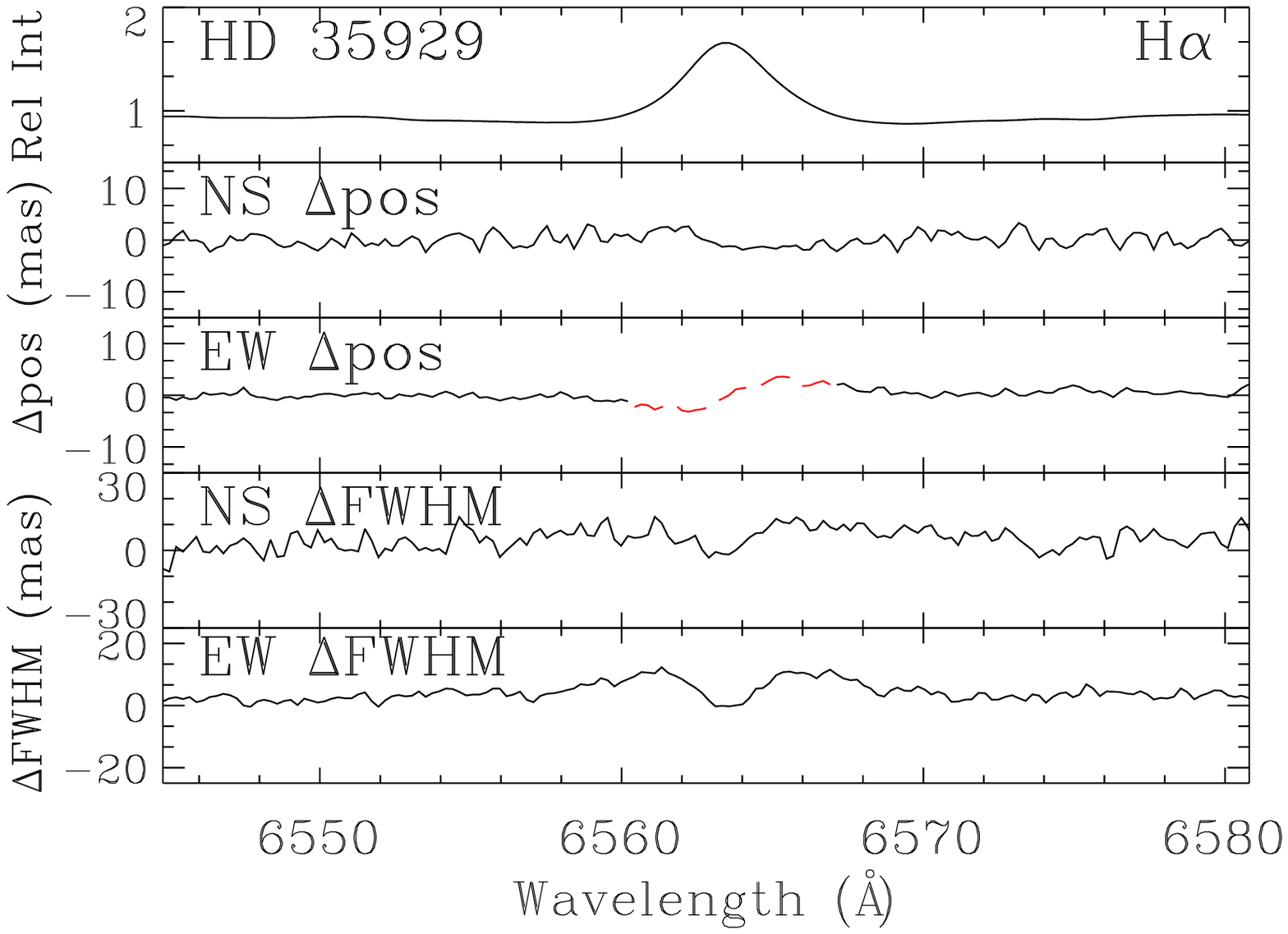} &
        \includegraphics[width=60mm,height=70mm]{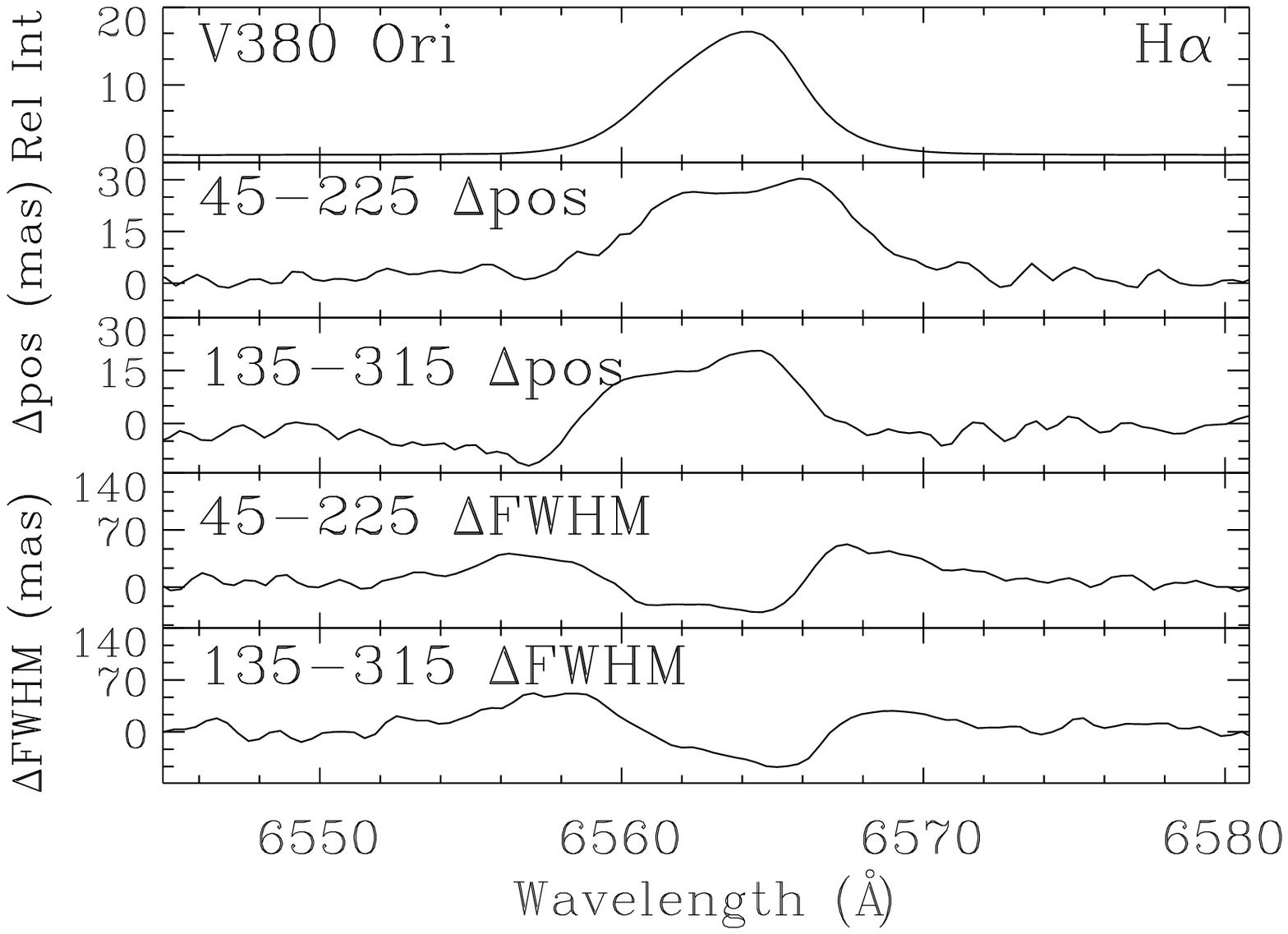} &
        \includegraphics[width=60mm,height=70mm]{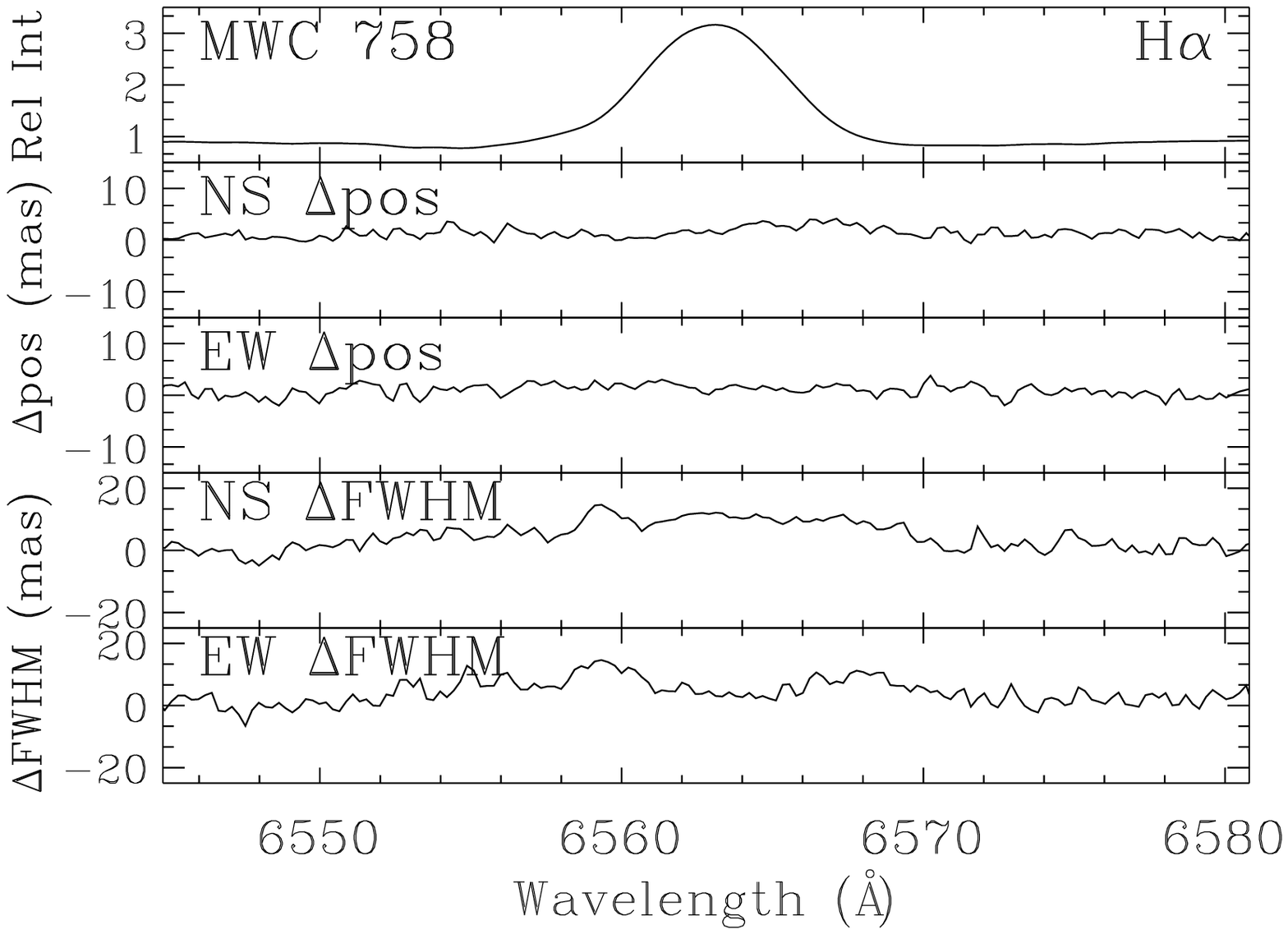} \\           
\end{tabular}
    
    \caption{H$\rm \alpha$ profiles and spectroastrometric signatures. From \textit{left} to \textit{right}: VX Cas (data from the WHT), VX Cas (data from the INT), IP Per, AB Aur, MWC 480, UX Ori, HD 35929, V380 Ori and MWC 708.}
\label{spec_ast_fig}
  \end{figure*}

\addtocounter{figure}{-1}

  \begin{figure*}
     \vspace*{9mm}
      \begin{tabular}{c c c}
	\includegraphics[width=60mm,height=70mm]{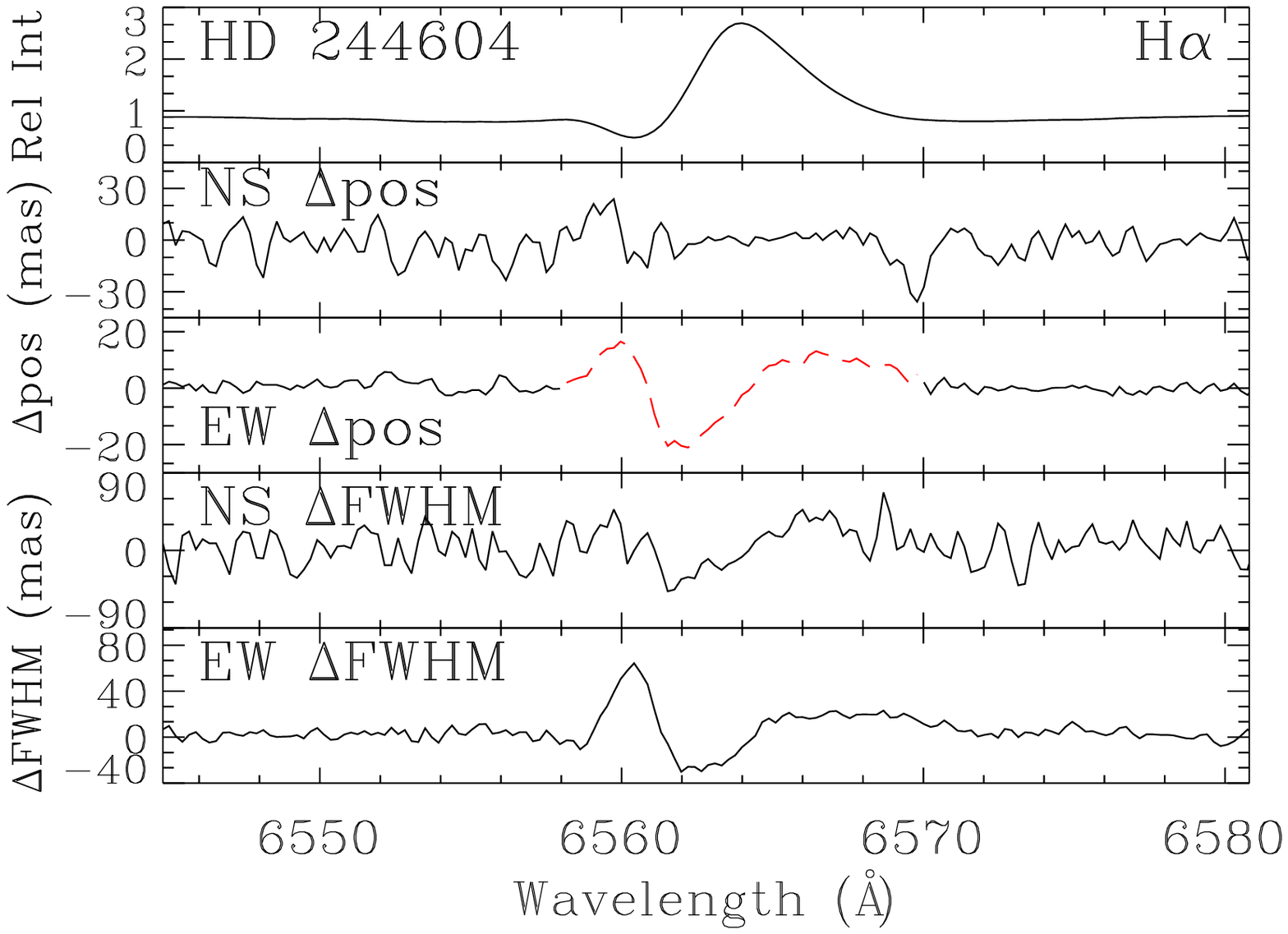} & 
	\includegraphics[width=60mm,height=70mm]{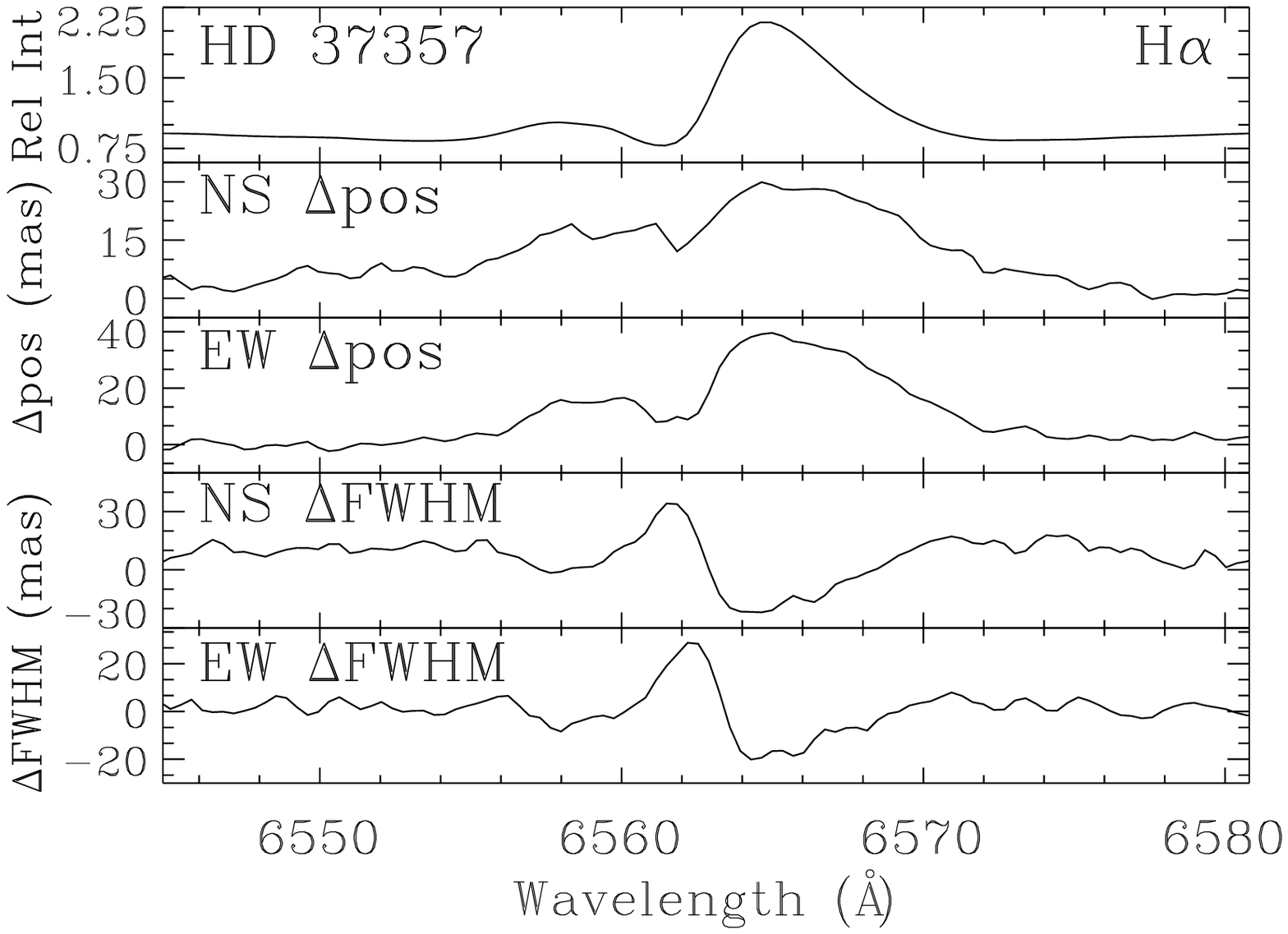} & 	\includegraphics[width=60mm,height=70mm]{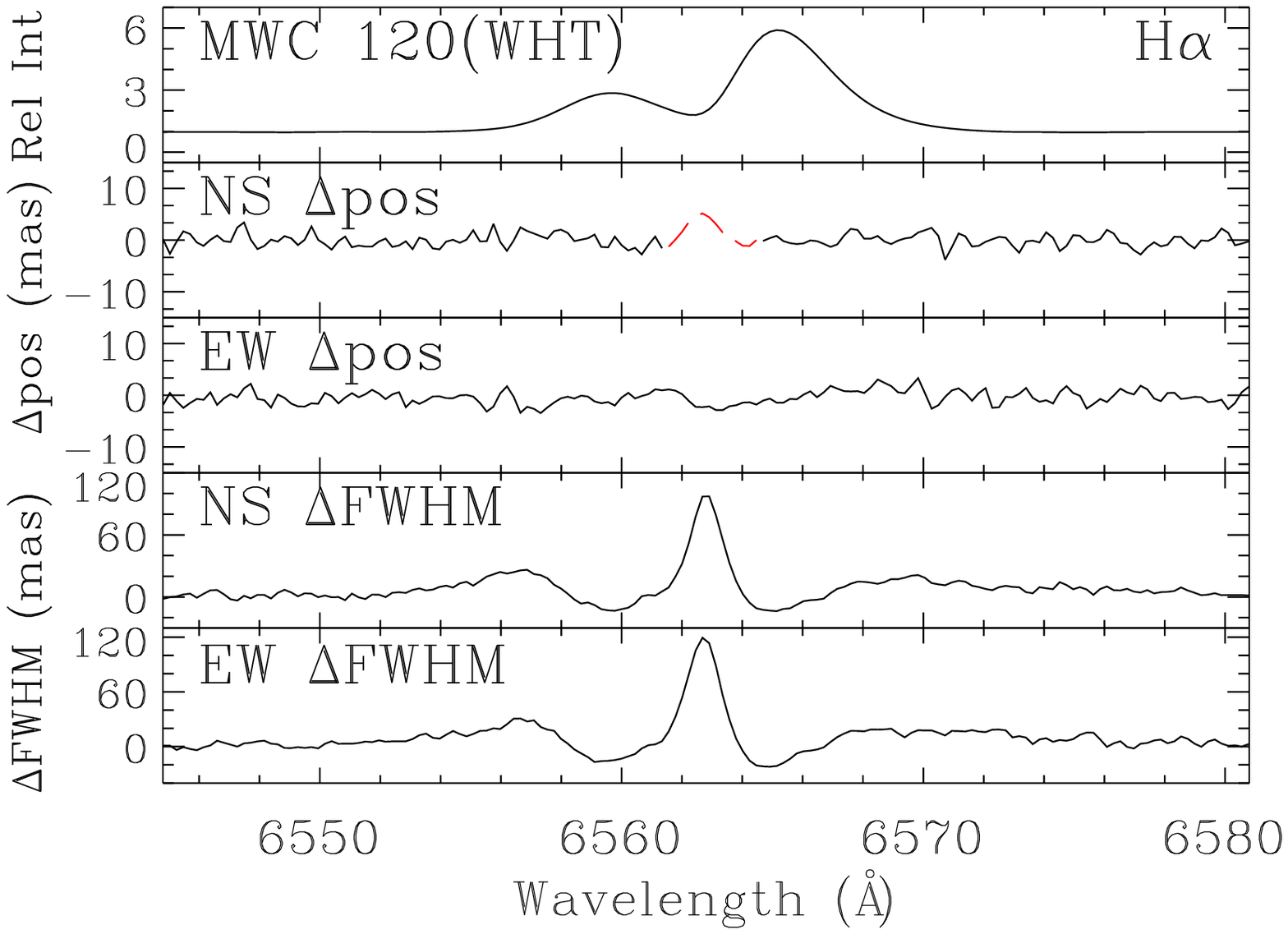}\\ 

      	\includegraphics[width=60mm,height=70mm]{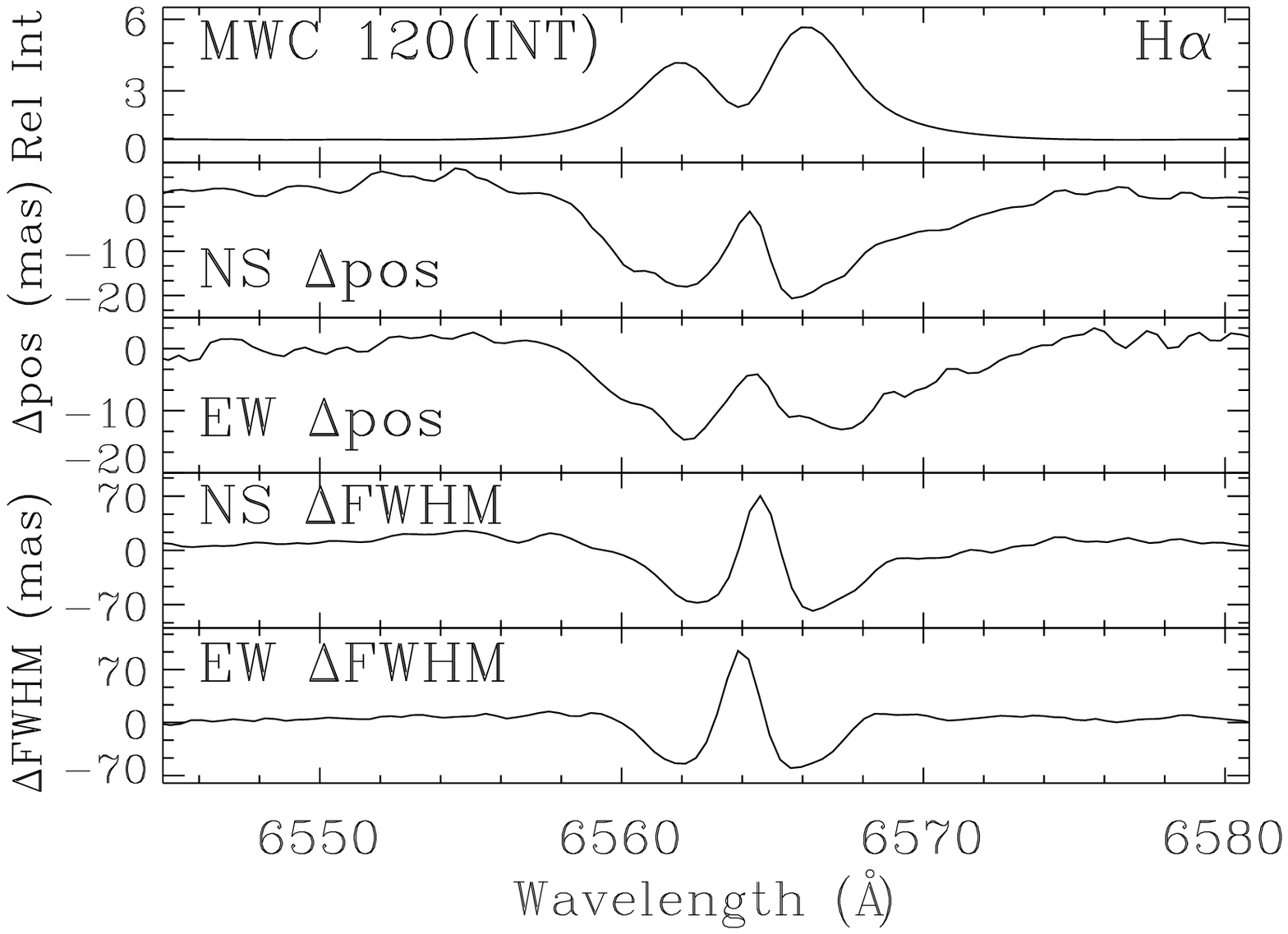} & 
	\includegraphics[width=60mm,height=70mm]{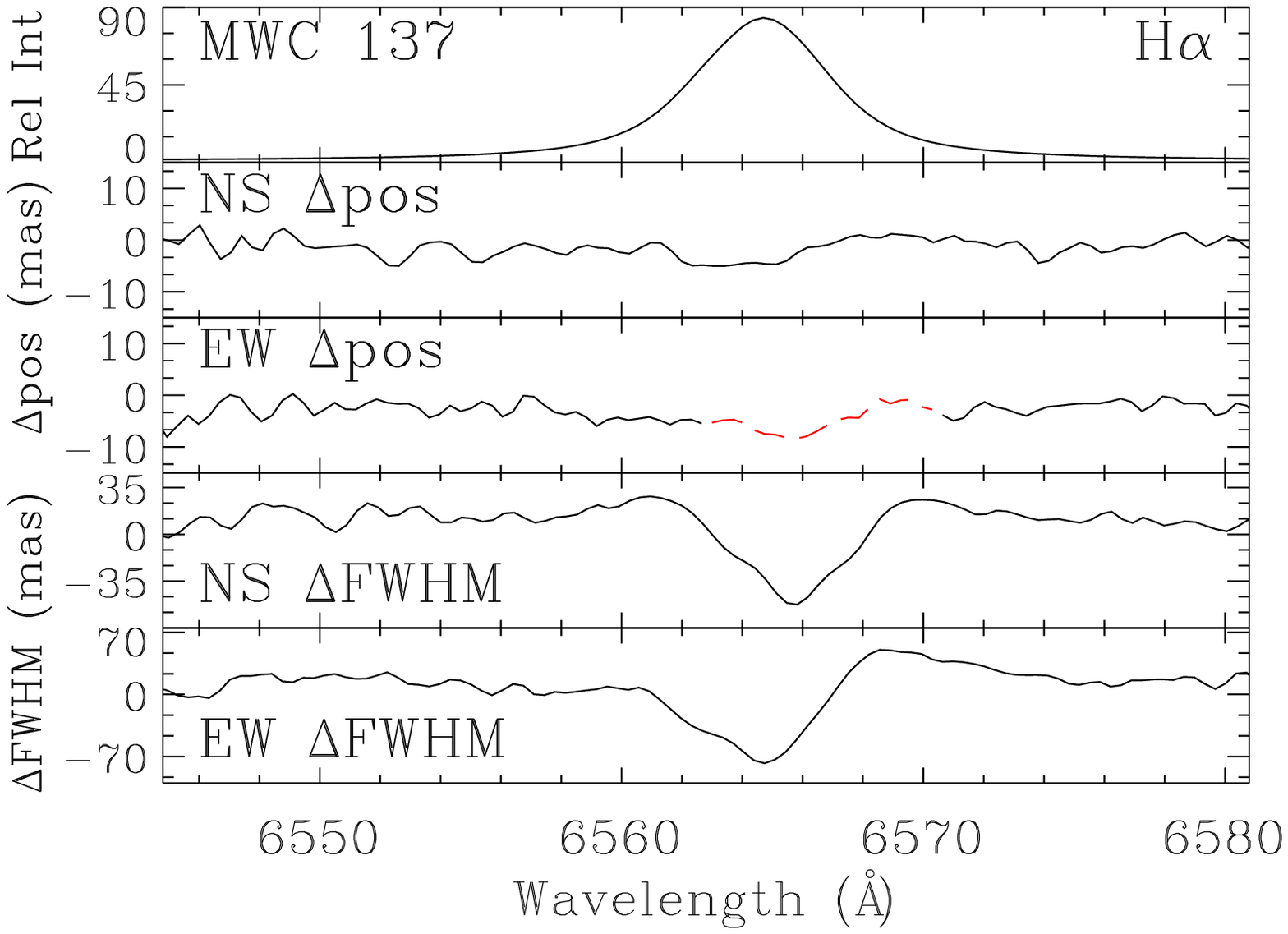} &
        \includegraphics[width=60mm,height=70mm]{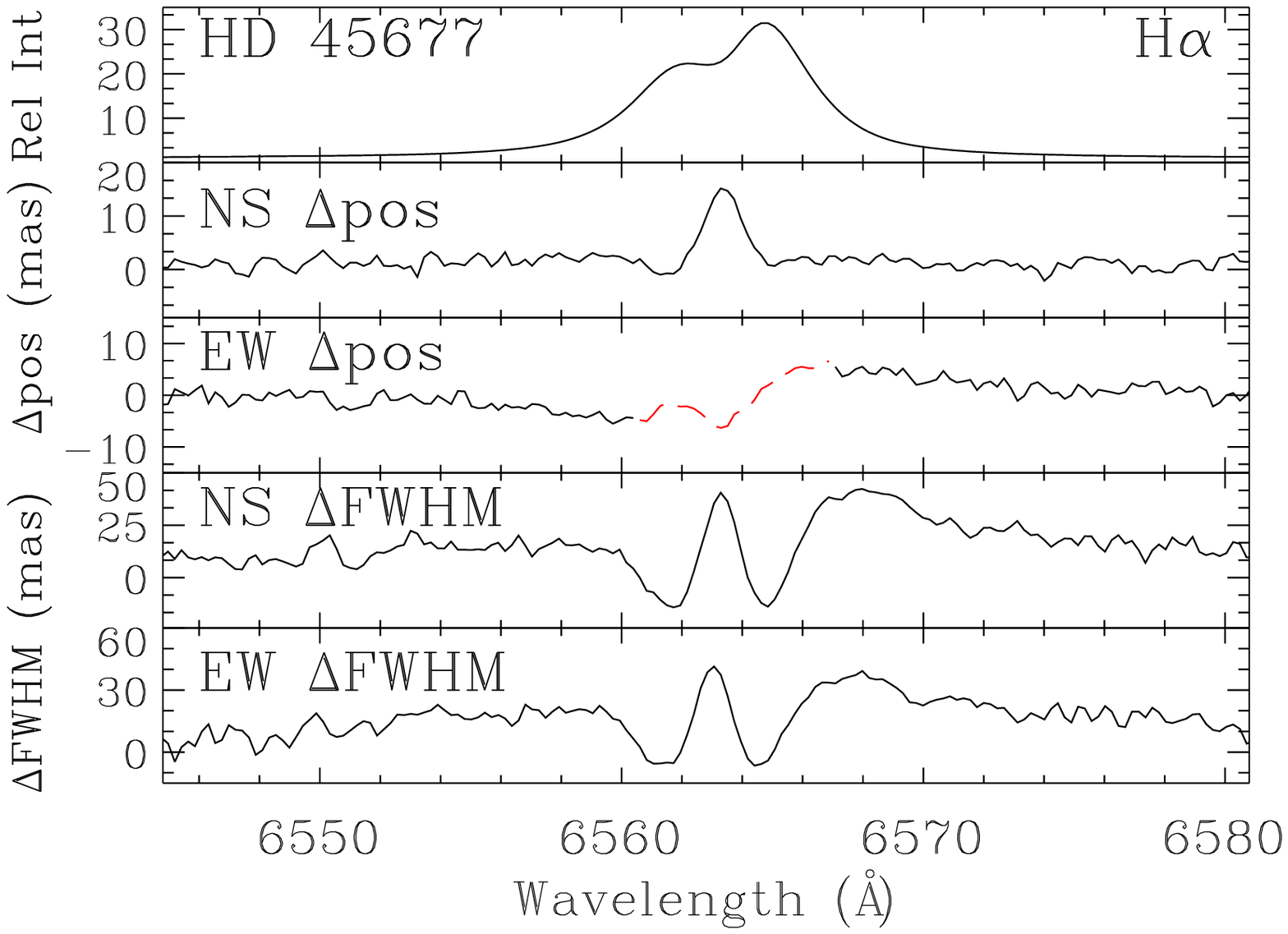} \\

	\includegraphics[width=60mm,height=70mm]{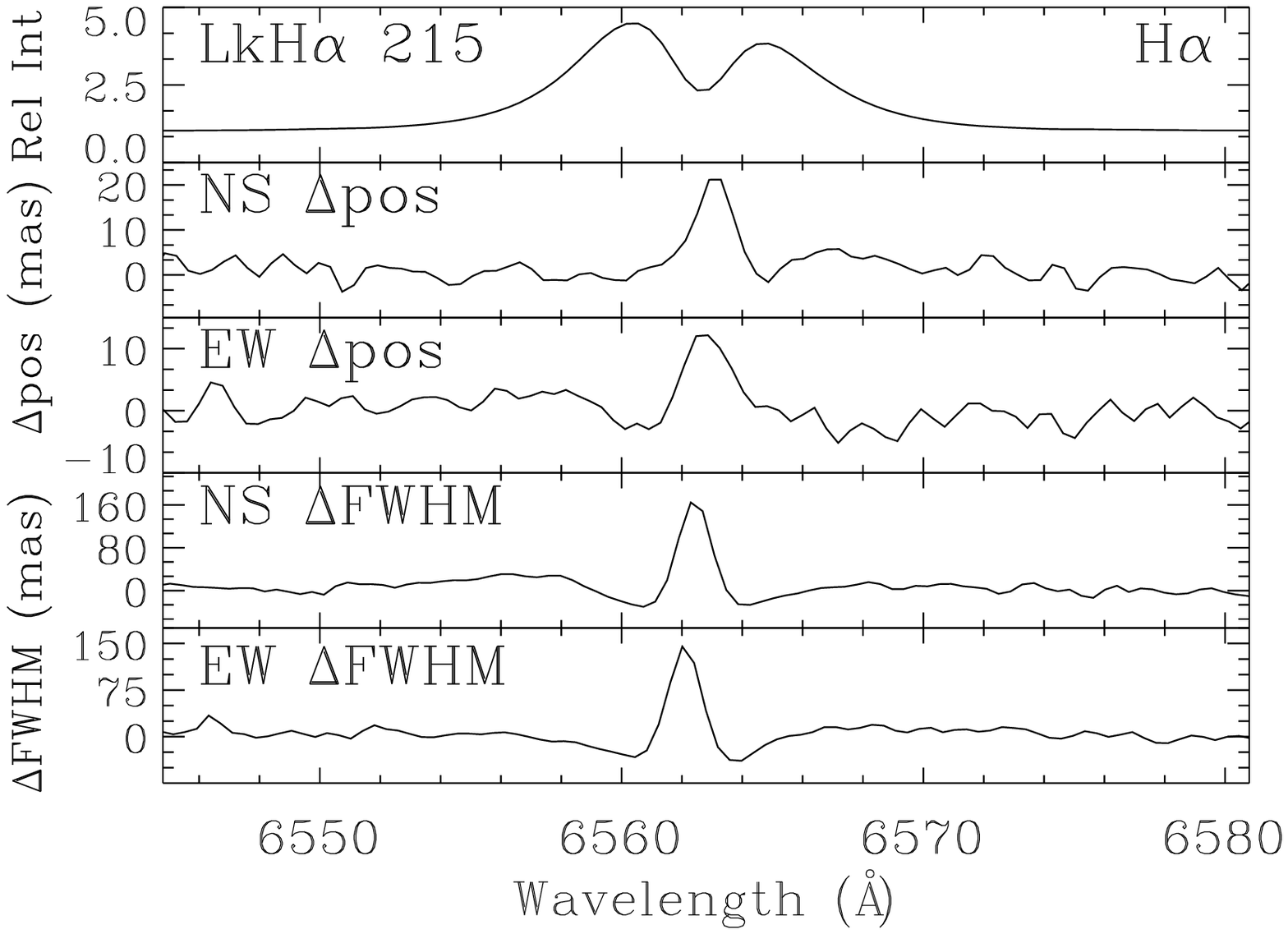} &
        \includegraphics[width=60mm,height=70mm]{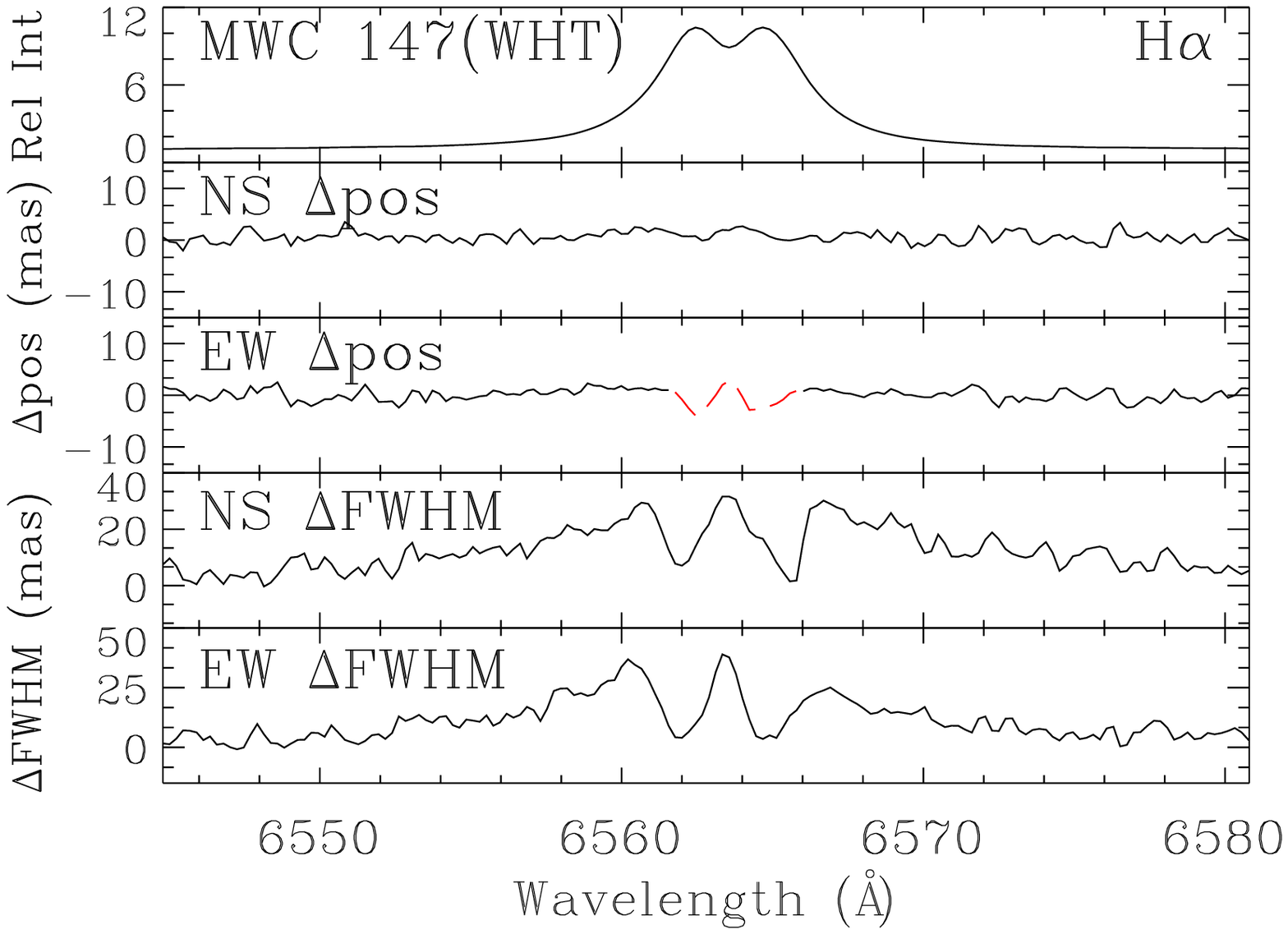} &
        \includegraphics[width=60mm,height=70mm]{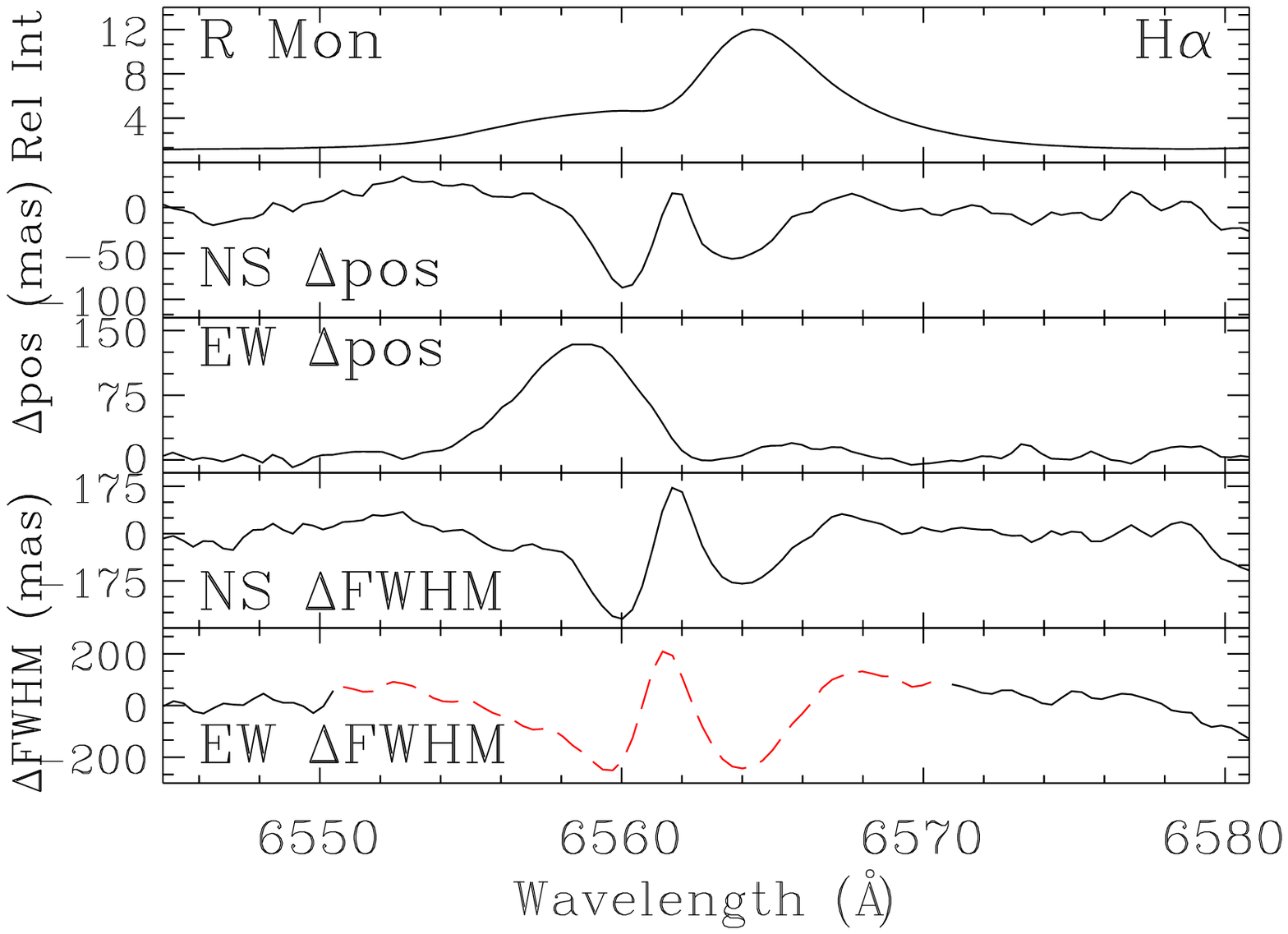} \\           
\end{tabular}
    
    \caption{H$\rm \alpha$ profiles and spectroastrometric signatures. From \textit{left} to \textit{right}: HD 244604, HD 37357, MWC 120 (data from the WHT), MWC 120 (data from the INT), MWC 137, HD 45677, LkH$\mathrm{\alpha}$ 215, MWC 147 (data from the WHT) and R Mon.}
\label{spec_ast_fig}
  \end{figure*}

\addtocounter{figure}{-1}

  \begin{figure*}
     \vspace*{9mm}
      \begin{tabular}{c c c}
	\includegraphics[width=60mm,height=70mm]{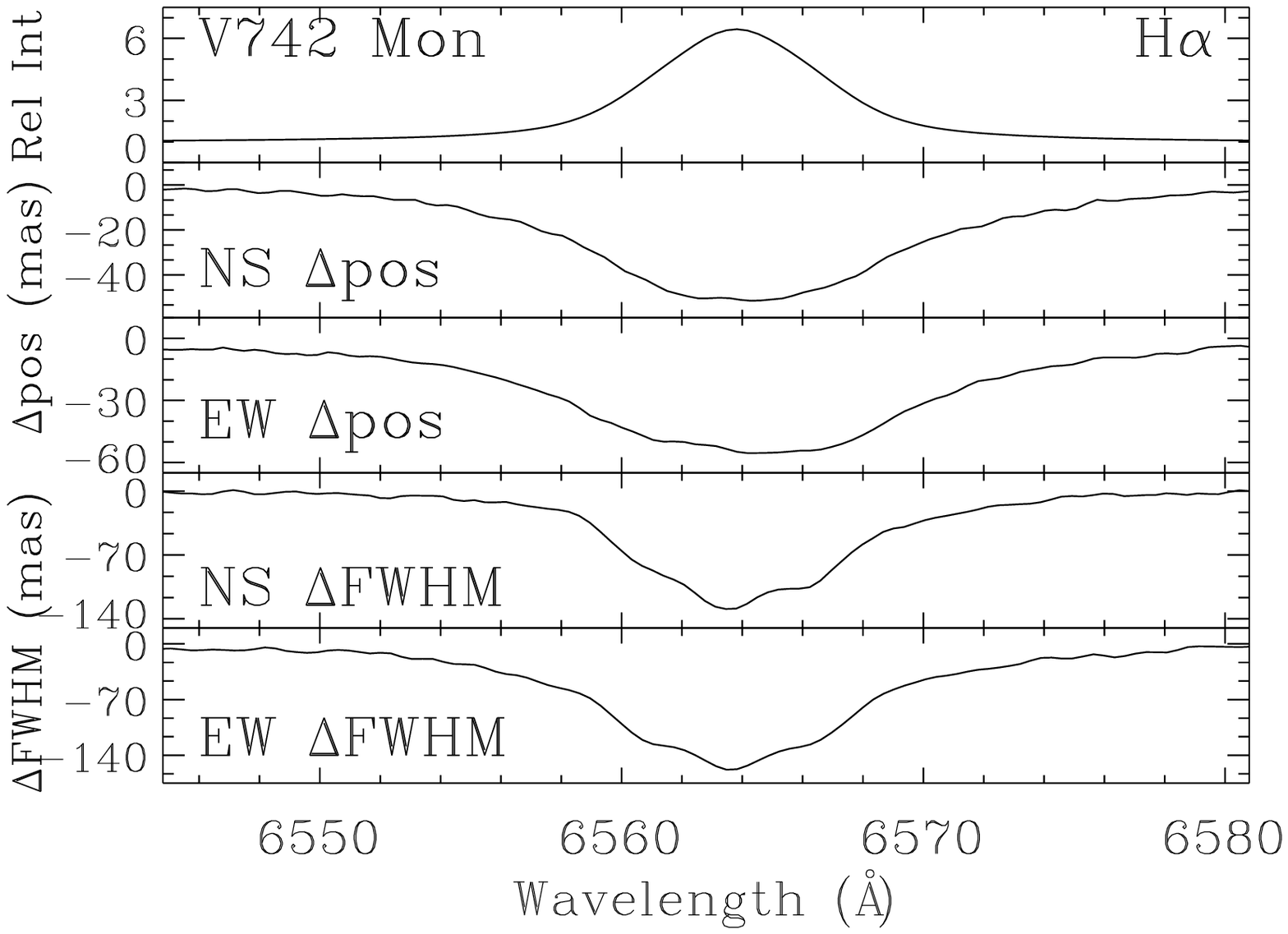} & 
	\includegraphics[width=60mm,height=70mm]{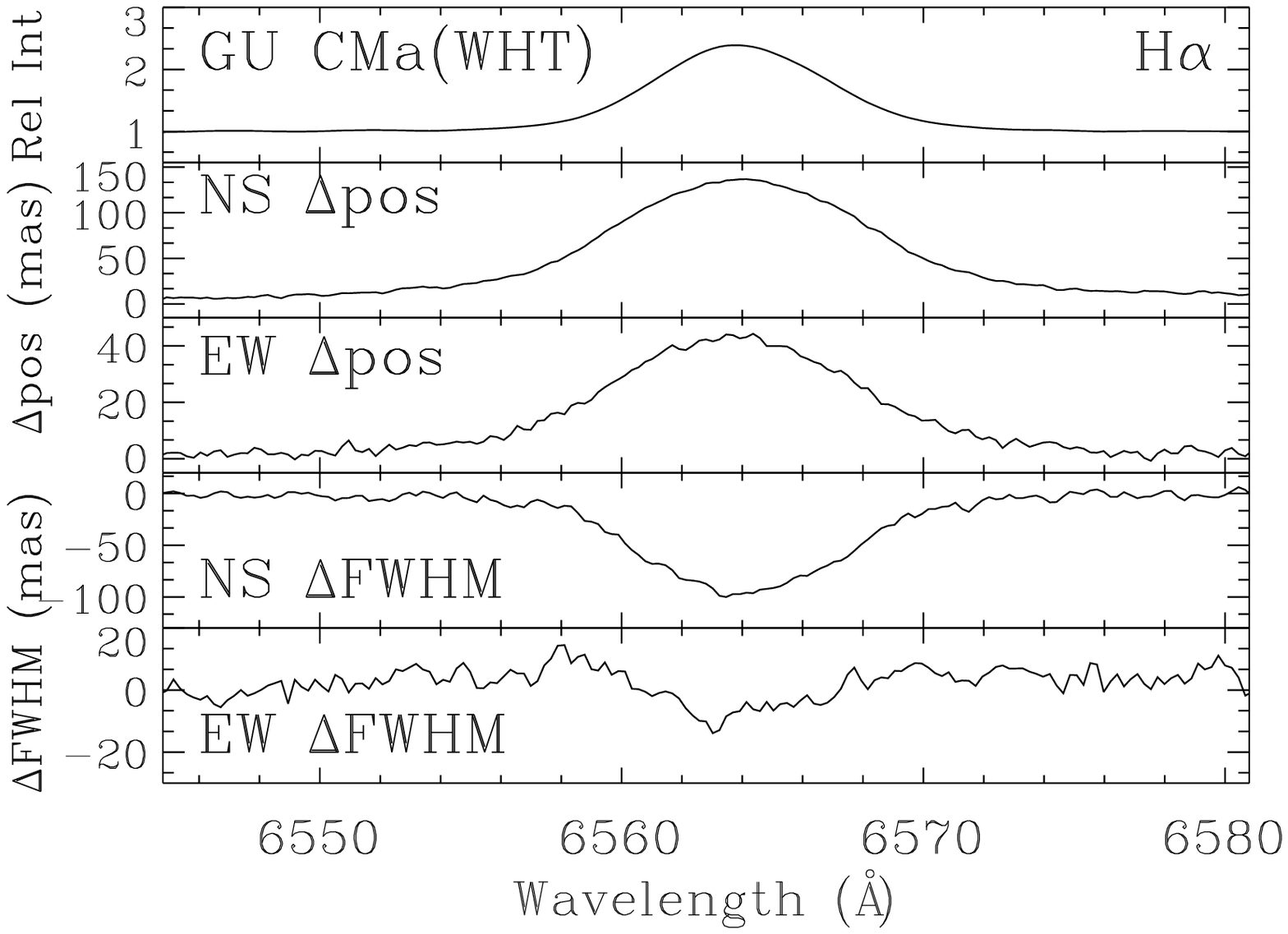} & 	\includegraphics[width=60mm,height=70mm]{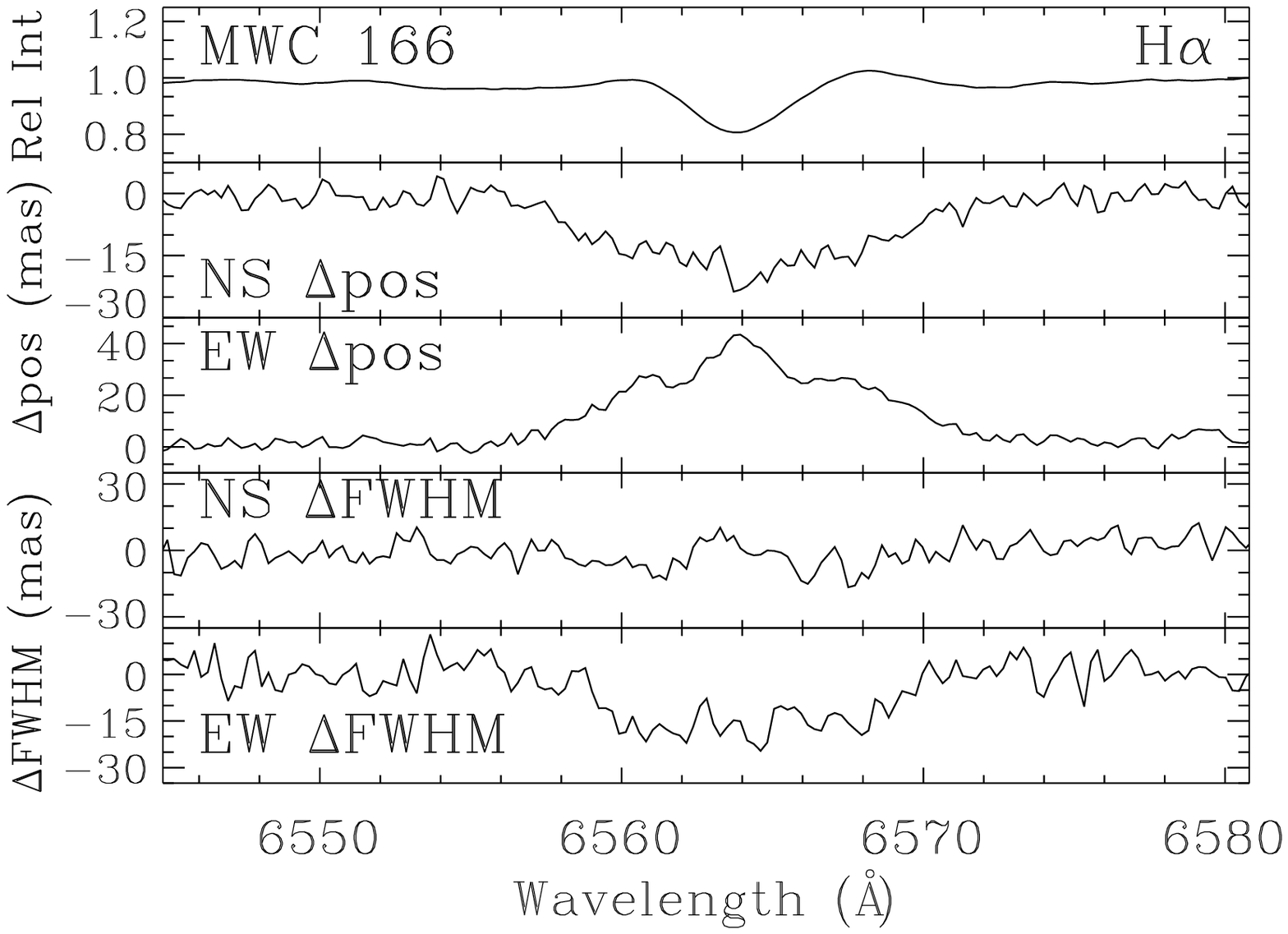}\\ 

      	\includegraphics[width=60mm,height=70mm]{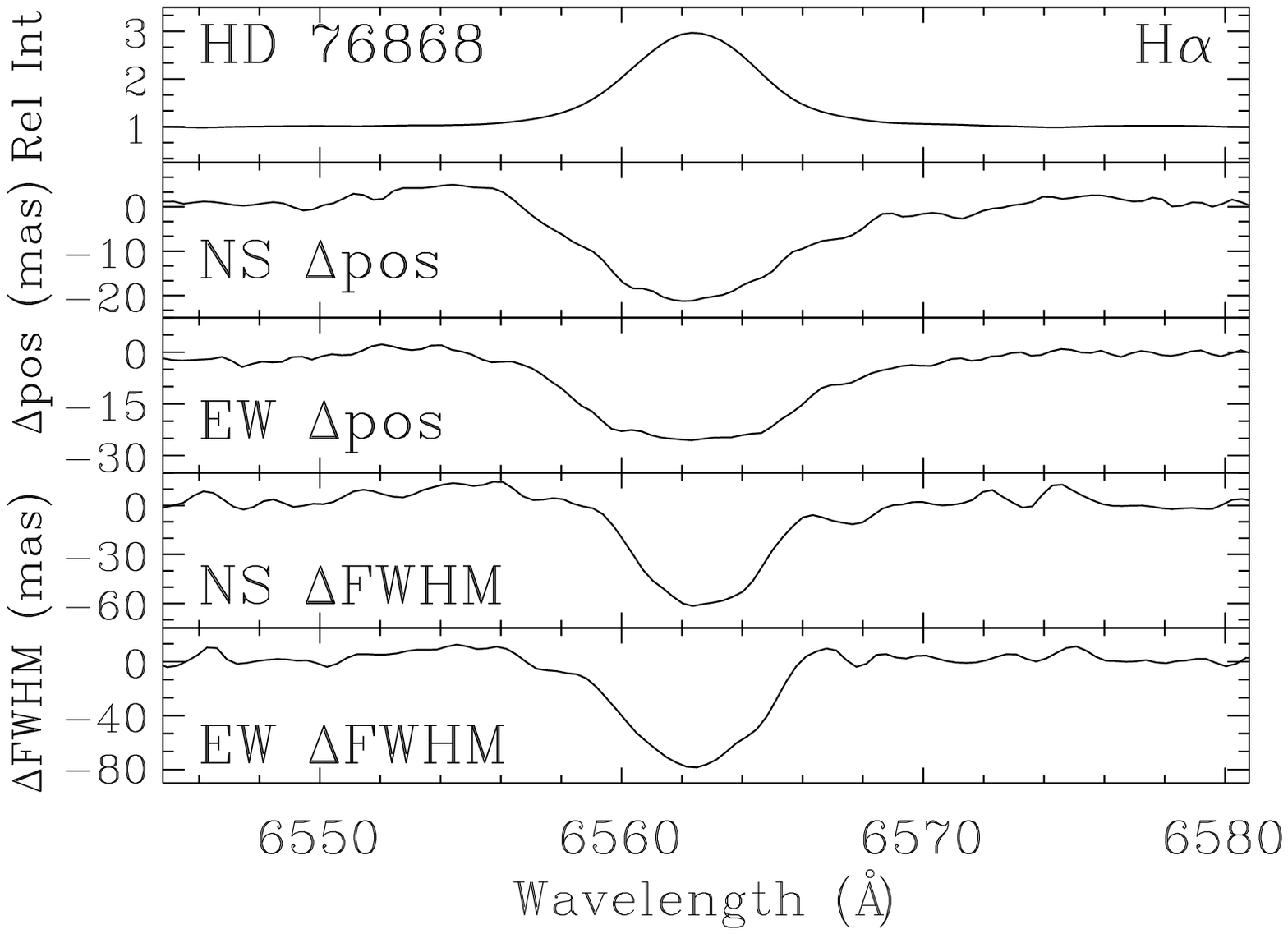} & 
	\includegraphics[width=60mm,height=70mm]{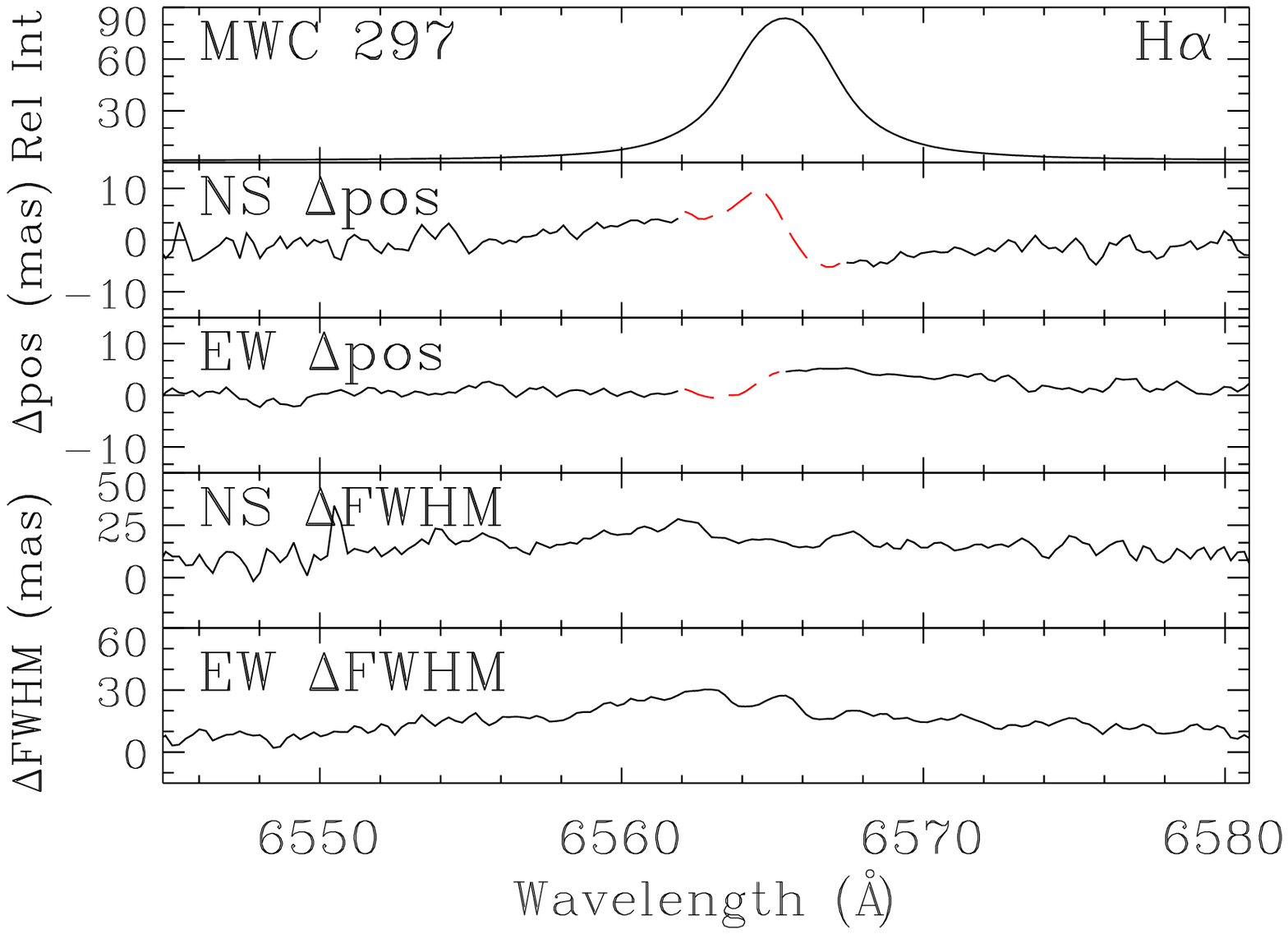} &
        \includegraphics[width=60mm,height=70mm]{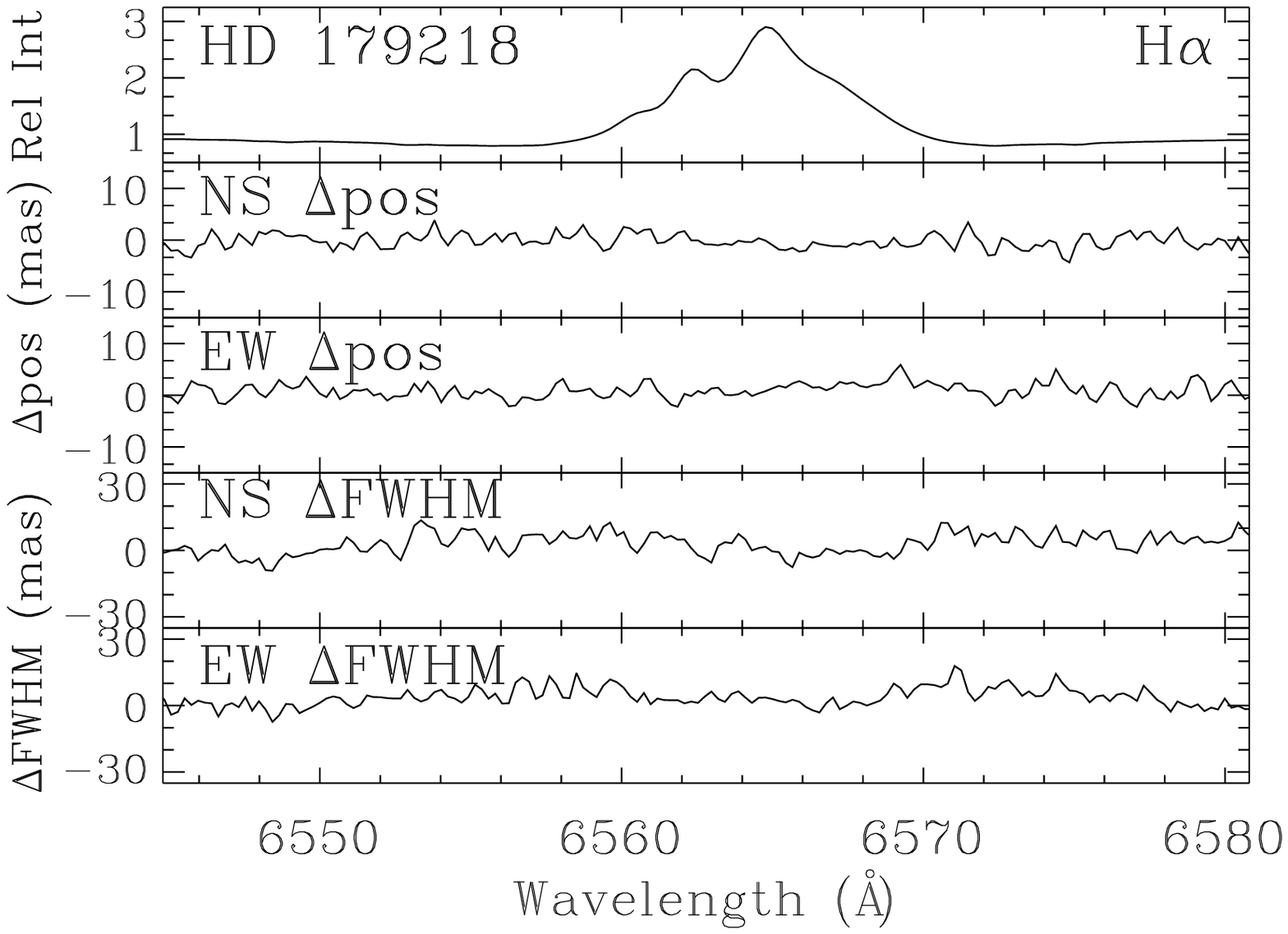} \\

	\includegraphics[width=60mm,height=70mm]{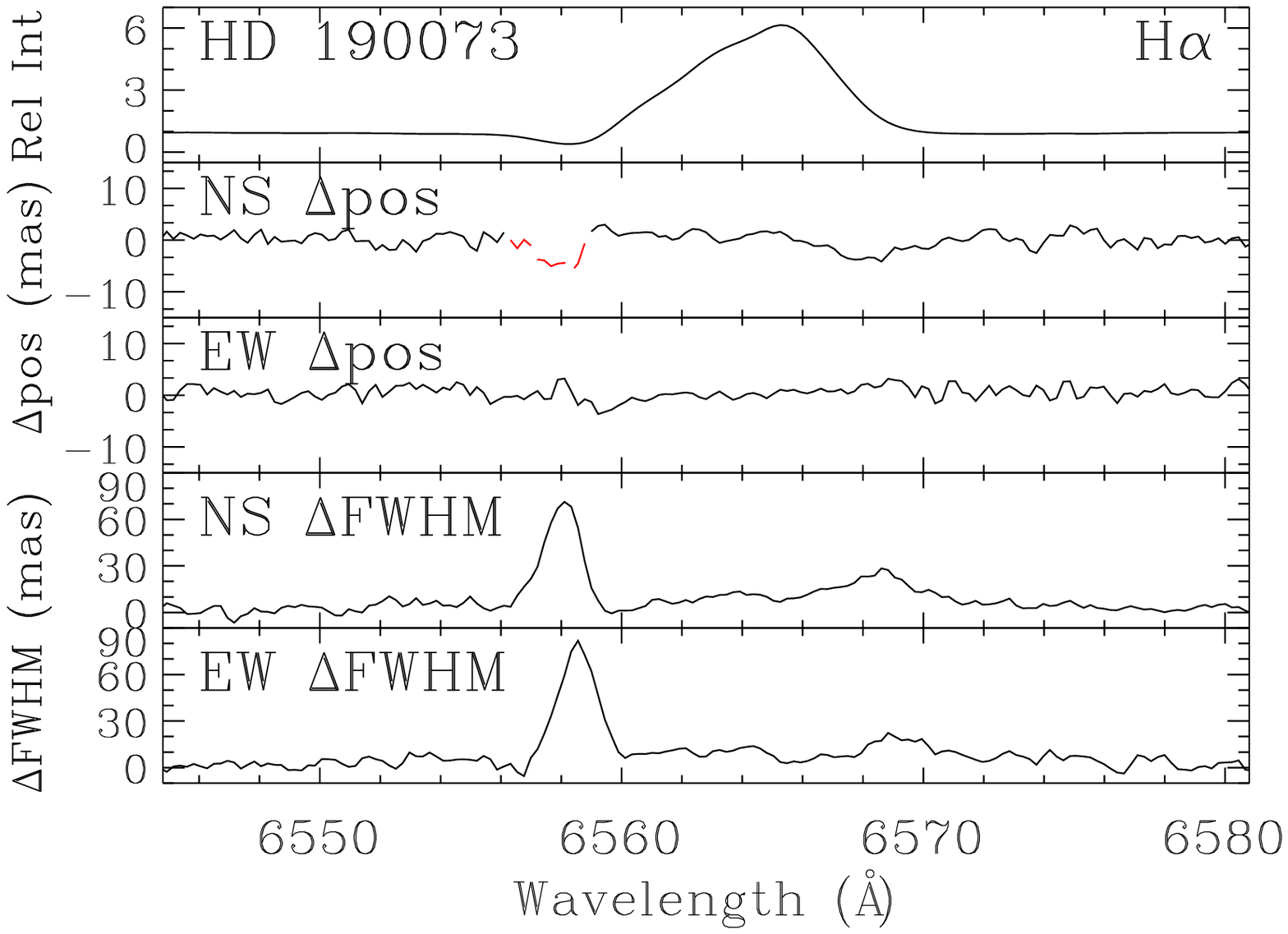} &
        \includegraphics[width=60mm,height=70mm]{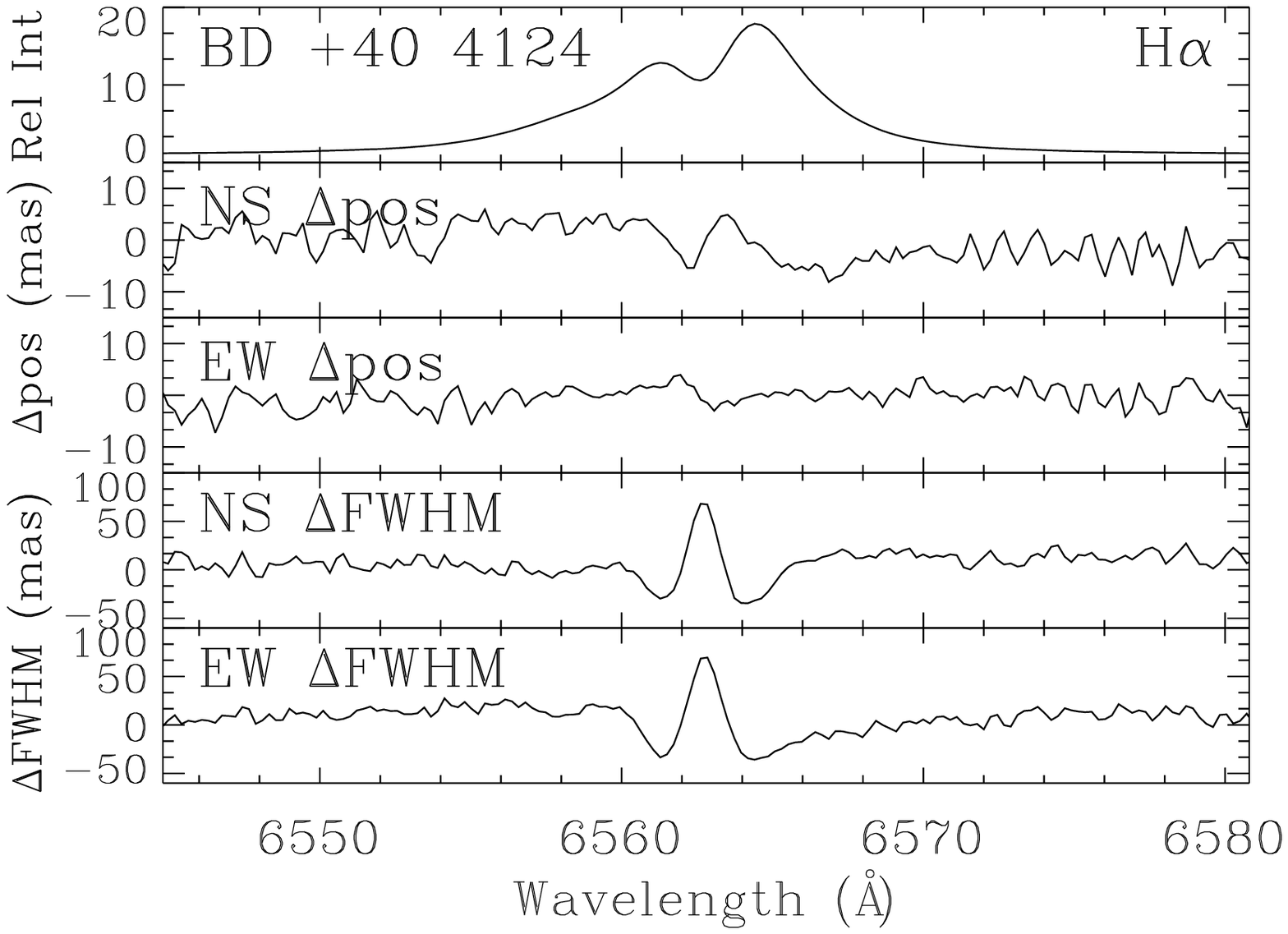} &
        \includegraphics[width=60mm,height=70mm]{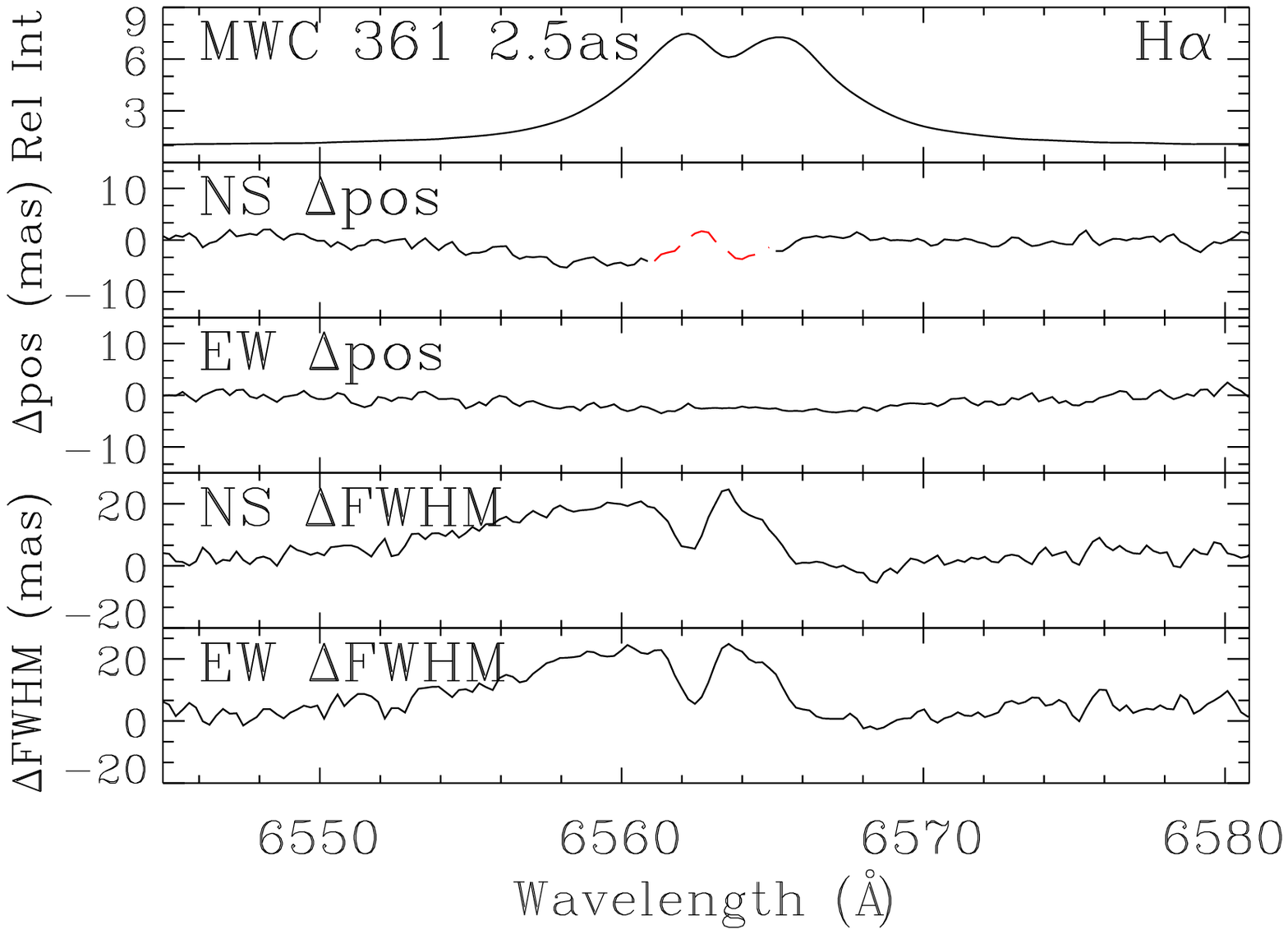} \\           
\end{tabular}
    
    \caption{H$\rm \alpha$ profiles and spectroastrometric signatures. From \textit{left} to \textit{right}: V742 Mon, GU CMa (data from the WHT), MWC 166, HD 76868, MWC 297, HD 179218, HD 190073, BD+40 4124 and MWC 361 (data obtained with a $\mathrm{2.5}$~arcsec slit).}
\label{spec_ast_fig}
  \end{figure*}

\addtocounter{figure}{-1}

  \begin{figure*}
     \vspace*{9mm}
      \begin{tabular}{c c c}
	\includegraphics[width=60mm,height=70mm]{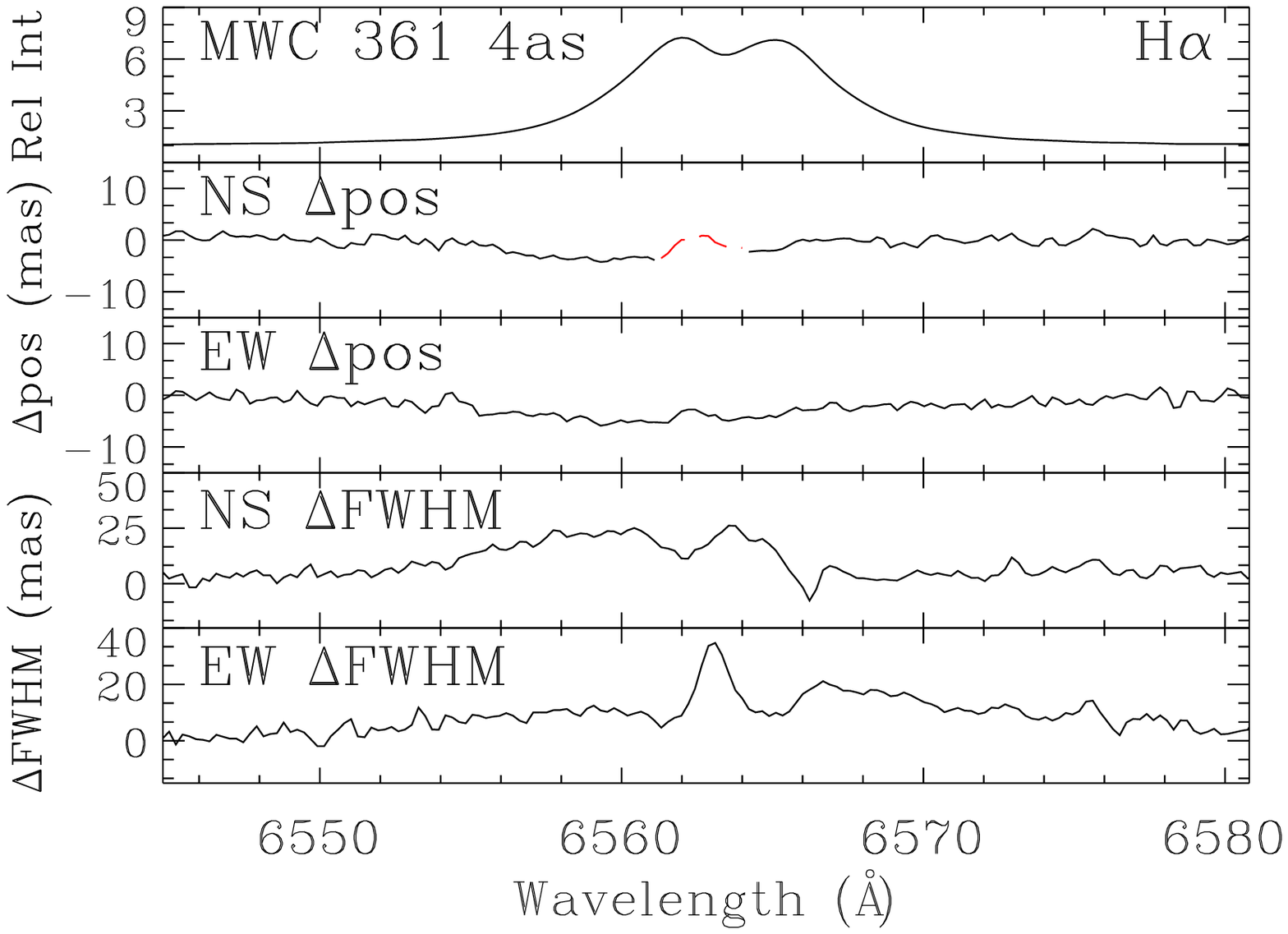} & 
	\includegraphics[width=60mm,height=70mm]{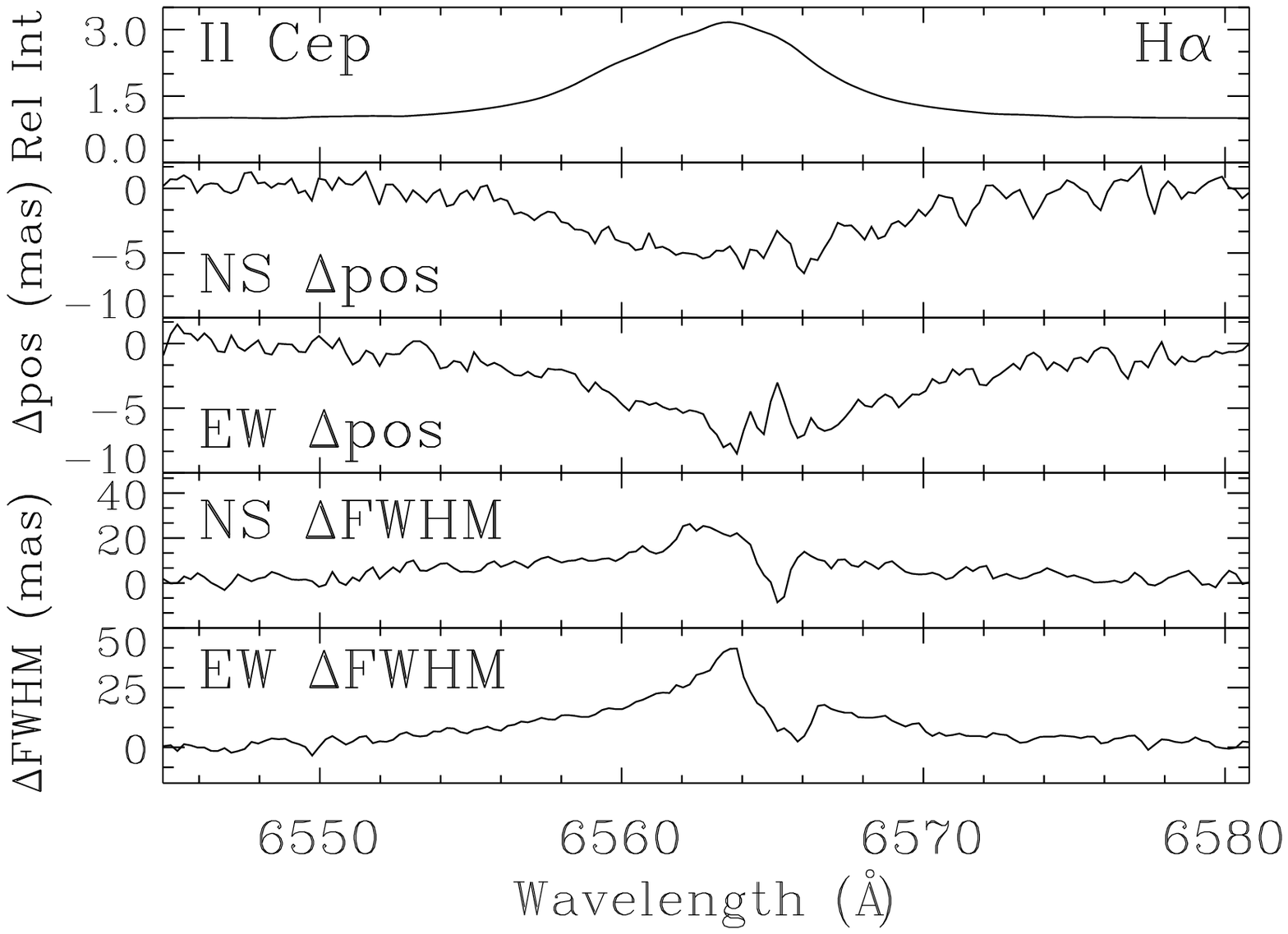} & 	\includegraphics[width=60mm,height=70mm]{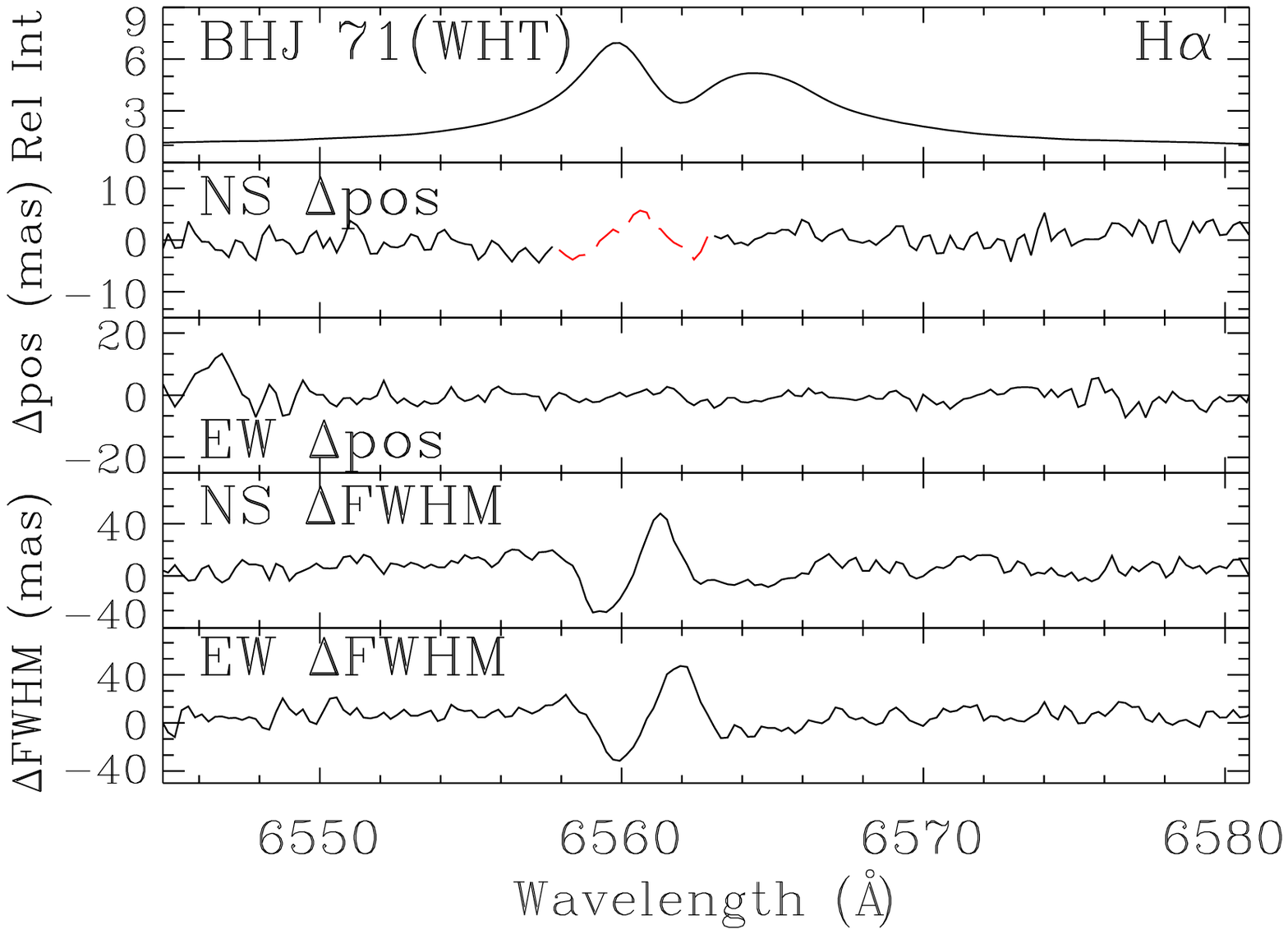}\\ 

      	\includegraphics[width=60mm,height=70mm]{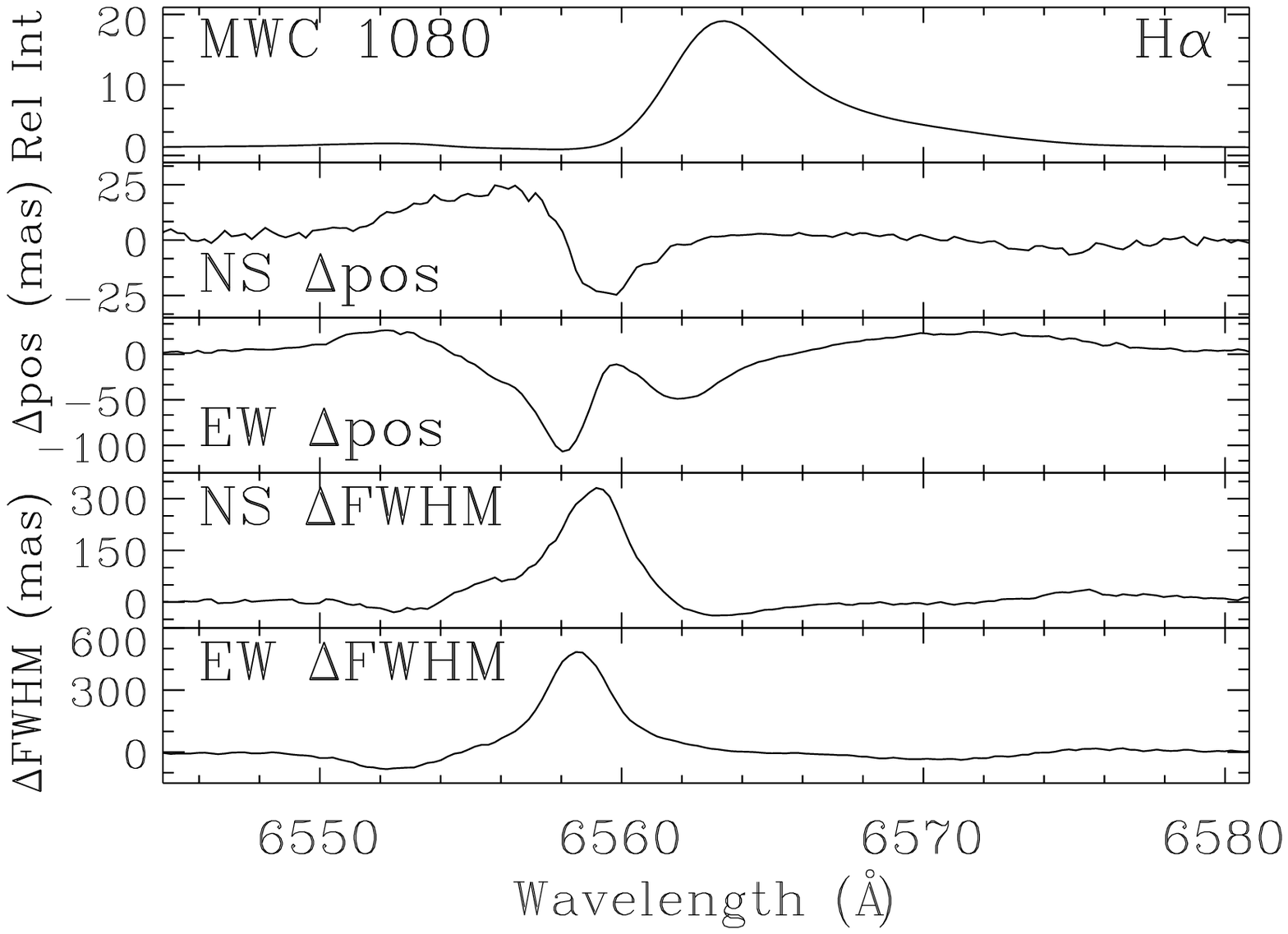} & 
	\includegraphics[width=60mm,height=70mm]{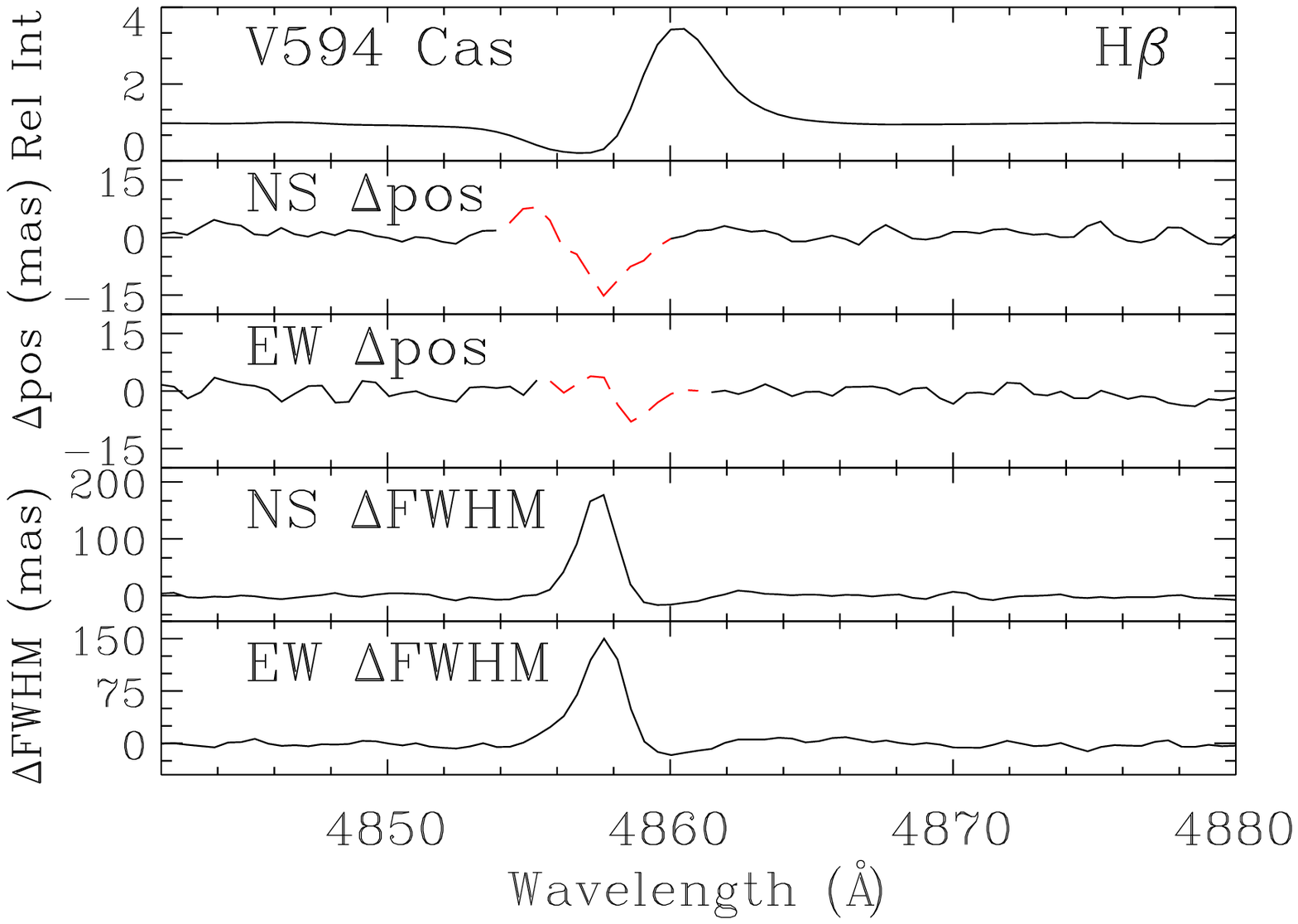} &
        \includegraphics[width=60mm,height=70mm]{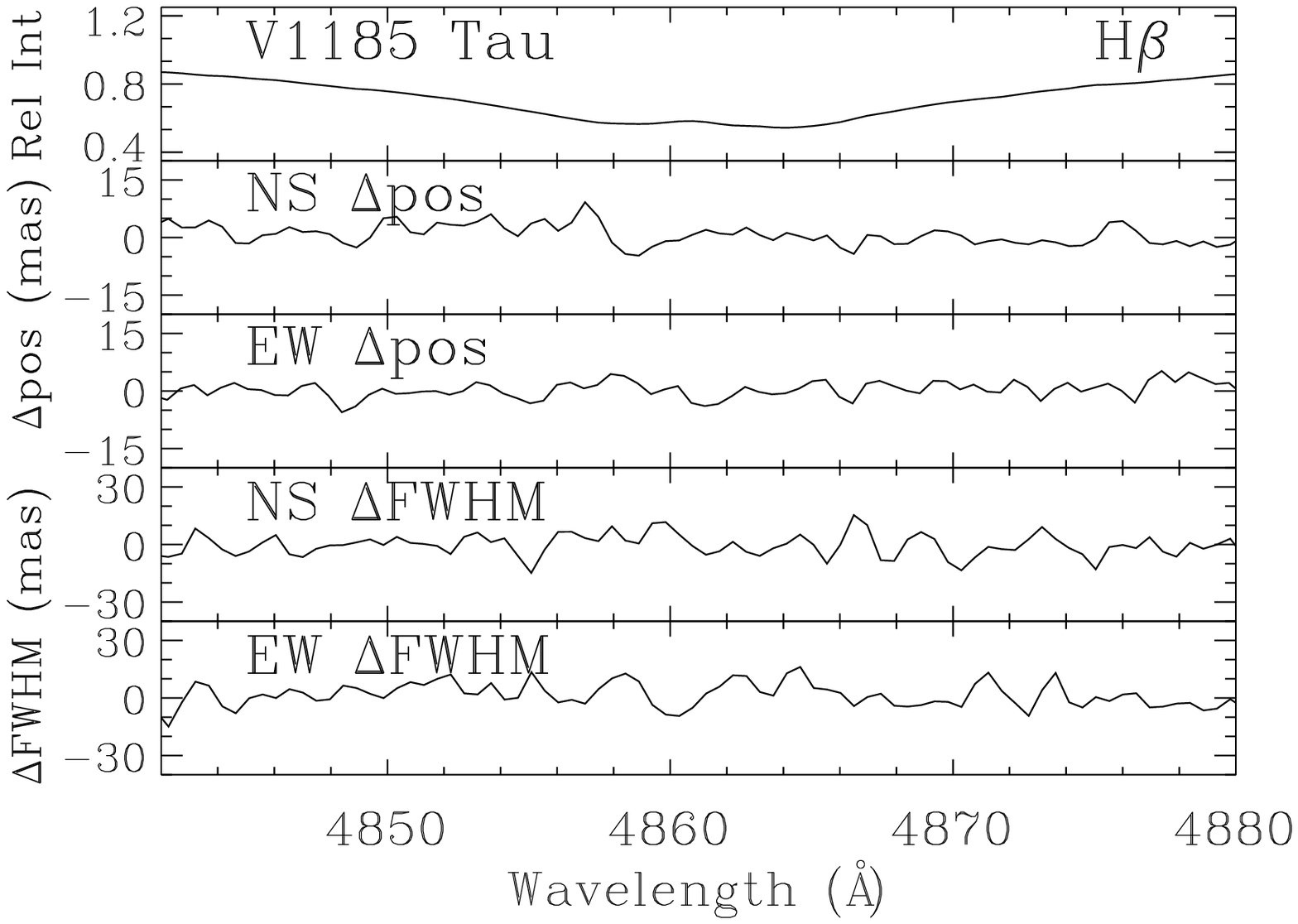} \\

	\includegraphics[width=60mm,height=70mm]{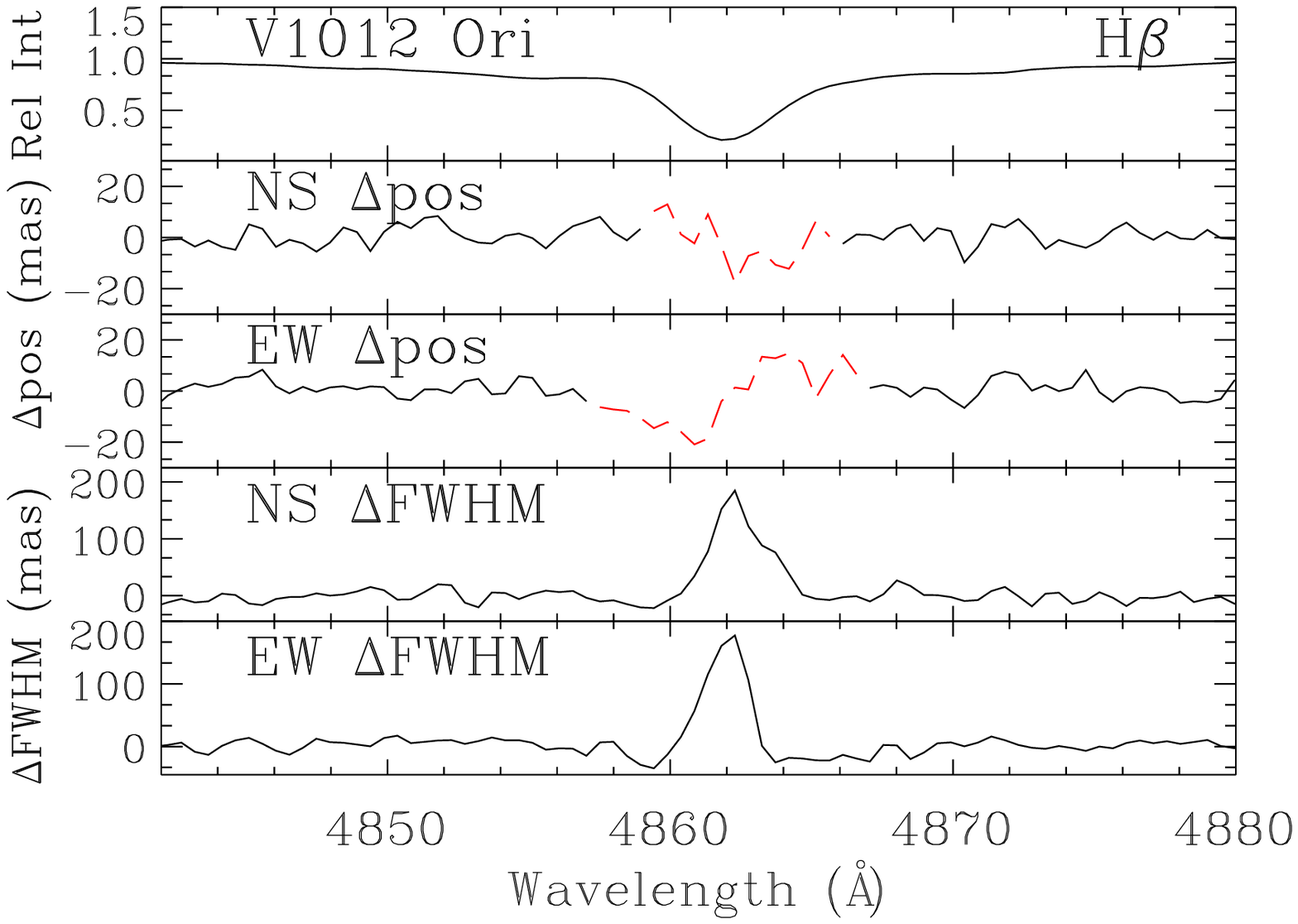} &
        \includegraphics[width=60mm,height=70mm]{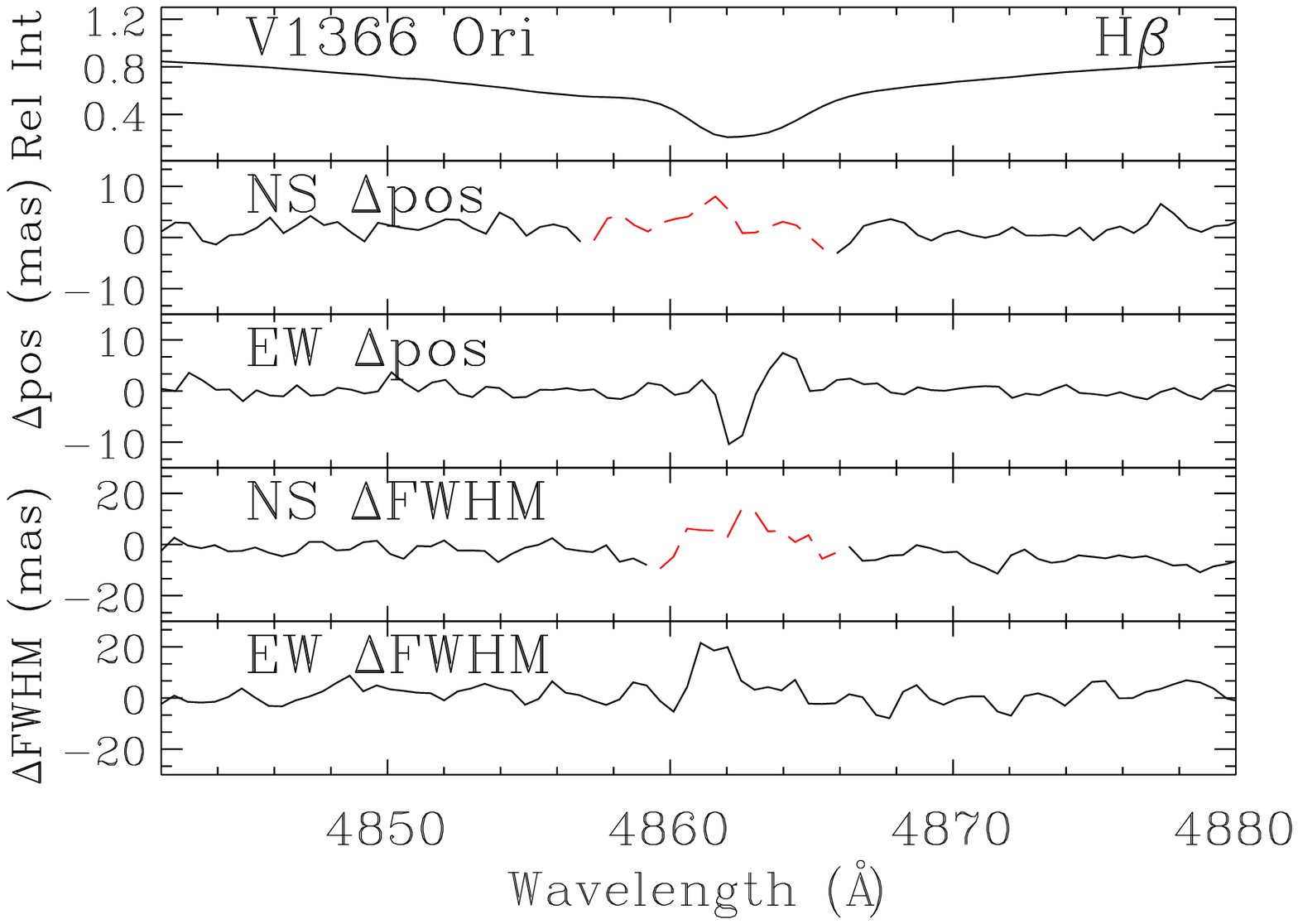} &
        \includegraphics[width=60mm,height=70mm]{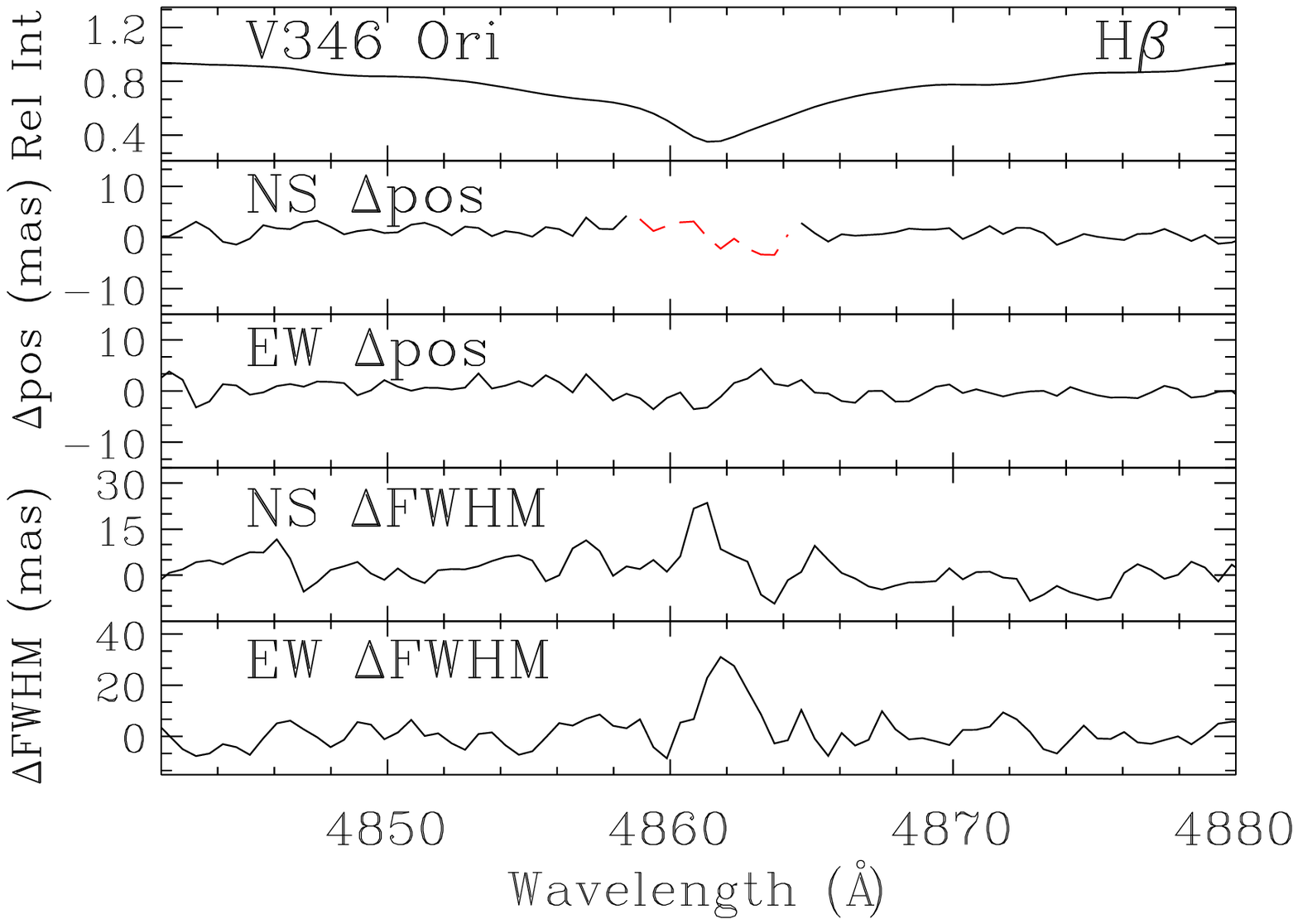} \\           
\end{tabular}
    
    \caption{H$\rm \alpha$ and H$\rm \beta$ profiles and spectroastrometric signatures. From \textit{left} to \textit{right}: MWC 361 (data obtained with a $\mathrm{4}$~arcsec slit), Il Cep, BHJ 71 (data from the WHT), MWC 1080, V594 Cas, V1185 Tau, V1012 Ori, V1366 Ori and V346 Ori.}
\label{spec_ast_fig}
  \end{figure*}

\addtocounter{figure}{-1}

  \begin{figure*}
     \vspace*{9mm}
      \begin{tabular}{c c c}
	\includegraphics[width=60mm,height=70mm]{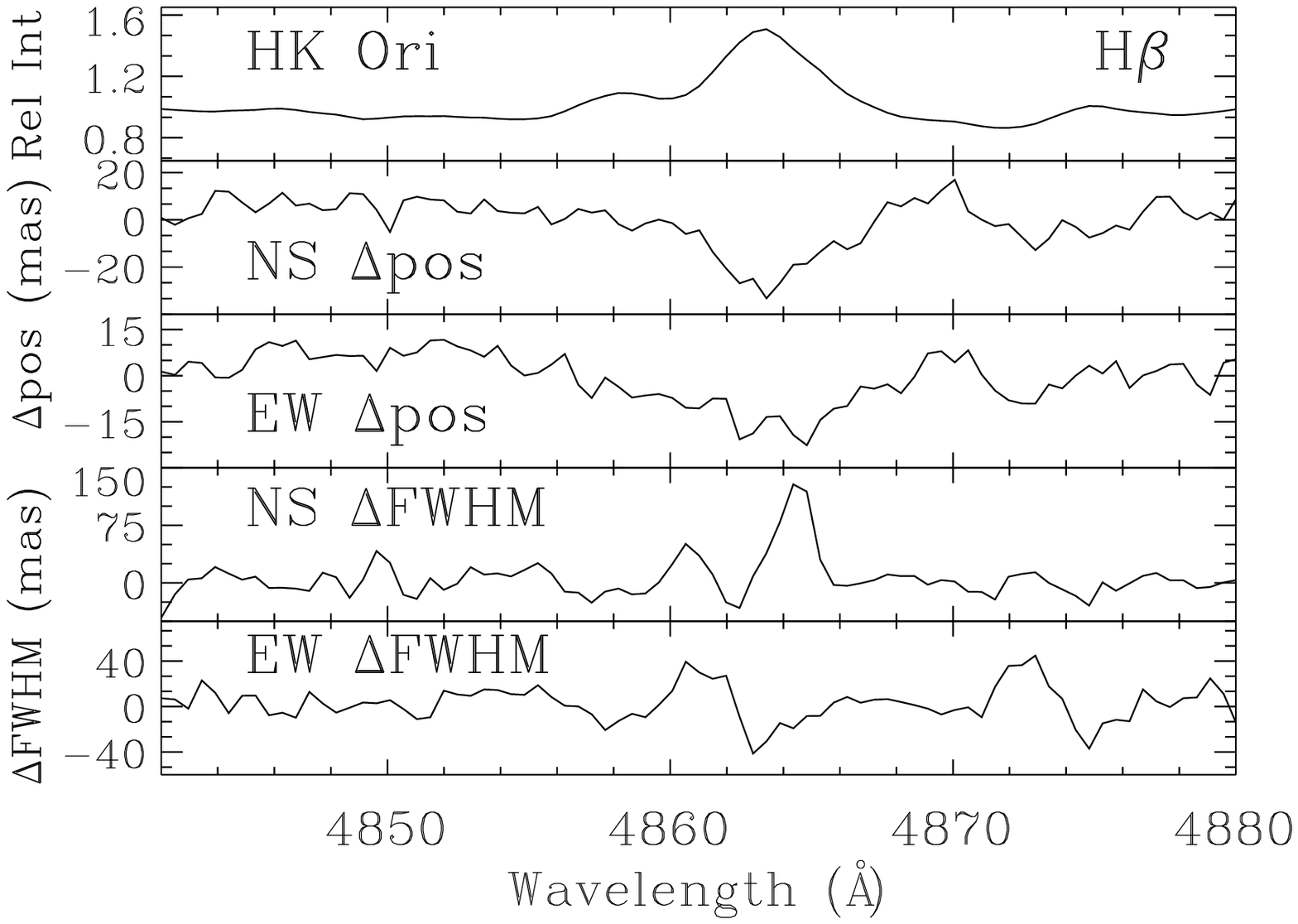} & 
	\includegraphics[width=60mm,height=70mm]{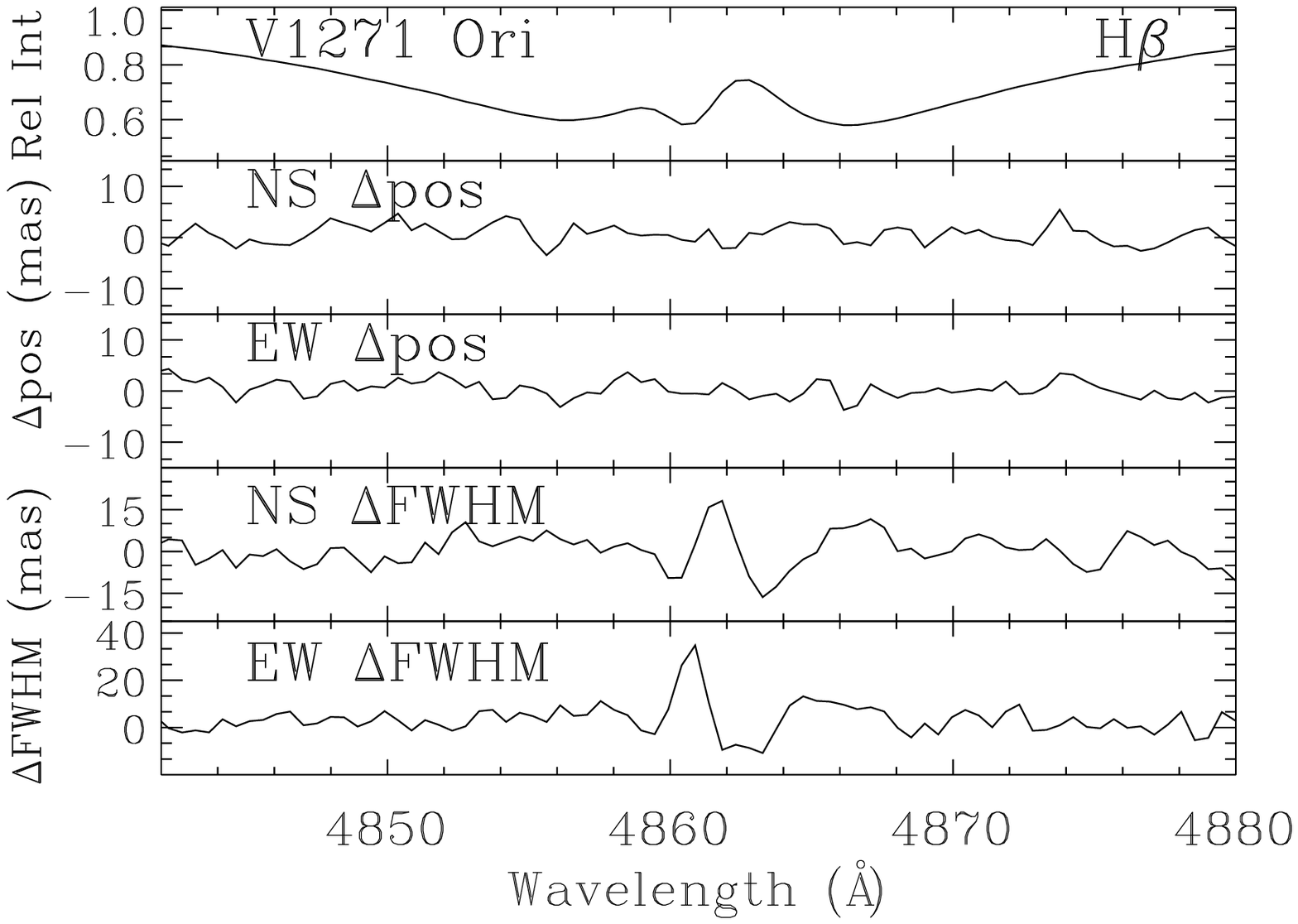} & 	\includegraphics[width=60mm,height=70mm]{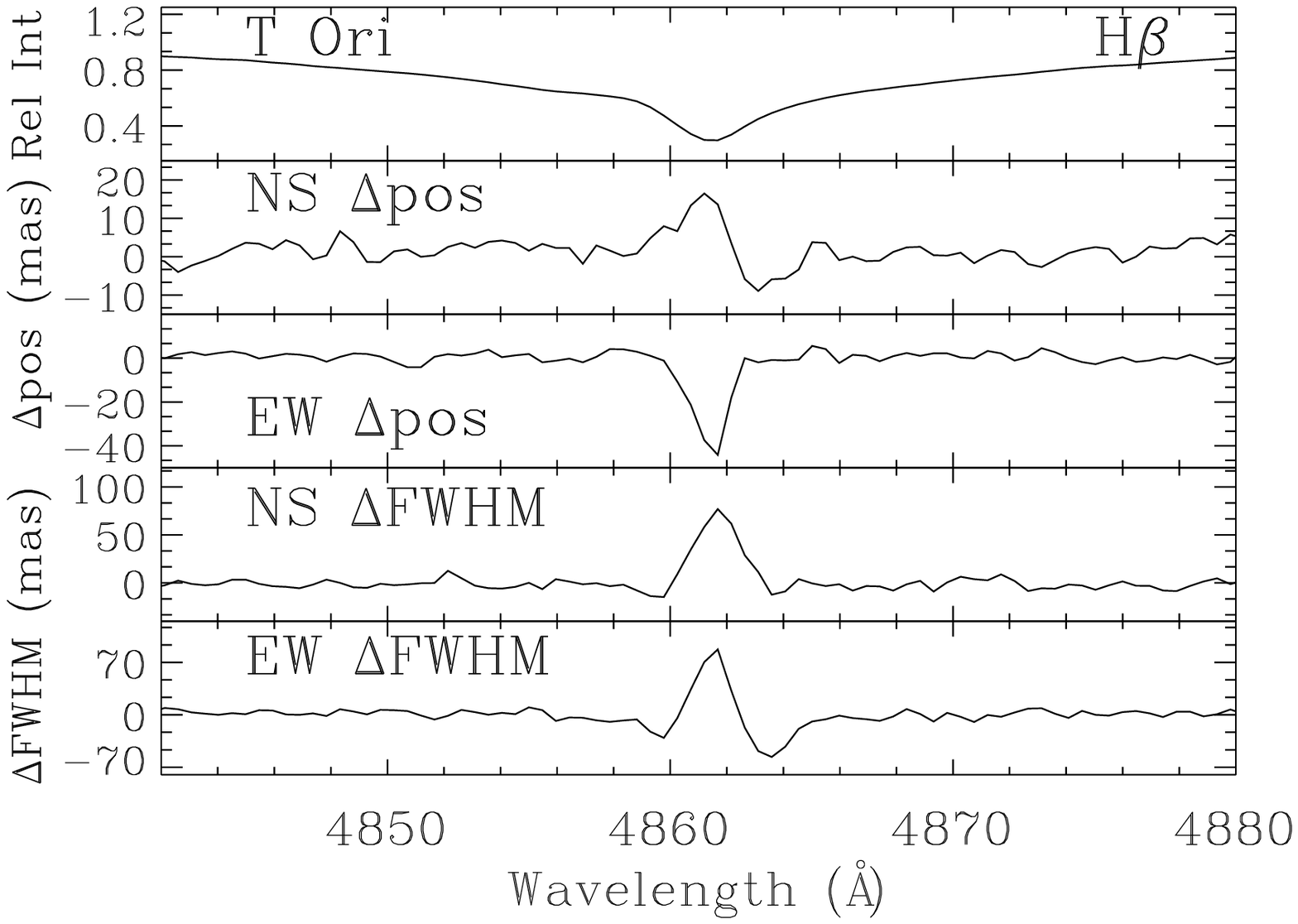}\\ 

      	\includegraphics[width=60mm,height=70mm]{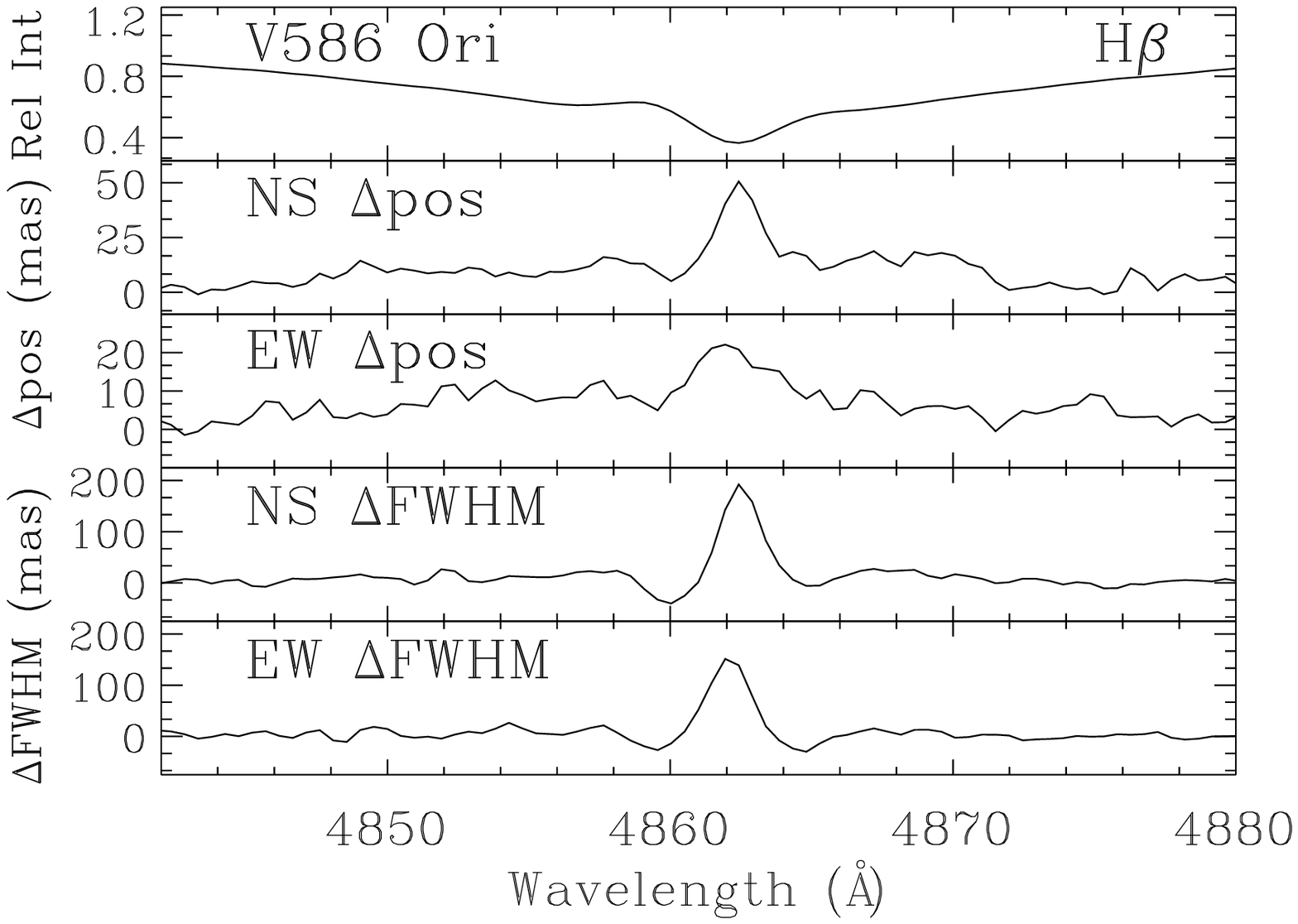} & 
	\includegraphics[width=60mm,height=70mm]{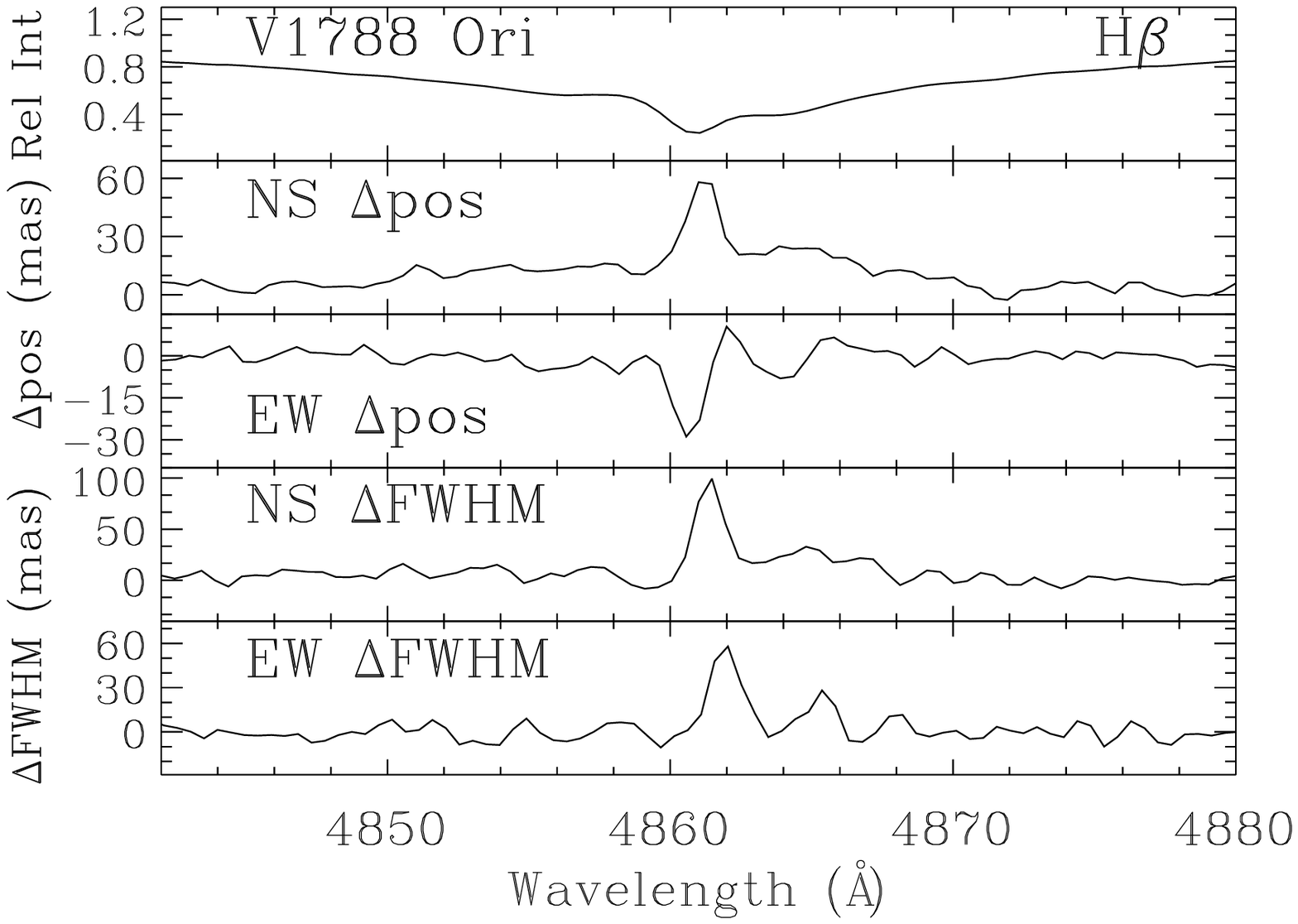} &
        \includegraphics[width=60mm,height=70mm]{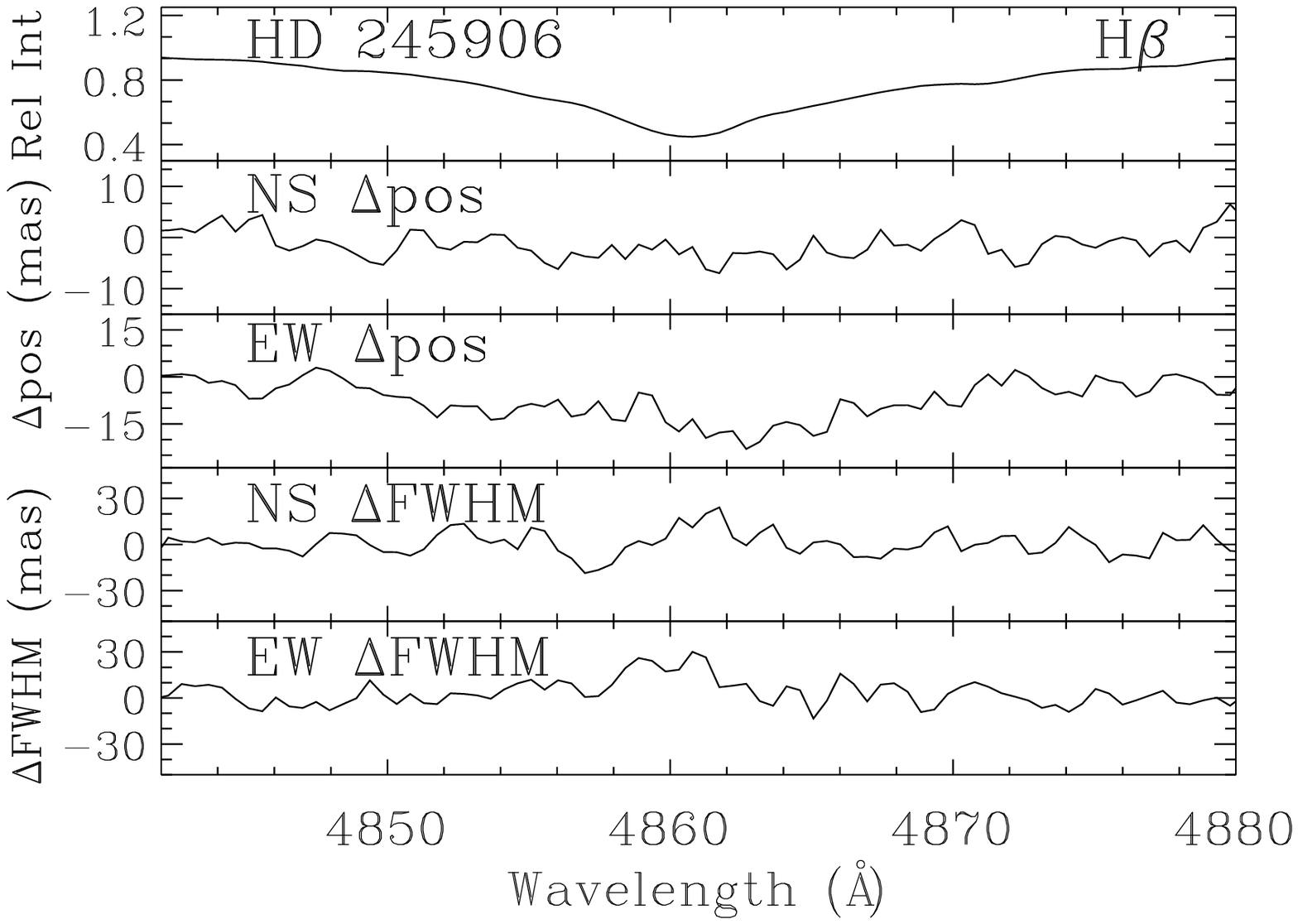} \\

	\includegraphics[width=60mm,height=70mm]{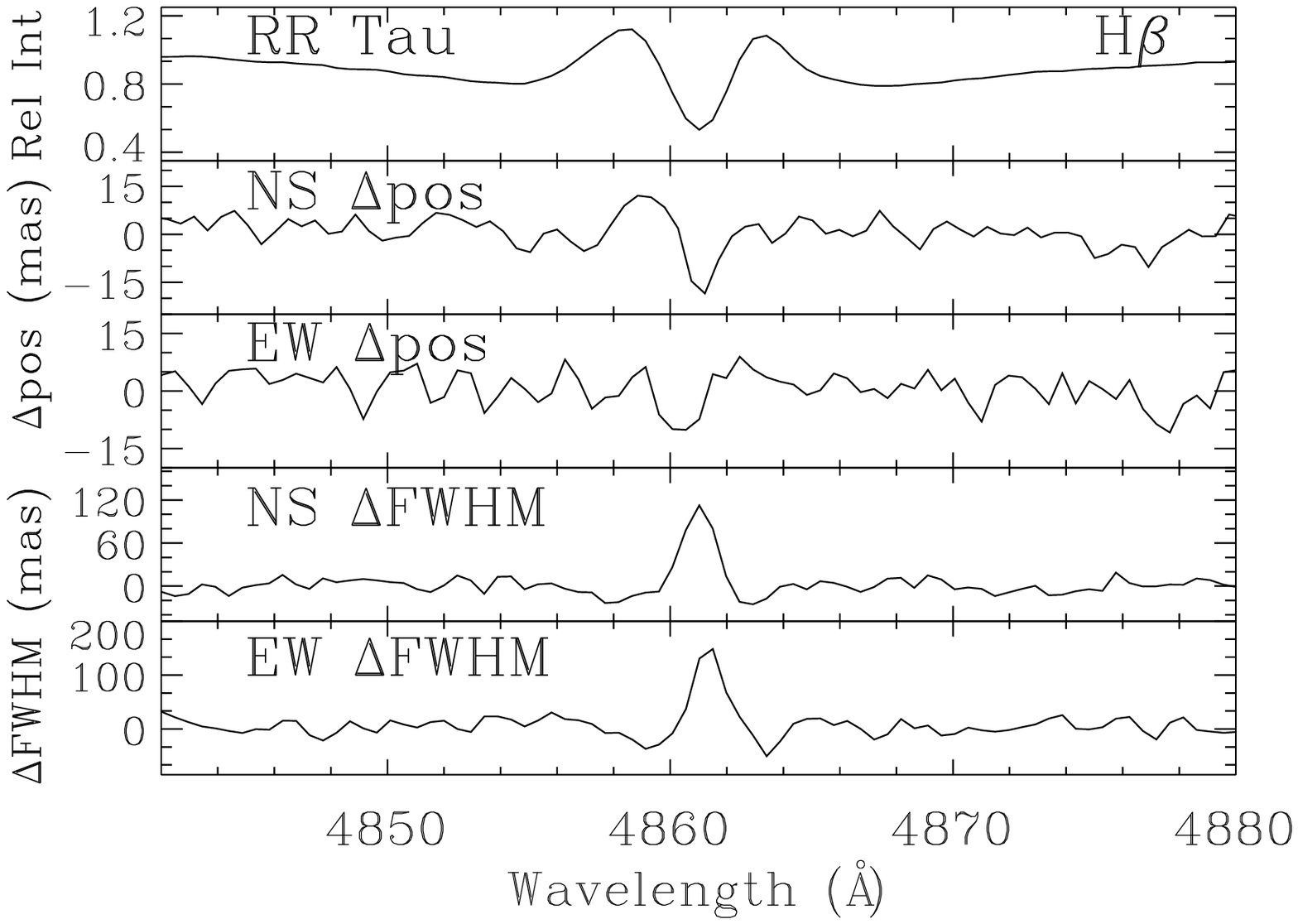} &
        \includegraphics[width=60mm,height=70mm]{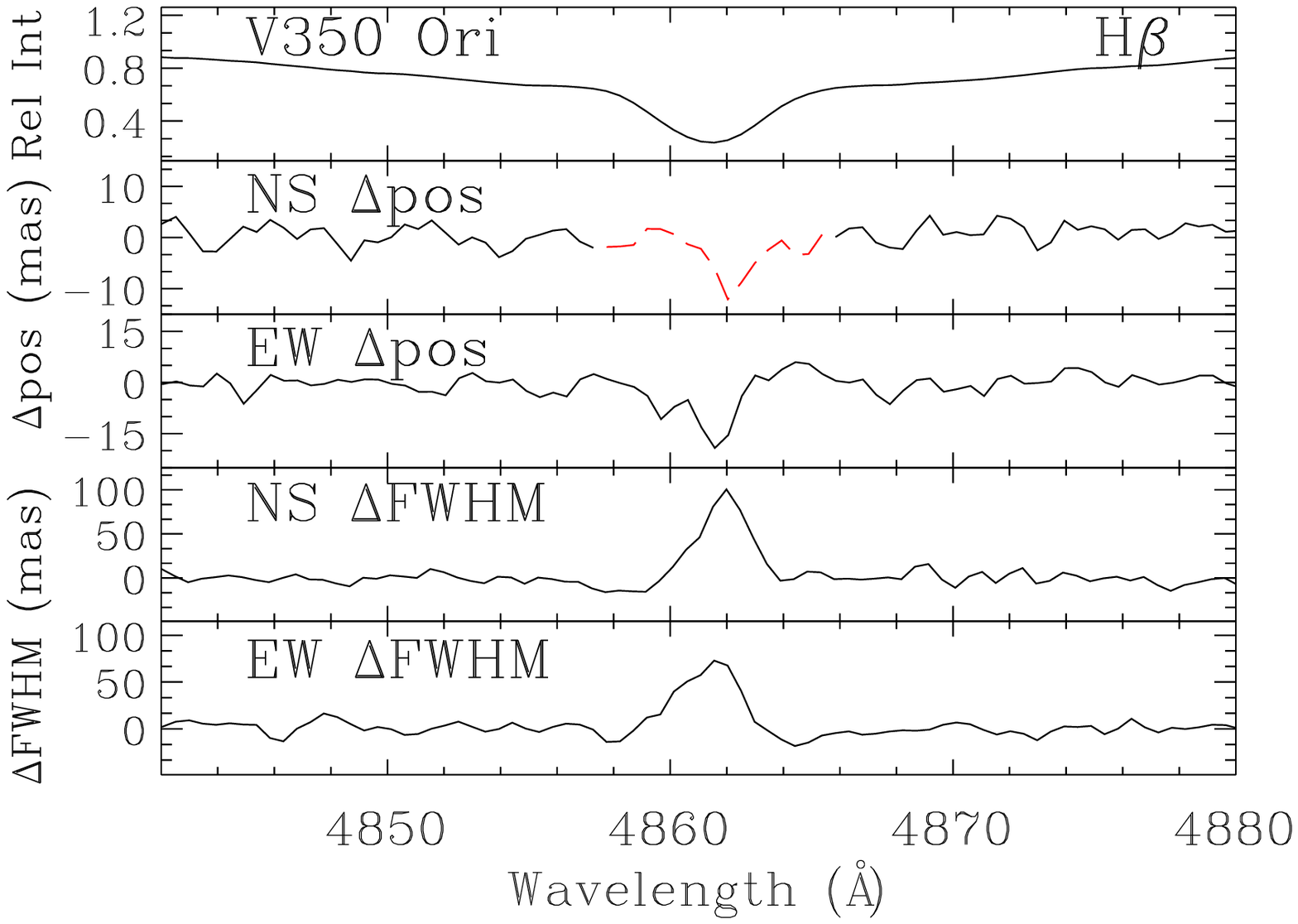} &
        \includegraphics[width=60mm,height=70mm]{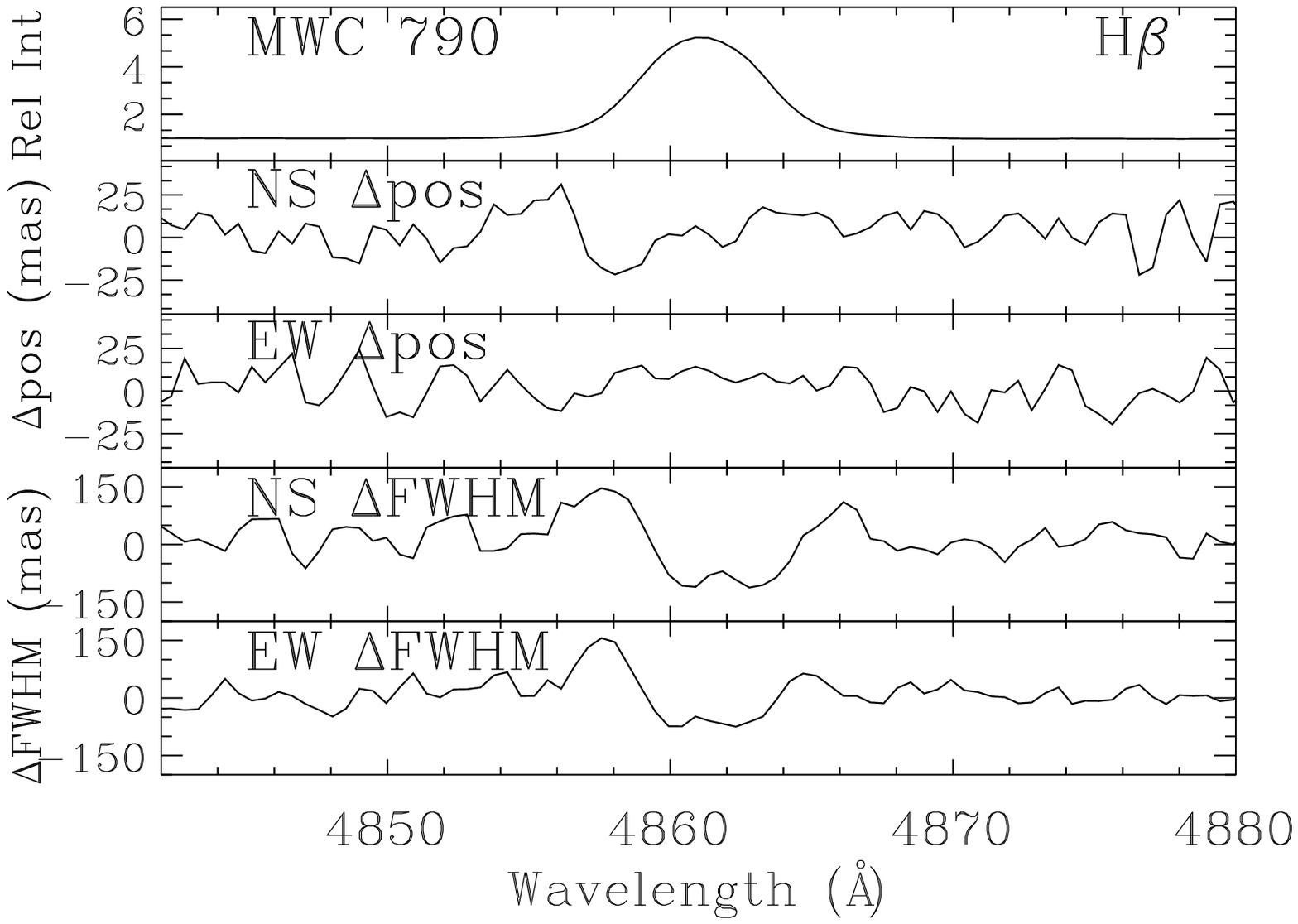} \\           
\end{tabular}
    
    \caption{H$\rm \beta$ profiles and spectroastrometric signatures. From \textit{left} to \textit{right}: HK Ori, V1271 Ori, T Ori, V586 Ori, V1788 Ori, HD 245906, RR Tau, V350 Ori and MWC 790.}
\label{spec_ast_fig}
  \end{figure*}

\addtocounter{figure}{-1}

  \begin{figure*}
     \vspace*{9mm}
      \begin{tabular}{c c c}
	\includegraphics[width=60mm,height=70mm]{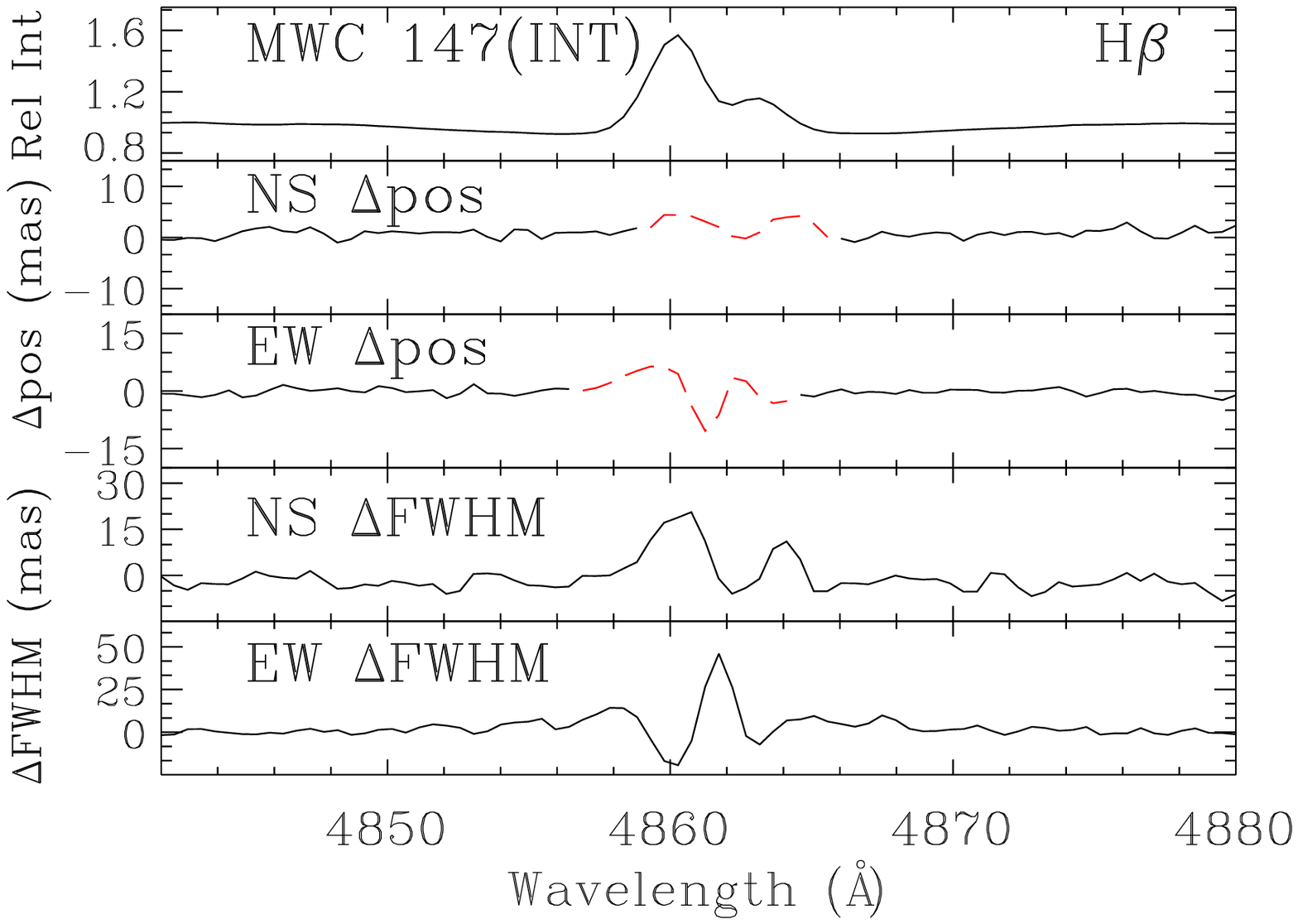} & 
	\includegraphics[width=60mm,height=70mm]{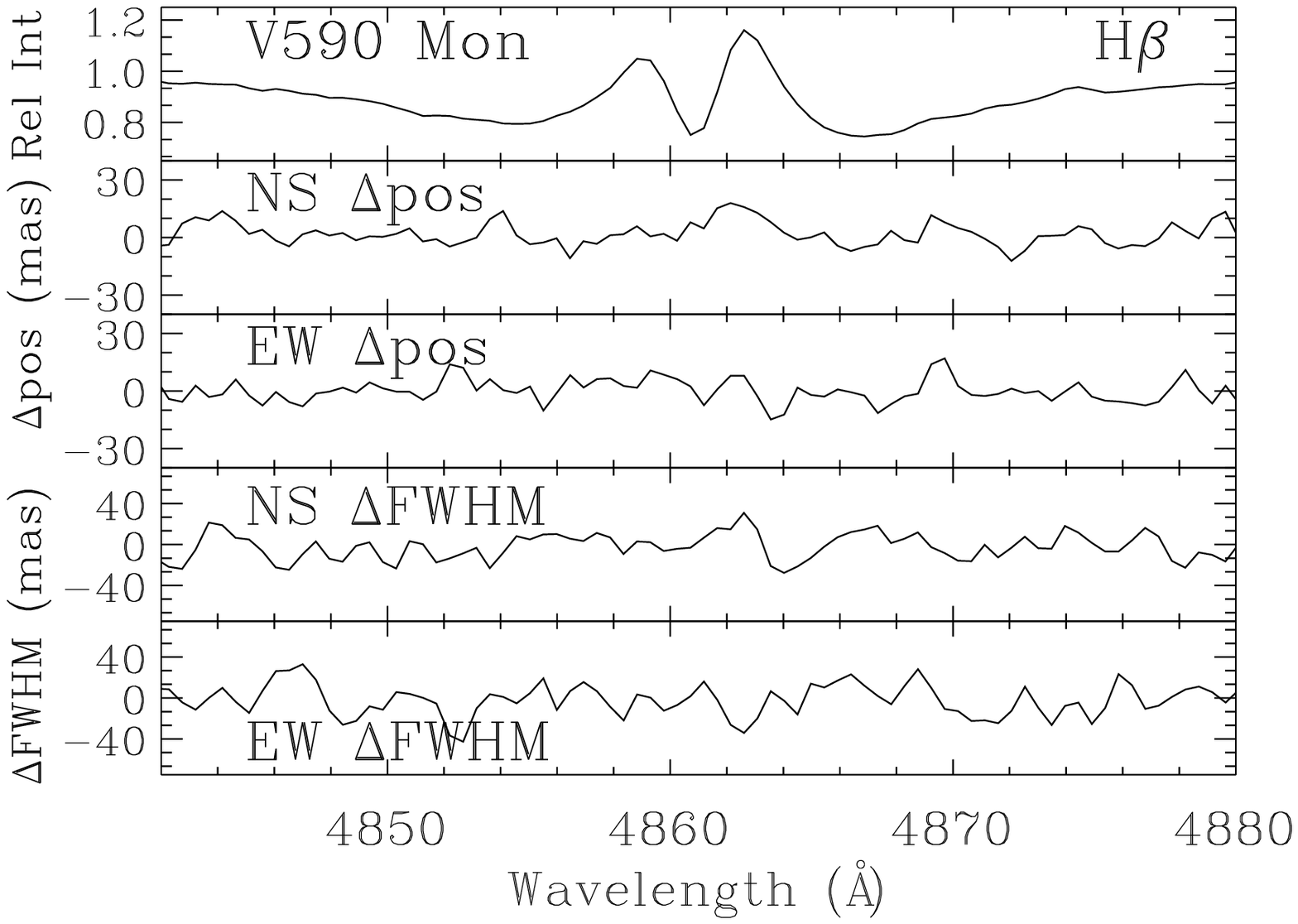} & 	\includegraphics[width=60mm,height=70mm]{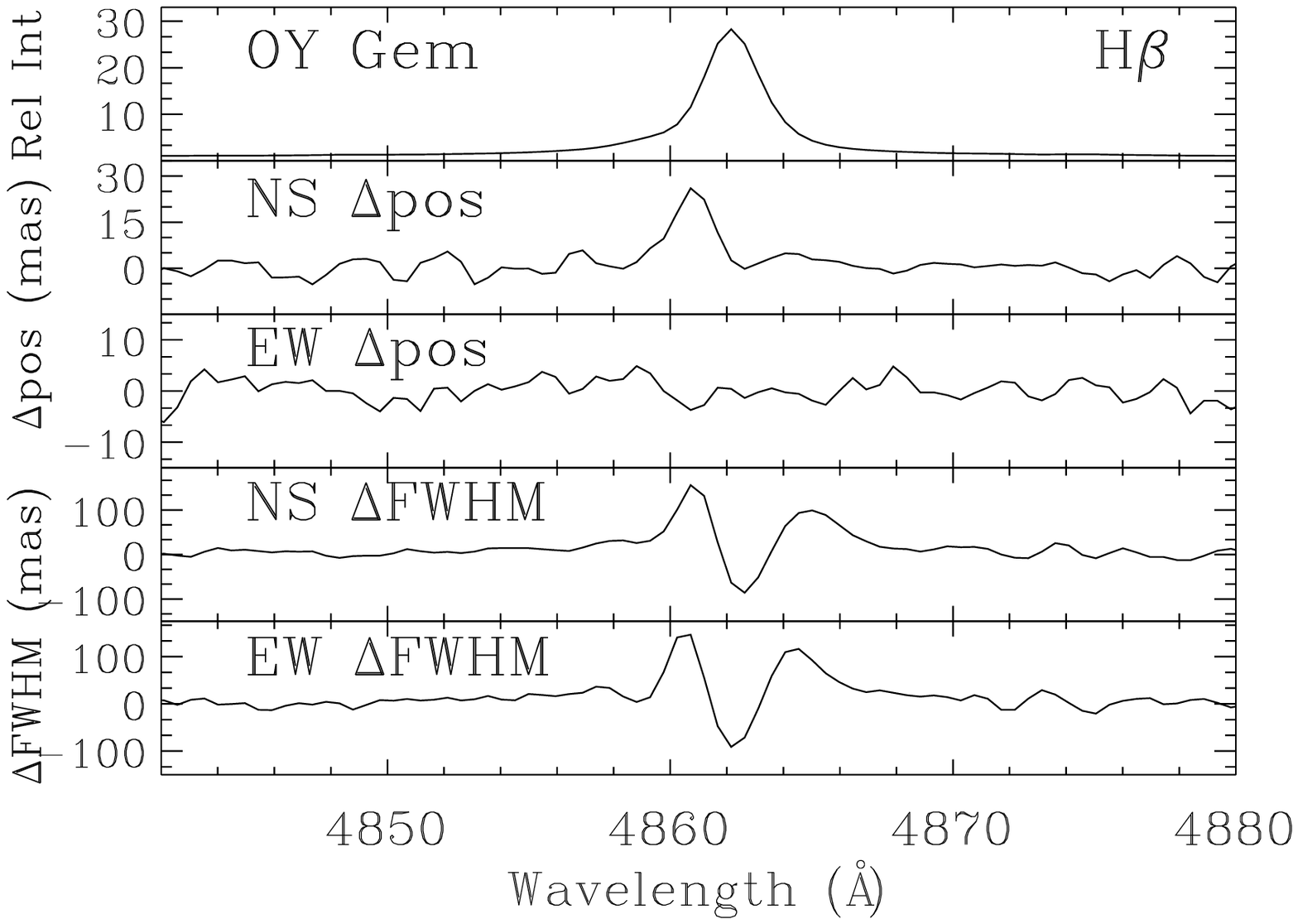}\\ 

      	\includegraphics[width=60mm,height=70mm]{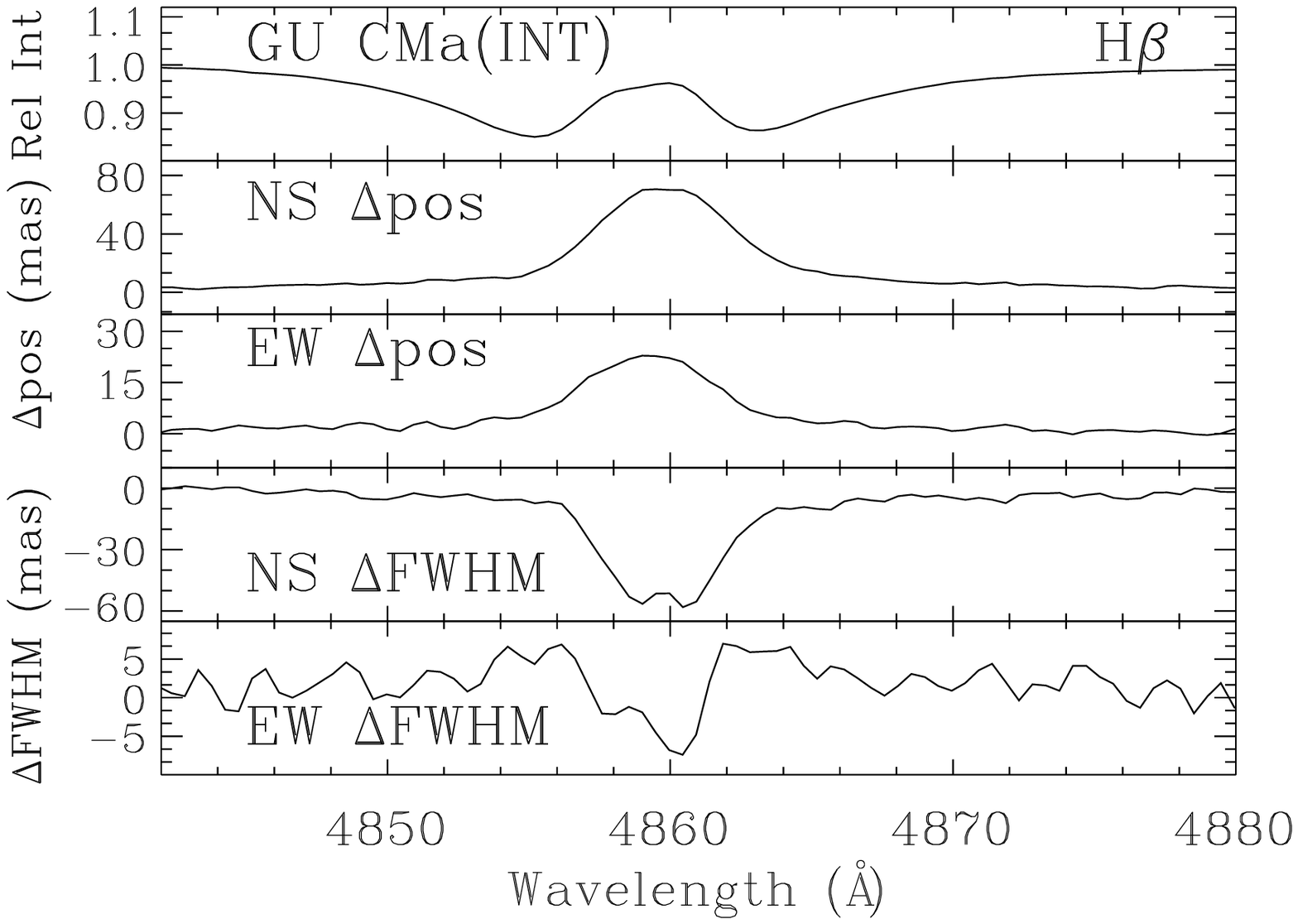} & 
	\includegraphics[width=60mm,height=70mm]{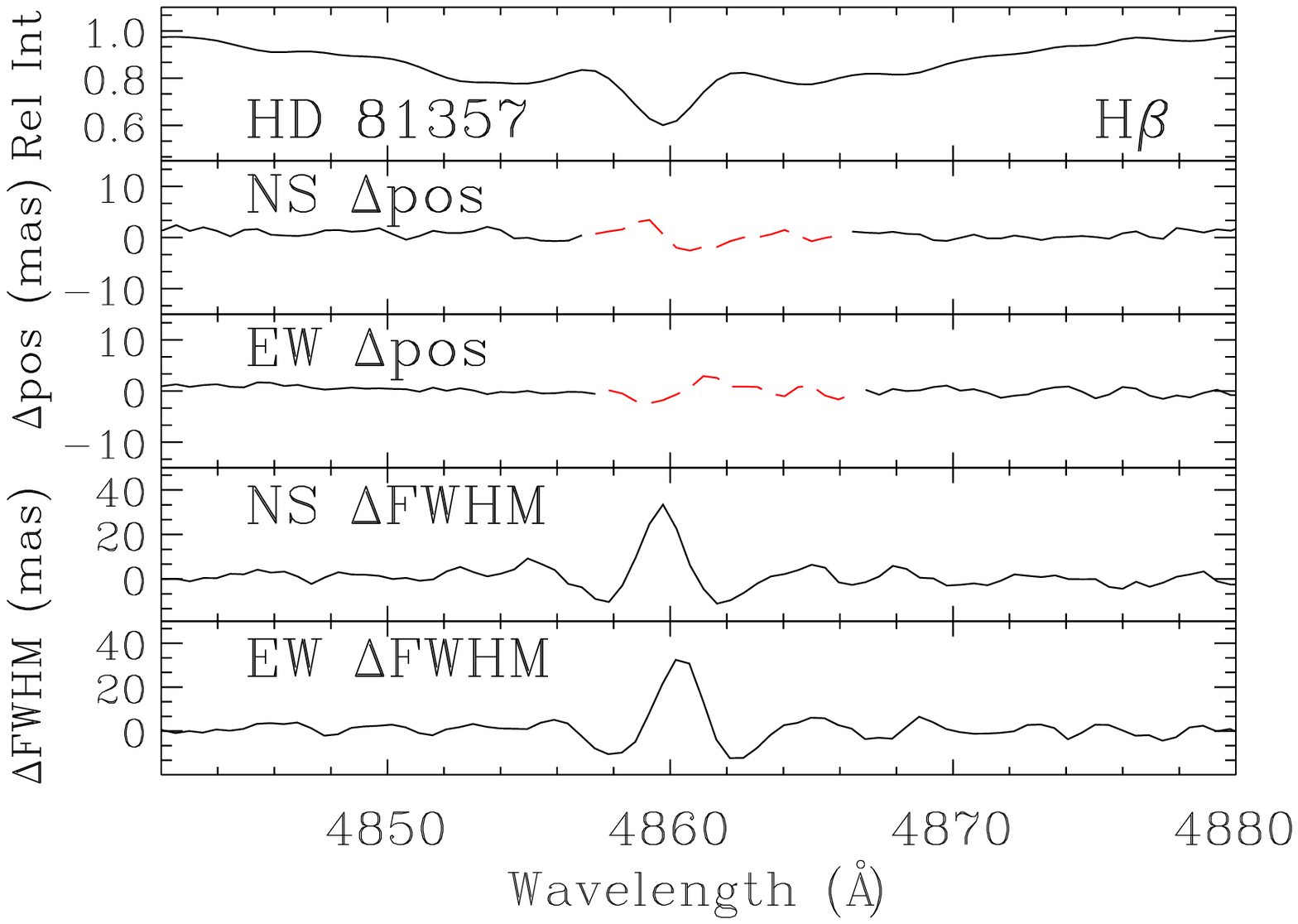} &
        \includegraphics[width=60mm,height=70mm]{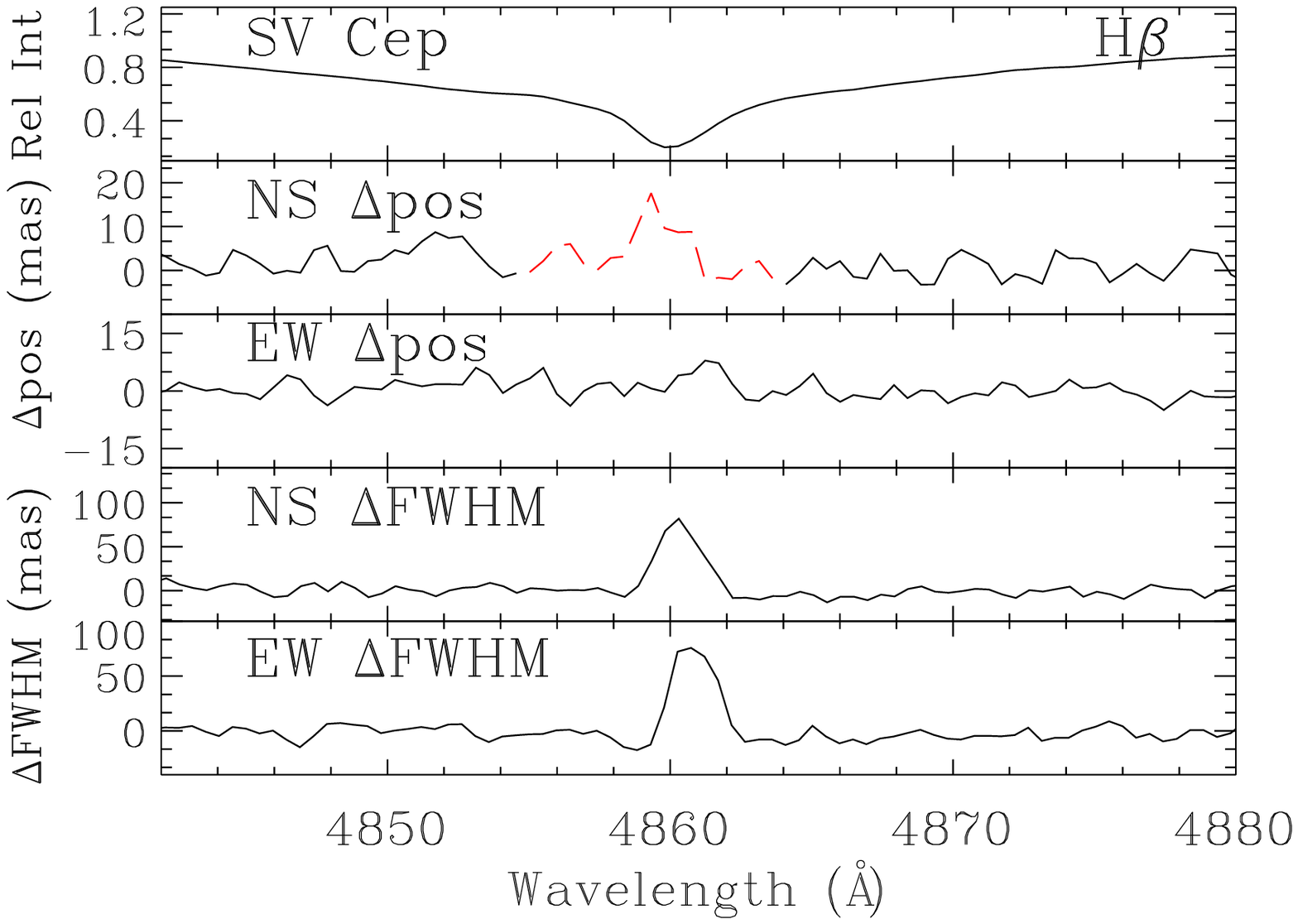} \\
\end{tabular}

\begin{tabular}{c c}

        \includegraphics[width=60mm,height=70mm]{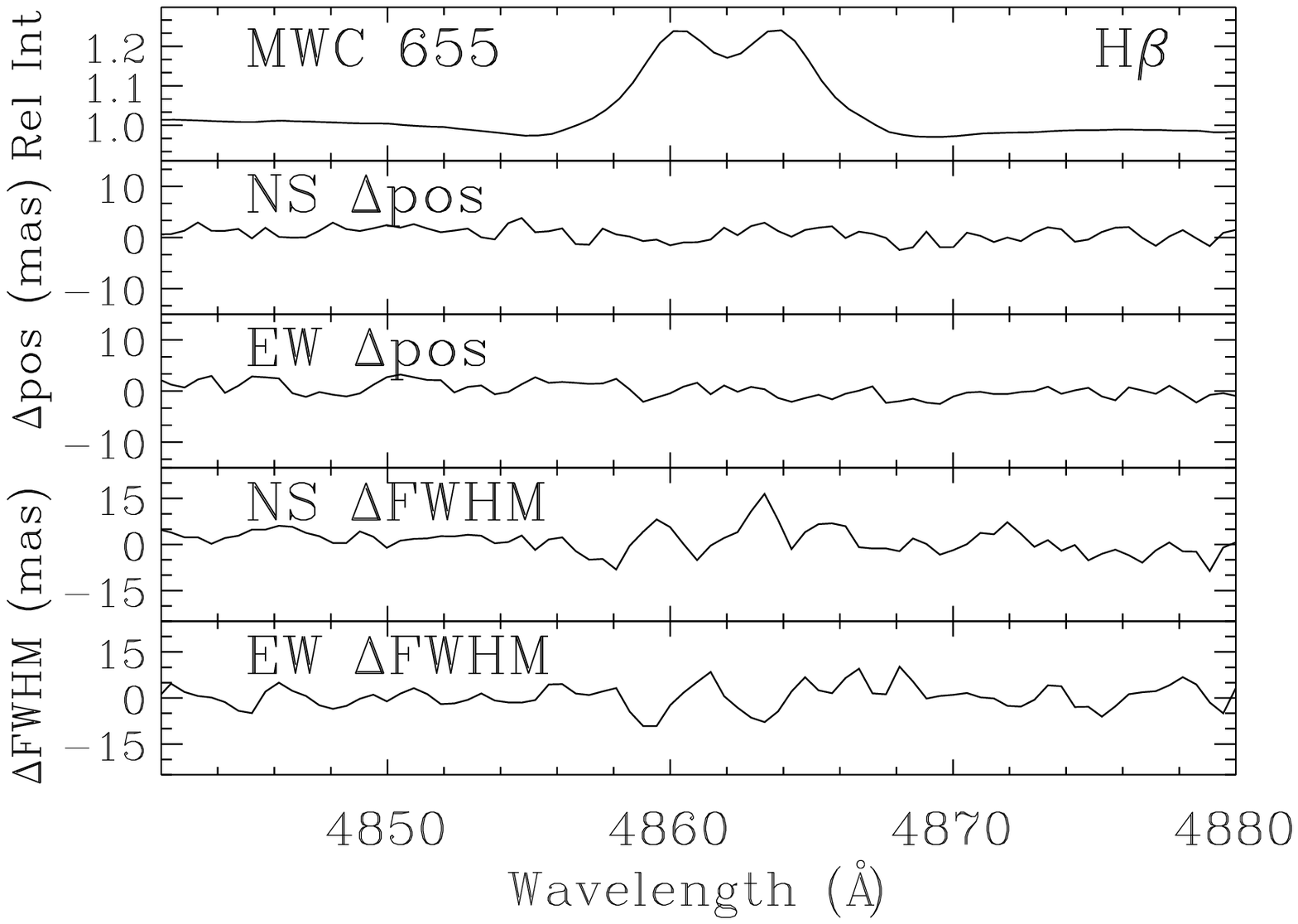} &
        \includegraphics[width=60mm,height=70mm]{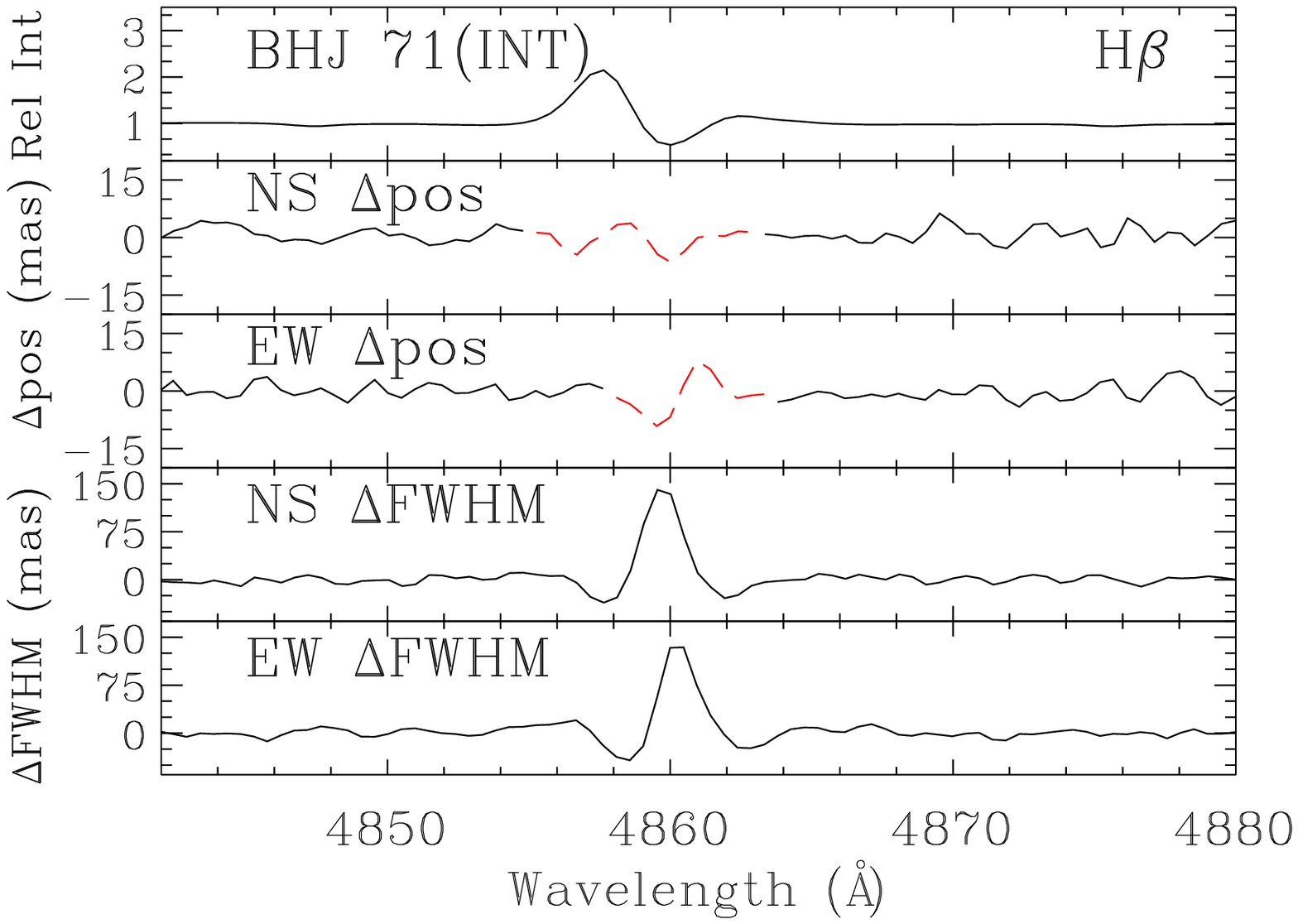} \\           
\end{tabular}
    
    \caption{H$\rm \beta$ profiles and spectroastrometric signatures. From \textit{left} to \textit{right}: MWC 147 (data from the INT), V590 Mon, OY Gem, GU CMa (data from the INT), HD 81357, SV Cep, MWC 655 and BHJ 71 (data from the INT).}
\label{spec_ast_fig}
  \end{figure*}

\addtocounter{figure}{-1}

\label{lastpage}
\end{document}